\documentclass[aps,prx,reprint,10pt,twocolumn,citeautoscript,superscriptaddress]{revtex4-2}
\pdfoutput=1

\usepackage{natbib}
\usepackage[english]{babel}
\usepackage{letltxmacro}
\usepackage{latexsym}
\usepackage[caption=false]{subfig}
\LetLtxMacro{\ORIGselectlanguage}{\selectlanguage}
\makeatletter
\DeclareRobustCommand{\selectlanguage}[1]{%
  \@ifundefined{alias@\string#1}
    {\ORIGselectlanguage{#1}}
    {\begingroup\edef\x{\endgroup
       \noexpand\ORIGselectlanguage{\@nameuse{alias@#1}}}\x}%
}
\newcommand{\definelanguagealias}[2]{%
  \@namedef{alias@#1}{#2}%
}
\makeatother
\definelanguagealias{en}{english}
\definelanguagealias{English}{english}
\usepackage{graphicx}
\usepackage{amsmath}
\usepackage{amsfonts}
\usepackage{amssymb}
\usepackage{bm}
\usepackage{color}
\usepackage[percent]{overpic}
\usepackage{soul} 
\usepackage{amssymb}
\usepackage{wasysym}
\usepackage{dsfont}
\usepackage{float}
\usepackage[mathscr]{euscript}
\usepackage{lipsum}
\usepackage{hyperref}
\hypersetup{
    pdfstartview={FitH},    
    colorlinks=true,       
    linkcolor=blue,          
    citecolor=blue,        
    filecolor=magenta,      
    urlcolor=blue           
}

\newcommand{\be}{\begin{equation}}
\newcommand{\ee}{\end{equation}}
\newcommand{\bea}{\begin{eqnarray}}
\newcommand{\eea}{\end{eqnarray}}

\usepackage{graphicx}
\usepackage[colorinlistoftodos]{todonotes}
\usepackage{verbatim}
\usepackage[normalem]{ulem}
\usepackage{enumitem}
\usepackage{multirow}

\begin{document}
\title{Correlations in interacting electron liquids: Many-body statistics and hyperuniformity}

\begin{abstract}
   Disordered hyperuniform many-body systems are exotic states of matter with novel optical, transport, and mechanical properties. 
   These systems are characterized by an anomalous suppression of large-scale density fluctuations compared to typical liquids, i.e., the structure factor obeys the scaling relation $S(k)\sim \mathcal{B}k^\alpha$ with $\mathcal{B}, \alpha>0$ in the limit $k$\,$\rightarrow$\,$ 0$. 
   Ground-state $d$-dimensional free fermionic gases, which are fundamental models for many metals and semiconductors, are key examples of \textit{quantum} disordered hyperuniform states with important connections to random matrix theory.
   However, the effects of electron-electron interactions as well as the polarization of the electron liquid on hyperuniformity have not been explored thus far. 
   In this work, we systematically address these questions by deriving the analytical small-$k$ behaviors (and associatedly, $\alpha$ and $\mathcal{B}$) of the total and spin-resolved structure factors of quasi-1D, 2D, and 3D electron liquids for varying polarizations and interaction parameters. 
   We validate that these equilibrium disordered ground states are hyperuniform, as dictated by the fluctuation-compressibility relation.
   Interestingly, free fermions, partially polarized interacting fermions, and fully polarized interacting fermions are characterized by different values of the small-$k$ scaling exponent $\alpha$ and coefficient $\mathcal{B}$. 
    In particular, partially polarized fermionic liquids exhibit a unique form of \textit{multihyperuniformity}, in which the net configuration exhibits a stronger form of hyperuniformity (i.e., larger $\alpha$) than each individual spin component. 
    The detailed theoretical analysis of such small-$k$ behaviors enables the construction of corresponding equilibrium classical systems under effective one- and two-body interactions that mimic the pair statistics of quantum electron liquids. 
    Our work thus reveals that highly unusual hyperuniform and multihyperuniform states can be achieved in simple fermionic systems and paves the way for harnessing the unique  hyperuniform scaling relations for applications, such as the construction of accurate density functionals.
\end{abstract}

\author{Haina Wang}
\affiliation{Department of Chemistry, Princeton University, Princeton,
NJ 08544}

\author{Rhine Samajdar}
\affiliation{Department of Physics, Princeton University, Princeton, NJ 08544}
\affiliation{Princeton Center for Theoretical Science, Princeton University, Princeton, NJ 08544}

\author{Salvatore Torquato}
\affiliation{Department of Chemistry, Princeton University, Princeton,
NJ 08544}
\affiliation{Department of Physics, Princeton University, Princeton, NJ 08544}
\affiliation{Princeton Materials Institute, Princeton University, Princeton, NJ 08540}
\affiliation{Program in Applied and Computational Mathematics, Princeton, NJ 08544}
\thanks{
Email:
\href{mailto:torquato@electron.princeton.edu}{torquato@electron.princeton.edu}}
\date{\today}
\maketitle
\newpage

\section{Introduction}
Large-scale density fluctuations in quantum and classical many-particle systems contain crucial structural and thermodynamic information and are of both fundamental and practical interest~\cite{Fi67, Ha73, Zi77, La80}.
In recent decades, much attention has been focused on the emerging field of disordered hyperuniform systems, which are exotic states of matter characterized by an anomalous suppression of large-scale density fluctuations compared to typical disordered systems, such as classical liquids~\cite{To03a, To18a}. Since disordered hyperuniform states combine the advantages of statistical isotropy and the suppression of density fluctuations on large scales as found in crystals, they are often endowed with novel optical~\cite{Fl09b, Ma13a, Kl22}, transport~\cite{To21d}, and mechanical~\cite{Do05d, Ma13a} properties.

A statistically homogeneous point configuration in $d$-dimensional Euclidean space $\mathbb{R}^d$ is hyperuniform if its structure factor~\cite{To03a, To18a} $S(\boldsymbol{k})$ vanishes as the wavenumber $k $\,$=$\,$ \lvert\boldsymbol{k}\rvert \rightarrow 0$.
In particular,
hyperuniform structure factors often obey a power-law scaling relation, i.e.,
\begin{equation}
    S(k) \sim \mathcal{B} \,k^\alpha, \quad \mathcal{B}, \alpha > 0.
    \label{scaling}
\end{equation}
The exponent $\alpha$ determines different large-$R$ scaling behaviors of the local number variance that characterizes three classes of hyperuniformity~\cite{To03a,Za09,To18a}: 
\begin{equation}
    \text{Var}(R)\sim \begin{cases}
          R^{d-1}, \quad &\alpha > 1 \text{ (class I)} \\
          R^{d-1}\ln R, \quad &\alpha=1 \text{ (class II)} \\
          R^{d-\alpha}, \quad &0<\alpha<1 \text{ (class III)}\\
     \end{cases}.
    \label{alpha}
\end{equation}
Classes I and III are the strongest and weakest forms of hyperuniformity, respectively.

While disordered hyperuniformity is observed in a large variety of classical systems, including one-component plasmas~\cite{Le00}, perfect glasses~\cite{Zh16a}, random organization models~\cite{He15, Co08}, maximally random jammed states~\cite{To00b} and biological congregates~\cite{Ji14, Hu21}, relatively less is known for hyperuniformity in quantum states.
Key among examples of quantum disordered hyperuniform states are ground states of $d$-dimensional  free fermionic gases~\cite{Wi34}.
Despite its simple nature, the Fermi gas model provides an excellent theoretical framework to describe electrons in metals and semiconductors~\cite{Le02, Go91, An97}.
The structure factor for a three-dimensional (3D) polarized free Fermi gas obeys the small-$k$ scaling behavior $S(k)$\,$\sim$\,$3k/(4k_F)$, where $k_F$ is the radius of the Fermi sphere~\cite{Wi34, Fe98}, implying that the system is class II hyperuniform.
Other notable hyperuniform quantum states include superfluid helium-4 (class II)~\cite{Fe56}, the 2D Laughlin states (class I), which are intimately connected to random matrices via the Ginibre ensemble~\cite{Gi65, Ro18}, and electrons in higher Landau levels (class I), the correlations of which are described by Weyl-Heisenberg ensembles~\cite{Ab17}.

The many-body statistics of Fermi gases are closely related to the eigenvalues of random Hermitian matrices and zeros of the Riemann zeta function~\cite{Dy62a, Mon73, Od87, Me91, To08b}. In one dimension, the suitably normalized pair correlation function $g_2(r)$ for the ground state of a polarized free Fermi gas with density $
\rho$ is given by~\cite{note8}
\begin{equation}
    g_2(r) = 1 - \left(\frac{\sin(k_F r)}{k_F r}\right)^2,
    \label{g2_1D_pol_free}
\end{equation}
which is the analogue of the well-known 3D result derived by Feynman~\cite{Fe98}, where $k_F$ is related to $\rho$ by~\cite{As76} 
\begin{equation}
    k_F \equiv 2 \sqrt{\pi } \left[\frac{\rho\, \Gamma \left({\frac{d}{2}}+1\right)}{2}\right]^{1/d}.
    \label{kF}
\end{equation}
The structure factor corresponding to Eq.~(\ref{g2_1D_pol_free}) is 
\begin{equation}
    S(k) = 
    \begin{cases}
    k/(2k_F), \quad k < 2k_F,\\
     1, \quad k \geq 2k_F.
    \end{cases}
    \label{s_1D_pol_free}
\end{equation}
Going beyond $d=1$, \citet{To08b} showed that the $d$-dimensional Fermi gas can be mapped to a determinantal point process, and for $d\geq 2$, this Fermi-sphere point process cannot be written as a Boltzmann factor of $N$ classical particles interacting  through only one- and two-body potentials at a finite temperature, implying that intrinsic $n$-particle interactions with $n\ge 3$ are generally necessary for describing such quantum systems. 
\footnotetext[8]{Dyson showed that the pair correlation function of the eigenvalues of the Gaussian Unitary Ensemble (GUE) is given by Eq.~(\ref{g2_1D_pol_free}). 
The GUE eigenvalues are mappable to a classical many-body system at positive temperature $T>0$ in which particles on a unit circle interact under a logarithmic repulsive potential~\cite{Dy62a}.}

The pair statistics $g_2(r)$ and $S(k)$ for fermionic systems are related to important thermodynamic properties, including the magnetic susceptibility, the ground state energy, and the isothermal compressibility~\cite{Pi66, Fe98}.
To model realistic metals and semiconductors, it is crucial to accurately ascertain pair statistics of fermionic systems as a function of the Wigner-Seitz radius $r_s$ and the polarization $\zeta$.
In practice, there exist numerous models that approximate interacting fermionic systems (thereby predicting pair statistics for different $r_s$ and $\zeta$), such as the random-phase approximation (RPA)~\cite{Bo51, Pi66}, the Hubbard approximation~\cite{Hu57}, and the Singwi--Tosi--Land--Sj{\"o}lander (STLS) model~\cite{Si68, Go93}. 
However, the small-$k$ behaviors of the structure factor have not been systematically and analytically explored across different values of $d, r_s$ and $\zeta$, nor has any work studied interacting fermions through the lens of hyperuniformity.
Therefore, it is highly desirable to theoretically analyze the small-$k$ scaling behaviors (\ref{scaling}) for $d$-dimensional free and interacting fermionic systems.
To our knowledge, explicit closed-form expressions for the small-$k$ behaviors for spin-resolved structure factors have only been derived for 3D fermionic systems within the RPA scheme~\cite{Wa91, Da03}.

In this work, we use both the RPA and the Hubbard approximation to derive analytical small-$k$ scaling behaviors of the total and spin-resolved structure factors for statistically homogeneous ground states of quasi-1D~\cite{Ta96, Ga08}, 2D, and 3D  electron liquids with varying interaction parameters and polarizations. 
Such small-$k$ expressions obtained via the Hubbard approximation are found to accurately match the numerical structure factors obtained from the STLS model~\cite{Si68, Lo69}, which is known to achieve excellent agreement with quantum Monte Carlo results for certain interaction parameters~\cite{Go93}.
The fact that these equilibrium disordered ground states are hyperuniform is dictated by the fluctuation-compressibility (FC) relation applied at zero temperature~\cite{To18a}.

Importantly, for the individual spin components, we find that free fermions, partially polarized interacting fermions and fully polarized interacting fermions are characterized by different hyperuniform scaling exponents $\alpha$ and coefficients $\mathcal{B}$. 
The free Fermi gases in one, two, and three dimensions are class-II hyperuniform ($\alpha = 1$) for both the total configuration and the individual spin components. 
On the other hand, fully polarized interacting fermions form class-I hyperuniform states with $\alpha = (d + 1)/2$ for $d=2,3$ and $S(k)\sim k/\sqrt{-\ln(ka)}$ for $d = 1$, where $a$ is a microscopic length scale.
Finally, we find that partially polarized interacting fermionic liquids exhibit \textit{multihyperuniformity}, in which each spin component as well as the net configuration is hyperuniform~\cite{Ji14}.
Remarkably, in these systems, the individual spin components are class-II hyperuniform, whereas the whole configuration exhibits a stronger class-I hyperuniformity.
These states provide the first example of multihyperuniform systems in which the total configuration is of a higher hyperuniformity class than its components.

Additionally, our detailed theoretical analyses of small-$k$ behaviors for fermionic systems enable us to construct equilibrium positive-temperature classical systems that precisely mimic the pair statistics of quantum electron liquids.
This is achieved by employing a variant of a recently developed inverse algorithm~\cite{To22, Wa24}, which accurately determines spin-specific effective one- and two-body potentials corresponding to the targeted fermionic pair statistics.
We show that the aforementioned unusual multihyperuniformity of partially polarized interacting fermions can be classically realized by classical many-particle systems at positive temperatures in which the particles interact with certain long-ranged pair potentials.
This quantum-classical mapping via inverse statistical mechanics also allows us to study the structural degeneracy problem~\cite{St19} for the pair statistics of free Fermi gases.
In contrast to the 1D free Fermi gas, which is exactly mappable to a classical equilibrium state under logarithmic interactions~\cite{Dy62a}, we find that the classical states corresponding to 2D and 3D free fermions possess distinct higher-order statistics compared to their quantum counterparts, confirming the predictions in Ref.~\cite{To08a} that higher-order intrinsic interactions are required to describe fermionic structural correlations in two and higher dimensions.
Our work thus reveals that interesting hyperuniform and multihyperuniform states can be achieved in simple quantum and classical systems and paves the way for harnessing the unique hyperuniform small-$k$ scaling relations for applications, such as the development of accurate density functionals, materials engineering, and computer vision.

This paper is organized as follows. Section \ref{def} outline some basic definitions and background, including those for pair statistics and models of fermionic systems.
Section \ref{res} then provides results for hyperuniform behaviors in free and interacting fermionic systems.
In Sec.~\ref{class}, we study classical equilibrium systems corresponding to free Fermi gases.
Lastly, we present our concluding remarks in Sec.~\ref{conc}.

\section{Definitions and preliminaries}
\label{def}
In this section, we introduce some fundamental relations for ground states of fermionic systems, as well as definitions of pair statistics.
We then describe some well-known models used in this work to determine the pair statistics of electron liquids, namely, the RPA~\cite{Bo51, Pi66}, Hubbard~\cite{Hu57} and STLS models~\cite{Si68, Lo69}.

\subsection{Fundamental relations for fermionic liquids}
We consider a $d$-dimensional fermionic system with number density $\rho$, embedded in a uniform neutralizing background; each particle has a mass $m$, charge $e$ and spin $\sigma$\,$=$\,$\pm 1$. 
Let $\rho_\uparrow$ and $\rho_\downarrow$ be the number densities of up and down spins, respectively.
The degree of spin polarization is defined as $\zeta \equiv |\rho_\uparrow - \rho_\downarrow|/\rho$. Note that $\zeta = 1$ for fully polarized fermions and $\zeta = 0$ for unpolarized fermions.
The Fermi wavenumber $k_F$ is given by Eq. (\ref{kF}), and the spin-specific Fermi wavenumber $k_{F\sigma}$ is given by
and
\begin{equation}
    k_{F\sigma} \equiv k_F(1 + \sigma\zeta)^{1/d}.
    \label{kFsigma}
\end{equation}

For a system of fermions interacting pairwise  via a repulsive long-ranged potential $V(r)$, a natural length scale is the radius of the $d$-dimensional sphere that encloses, on average, exactly
one particle. This typical distance between two fermions can be written as
\begin{equation}
    r_sa_B = [v_1(1) \rho]^{-1/d},
\end{equation}
where $ r_s$ is a dimensionless number, $a_B \equiv \hbar^2/(m e^2)$ is the Bohr radius, and 
\begin{equation}
v_1(r) \equiv \frac{\pi^{d/2} r^d}{\Gamma(1+d/2)}
\end{equation}
is the volume of a $d$-dimensional sphere of radius $r$. 
The parameter $r_s$ represents the ratio of the average
potential energy to the average kinetic energy of the system, and for free fermions, $r_s = 0$. As a function of $r_s$, the fermionic liquid exhibits two distinct regimes: the weak-coupling regime (for low $r_s$ values), where the kinetic energy dictates the system's behavior and the system resembles a noninteracting gas; and the strong-coupling (large $r_s$) regime, where the potential energy prevails, resulting in collective behavior akin to that of a crystal.

In this work, we consider electrons in $\mathbb{R}^d$ interacting via a Coulomb potential
\begin{equation}
    V(r) = \frac{e^2}{r},
\end{equation}
whose Fourier transforms, in the first three spatial dimensions, are given by
\begin{equation}
    \tilde{V}(k) = e^2 \times
\begin{cases}
    -2 (\gamma + \ln{k a}), \quad &d = 1\\
    {\displaystyle\frac{2\pi}{k}}, \quad &d = 2\smallskip\\
    {\displaystyle\frac{4\pi}{k^2}}, \quad &d = 3
\end{cases},
\label{vq}
\end{equation}
where $\gamma$ is the Euler-Mascheroni constant and $a$ is some microscopic length scale 
set by the diameter of the one-dimensional channel.

The case of $d$\,$=$\,$1$ is rather special: the ground state, for any strength of the interaction, is a Tomonaga-Luttinger liquid~\cite{tomonaga1950remarks,luttinger1963exactly} characterized by collective bosonic low-energy excitations (unlike the Fermi-liquid ground states in two and three spatial dimensions)~\cite{sarma1985screening,li1992elementary,schulz1993wigner,iucci2000exact,wang2001coulomb}. Moreover,
it has been noted \cite{Ta96} that the unboundedness of the bare 1D potential (which corresponds to the case of an infinintely thin wire) $\tilde{V}(k)/e^2 \sim -2(\ln ka)$ as $k\rightarrow \infty$ causes the structure factors to diverge at large $k$ in the models considered in Sec.~\ref{sec:models}.
Thus, it is more practical to consider a \textit{quasi}-1D system, in which electrons are allowed to move in a wire with a small but positive cross section, such that the $1/r$ repulsive interaction between electrons is cut off at short distances by the transverse width of the lowest subband wavefunction~\cite{creffield2001spin,fogler2005ground,casula2006ground,shulenburger2008correlation}. This approximation retains the long-ranged nature of the Coulomb interaction but
$\tilde{V}(k)$ is now bounded at large $k$. Theoretically, several confinement models have been employed to describe the quasi-1D system, including hard wall~\cite{gold1990analytical,hu1993many}, Coulombic~\cite{sun1993density} or harmonic~\cite{friesen1980dielectric,Hu90} potentials.
In this work, we consider a commonly used quasi-1D approximation in which the electrons are laterally confined by a harmonic transversal potential. 
If the confinement is sufficiently strong, so that the electrons occupy only the lowest energy subband in the transverse direction, the corresponding potential is given by~\cite{Hu90, Ta96}
\begin{equation}
    \frac{\tilde{V}(k)}{e^2} = \exp \left(\frac{k^2 \ell^2 }{4}\right) K_0\left(\frac{k^2 \ell^2 }{4}\right),
    \label{Vq1D}
    \end{equation}
where $\ell>0$ is a parameter indicating the lateral width of the quantum wire, and $K_0(x)$ is the zeroth-order Bessel-$K$ function. We note that while other forms of the confining harmonic potential have also been suggested~\cite{casula2006ground,Ga08b}, at small $k$, all the models behave as $\tilde{V}(k)\sim -2\ln(k\ell)$ independent of the microscopic details \cite{li1991elementary}.

In what follows, we will work in units where $\hbar = e =1$. For notational convenience, we also introduce a dimensionless wavevector as $q = k/k_F$.

\subsection{Pair statistics}
A many-particle system in $\mathbb{R}^d$ is completely statistically characterized by the $n$-particle probability density functions $\rho_n(\boldsymbol{r}_1,...,\boldsymbol{r}_n)$ for all $n\geq 1$~\cite{Ha86}. We define the pair correlation function as
\begin{equation}
    g_2(\boldsymbol{r}_1,\boldsymbol{r}_2) = \frac{1}{\rho_1(\boldsymbol{r}_1)\rho_1(\boldsymbol{r}_2)} \left \langle \sum_{i\ne j} \delta (\boldsymbol{r}_1-\boldsymbol{r}_i) \delta (\boldsymbol{r}_2-\boldsymbol{r}_j)\right \rangle.
\end{equation}
In the case of statistically homogeneous systems, $\rho_1(\boldsymbol{r})=\rho$, and $\rho_2(\boldsymbol{r}_1,\boldsymbol{r}_2)=\rho^2 g_2(\boldsymbol{r})$, where $g_2(\boldsymbol{r}_1,\boldsymbol{r}_2) = g_2(\boldsymbol{r}_1-\boldsymbol{r}_2,0) \equiv g_2(\boldsymbol{r})$.
If the system is also statistically isotropic, then $g_2(\boldsymbol{r})$ is simply a radial function $g_2(r)$, with $r\equiv|\boldsymbol{r}|$. 
The ensemble-averaged structure factor $S(\boldsymbol{k})$ is defined as
	\begin{equation}
		S(\boldsymbol{k})=1+\rho \tilde{h}(\boldsymbol{k}),
		\label{skdef}
	\end{equation}
where $h(\boldsymbol{r})=g_2(\boldsymbol{r})-1$ is the total correlation function, and $\tilde{h}(\boldsymbol{k})$ is the Fourier transform of $h(\boldsymbol{r})$. 
For statistically isotropic systems, the structure factor is radially symmetric and can be written as $S(k)$ with $k \equiv |\boldsymbol{k}|$. If $S(k)$ is analytic at the origin (i.e., the Taylor expansion of $S(k) = \sum_{i = 0}^\infty s_i k^i$ at $k = 0$ contains only even powers in $k$), then the total correlation function $h(r) = g_2(r) - 1$ decays exponentially or superexponentially fast at large $r$. 
On the other hand, if $S(k)$ is nonanalytic at the origin, then $h(r)$ decays as a power law dictated by the  leading-order term in the expansion about $k=0$ with an odd integer or real noninteger power~\cite{To18a, To22}.

For a single periodic configuration containing $N$ point particles at positions ${\boldsymbol{r}}_1,{\boldsymbol{r}}_2,\ldots,{\boldsymbol{r}}_N$ within a fundamental cell $F$ of a lattice $\Lambda$, the {\it scattering intensity} $\mathcal{I}(\boldsymbol{k})$  is defined as
	\begin{equation}
		\mathcal{I}(\boldsymbol{k})=\frac{1}{N}\left|\sum_{i=1}^{N}e^{-i\boldsymbol{k}\cdot\boldsymbol{r}_i}\right|^2.
		\label{scattering}
	\end{equation}
Now, for an ensemble of periodic configurations of $N$ particles within the fundamental cell $F$, the ensemble average of the scattering intensity in the thermodynamic limit, is directly related to the structure factor $S(\boldsymbol{k})$ (\ref{skdef}) by
	\begin{equation}
        \lim_{\substack{N,V_F\rightarrow+\infty,\\ \text{constant }\rho}}\langle\mathcal{I}(\boldsymbol{k})\rangle=(2\pi)^d\rho\delta(\boldsymbol{k})+S(\boldsymbol{k}),
	\end{equation}
where $V_F$ is the volume of the fundamental cell~\cite{To18a}. 
In simulations of many-body systems with finite $N$ and periodic boundary conditions, Eq.~(\ref{scattering}) is used to compute $S(\boldsymbol{k})$ directly by averaging over configurations.

Since the electron liquid is, in general, a two-component system consisting of both up and down spins, it is crucial to study the spin-resolved pair correlation function $g_{2, \sigma\sigma'}(r)$, defined such that $s_1(r)\rho_{\sigma'}g_{2,\sigma\sigma'}(r)dr$ gives the average number of particles with spin $\sigma'$ within a spherical shell of volume $s_1(r)dr$ centered at a particle with spin $\sigma$, where
\begin{equation}
    s_1(r) = \frac{2\pi^{d/2}r^{d-1}}{\Gamma(d/2)}
\end{equation}
is the surface area of a $d$-dimensional sphere of radius $r$. 
The corresponding spin-resolved structure factors are given by 
\begin{equation}
    S_{\sigma\sigma'}(k) = \delta_{\sigma\sigma'} + \sqrt{\rho_\sigma\rho_{\sigma'}}\tilde{h}_{\sigma\sigma'}(k),
    \label{g2-S}
\end{equation}
where $\tilde{h}_{\sigma\sigma'}(k)$ is the Fourier transform of the spin-resolved total correlation function $h_{\sigma\sigma'}(r) = g_{2,\sigma\sigma'}(r) - 1$. 
The total (i.e., spin-unresolved) pair statistics are related to the spin-resolved ones via~\cite{Ng87, Da03, Ku09}
\begin{widetext}
\begin{alignat}{2}
    g^{}_2(r) &= \sum_{\sigma, \sigma'} \frac{\rho_\sigma\rho_{\sigma'}}{\rho^2} g^{}_{2,\sigma\sigma'}(r)&&= \left(\frac{1 + \zeta}{2}\right)^2g^{}_{2,\uparrow\uparrow}(r) + \left(\frac{1 - \zeta}{2}\right)^2g^{}_{2,\downarrow\downarrow}(r) + \frac{1 - \zeta^2}{2}g^{}_{2,\uparrow\downarrow}(r),\\
    S(k) &= \sum_{\sigma, \sigma'} \frac{\sqrt{\rho_\sigma\rho_{\sigma'}}}{\rho}S_{\sigma\sigma'}(k) 
    &&= \frac{1 + \zeta}{2}S_{\uparrow\uparrow}(k) + \frac{1 - \zeta}{2}S_{\downarrow\downarrow}(k) + \sqrt{1 - \zeta^2}S_{\uparrow\downarrow}(k).
    \label{Stotal_from_spinspec}
\end{alignat}
\end{widetext}
It is also useful to define the \textit{magnetic structure factor} corresponding to fluctuations of the spin density:
\begin{alignat}{1}
    \label{tildeS-spinspec}
    \tilde{S}(\boldsymbol{k}) &= \frac{|\sum_{j}^N \sigma_i\exp(-i\boldsymbol{k}\cdot\boldsymbol{r}_i)|^2}{N}\\ 
    \nonumber
    &= \frac{1 + \zeta}{2}S_{\uparrow\uparrow}(k) + \frac{1 - \zeta}{2}S_{\downarrow\downarrow}(k) - \sqrt{1 - \zeta^2}S_{\uparrow\downarrow}(k),
\end{alignat}
where $\sigma_i$ is the spin of fermion $i$ along the quantization axis, and we have assumed a statistically isotropic system in the second line above.
It is clear from Eqs.~(\ref{Stotal_from_spinspec}) and (\ref{tildeS-spinspec}) that for an unpolarized electron liquid ($\zeta = 0$), the spin-specific structure factors are given by~\cite{Lo69}
\begin{alignat}{1}
    S_{\uparrow\uparrow}(k) &= S_{\downarrow\downarrow}(k) =  \frac{S(k) + \tilde{S}(k)}{2},
    \label{Supup_unp}\\
    S_{\uparrow\downarrow}(k) &= \frac{S(k) - \Tilde{S}(k)}{2}.
    \label{Supdown_unp}
\end{alignat}

Evidently, for a free Fermi gas ($r_s$\,$=$\,$0$), the oppositely aligned spins are uncorrelated, and the corresponding pair statistics are simply $g_{2,\uparrow\downarrow}^0(r) = 1$ and $S_{\uparrow\downarrow}^0(k) = 0$. 
Owing to the correspondence between free Fermi gases and a determinantal point process identified in Ref.~\cite{To08a}, the pair correlation function for parallel-spin free fermions in any dimension is given by~\cite{Wi34, giuliani2008quantum, To08b}
\begin{equation}
    g_{2, \sigma\sigma}^0(r) = 1 - 2^d\Gamma(1 + d/2)^2\frac{J^2_{d/2}(k_{F\sigma} r)}{(k_{F\sigma} r)^d},
    \label{g2_free}
\end{equation}
where $J_\nu(x)$ is the Bessel function of the first kind of order $\nu$ (note that in 1D, Eq.~(\ref{g2_free}) reduces to Eq.~(\ref{g2_1D_pol_free}) above).
The corresponding parallel-spin structure factor is
\begin{equation}
    S_{\sigma\sigma}^0(k) = 1 - \alpha(k; k_{F\sigma}),
    \label{S_free}
\end{equation}
where $\alpha(k; k_{F\sigma})$ is the volume common to two spherical windows of radius $k_{F\sigma}$ whose centers are separated by a distance $k$ divided by $v_1(k_{F\sigma})$. 
In the first three spatial dimensions, one therefore has
\begin{widetext}
\begin{equation}
    S_{\sigma\sigma}^0(k) =
    \begin{cases}
        {\displaystyle\frac{k}{2k_{F\sigma}}} \quad & k < 2k_{F\sigma} \qquad (d = 1)\smallskip\\
        {\displaystyle\frac{2}{\pi}\left(\arcsin{\frac{k}{2k_{F\sigma}}} + \frac{k}{2k_{F\sigma}}\sqrt{1-\frac{k^2}{4k_{F\sigma}^2}}\right)}, \quad & k < 2k_{F\sigma} \qquad (d = 2)\smallskip\\
        {\displaystyle\frac{3k}{4k_{F\sigma}} - \frac{k^3}{16 k_{F\sigma}^3}}, \quad & k < 2k_{F\sigma} \qquad (d = 3)\smallskip\\
        1, \quad & k \geq 2k_{F\sigma}
    \end{cases}.
\end{equation}
\end{widetext}

\subsection{Models of interacting electron liquids}
\label{sec:models}
Having established the requisite definitions, we now outline, in increasing order of complexity, the three models that we employ to study the structure factors of interacting unpolarized fermionic systems, i.e., $\zeta = 0$.
The generalization to arbitrary polarizations for the small-$k$ behaviors of $S(k)$ will be discussed in Sec.~\ref{smallk_analytic}.

For an accurate description of an interacting electron liquid, there are two effects that have to be taken into account~\cite{giuliani2008quantum}. First, due to the antisymmetric nature of the $N$-particle wavefunction, if an electron occupies a particular position, it prevents another with the same spin from occupying the same site. This exclusion constraint generally decreases the likelihood of finding a second electron with the same spin in proximity to the first one. This phenomenon, known as the ``exchange hole," persists even in an electron gas where interactions between electrons are absent. Secondly, the likelihood of discovering another electron within a certain distance from an electron being observed is additionally influenced by the Coulomb repulsion between electrons. This redistribution of the electron density is commonly referred to as a ``correlation hole''.

Mathematically, our starting point is the fluctuation-dissipation theorem, which, in equilibrium, relates $S(q)$ and $\tilde{S}(q)$ to the density and spin-density susceptibilities, $\chi^d(q, \omega)$ and $\chi^s(q, \omega)$, respectively~\cite{Pi66}:
\begin{alignat}{1}
    S(q) &= -\frac{1}{\rho\pi}\int_0^\infty \text{Im}[\chi^d(q, \omega)] d\omega,
    \label{S-chid}\\
    \tilde{S}(q) &= \frac{1}{\rho\pi g^2\mu_B^2}\int_0^\infty \text{Im}[\chi^s(q, \omega)]d\omega,
    \label{tildeS-chis}
\end{alignat}
where $g$ is the Land{\'e} factor and $\mu_B$ is the Bohr magneton.
The response functions appearing in the equations above are given by~\cite{Si68, Lo69}
\begin{alignat}{1}
    \chi^d(q, \omega) &= \frac{\chi_0(q,\omega)}{1 - \tilde{V}(q)\chi_0(q,\omega)[1 - G(q)]},
    \label{chid-G}\\
    \chi^s(q, \omega) &= -g^2\mu_B^2 \frac{\chi_0(q,\omega)}{1 - \tilde{V}(q)\chi_0(q,\omega)I(q)},
    \label{chis-I}
\end{alignat}
where $\chi_0(q, \omega)$ is the $d$-dimensional Lindhard function~\cite{Li54, Mi11}, and $G(q)$ and $I(q)$ are the so-called many-body local-field corrections (LFC) arising from the short-range Coulomb correlations and the exchange-correlation effects for the density and spin-density responses, respectively~\cite{Si68, Lo69, Ta96}. The differences between the various models used to describe the electron liquid lie precisely in their expressions for $G(q)$ and $I(q)$. 

The simplest description of an interacting electron gas is the random phase approximation (RPA), in which electrons respond to a time-dependent Hartree potential: this is the effective field
perceived by a classical test charge embedded in the electron gas~\cite{giuliani2008quantum}. Since RPA ignores the correlations between an electron and its surrounding medium, there is no LFC, and $G(q)=I(q)=0$. 
RPA is known to be exact in the limit of a dense gas, $r_s \rightarrow 0$; however, for finite $r_s$, it overestimates the exchange correlation energy to the point of producing negative (unphysical) values for the positive-definite pair correlation function. This serious shortcoming is remedied by the introduction of nonzero local field factors.
The most straightforward of such corrections is the Hubbard approximation, in which one accounts for the Pauli hole around electrons; this gives \cite{Si68, Ta96}
\begin{equation}
    G_{\text{Hubbard}}(q) = - I_{\text{Hubbard}}(q) = \frac{1}{2}\frac{\tilde{V}(\sqrt{q^2 + 1})}{\tilde{V}(q)}.
    \label{HubbardG}
\end{equation}
While this modification improves upon the RPA scheme, the Hubbard approximation still does not take into account the correlation hole.

The next advance beyond the RPA and Hubbard approximations is the scheme developed by Singwi, Tosi, Land, and Sj\"olander (STLS) \cite{Si68}.
The key observation here is that the local field factors introduced above allow us to calculate the response functions of an electron liquid. However, once we  know the linear response functions, we can employ them to determine the exchange-correlation hole, which, in turn, influences the local field factors and closes the loop~\cite{giuliani2008quantum}.
Hence, in the STLS approximation, $G(q)$ and $I(q)$ are solved iteratively via the self-consistent field relations~\cite{Si68, Lo69}.
\begin{alignat}{1}
    G_{\text{STLS}}(q) &= -\frac{1}{\rho (2
    \pi)^d}\int_{\mathbb{R}^d}\frac{\boldsymbol{q}\cdot \boldsymbol{q}'}{|\boldsymbol{q}||\boldsymbol{q}'|}[S(\boldsymbol{q}' - \boldsymbol{q}) - 1]\, d\boldsymbol{q}',
    \label{GSTLS}\\
    I_{\text{STLS}}(q) &= \frac{1}{\rho (2
    \pi)^d}\int_{\mathbb{R}^d}\frac{\boldsymbol{q}\cdot \boldsymbol{q}'}{|\boldsymbol{q}||\boldsymbol{q}'|}[\tilde{S}(\boldsymbol{q}' - \boldsymbol{q}) - 1]\, d\boldsymbol{q}',
    \label{ISTLS}
\end{alignat}
where $\boldsymbol{q}$ is any wavevector with norm $q$.
Equations (\ref{GSTLS}) and (\ref{ISTLS}) are then inserted back into Eqs.~(\ref{S-chid}) and (\ref{tildeS-chis}), respectively, to obtain $S(q)$ and $\tilde{S}(q)$ in the subsequent iteration. 
This process is repeated until convergence is reached.
Note that in the STLS model, $\lim_{q\rightarrow 0} G(q) = \lim_{q\rightarrow 0} I(q) = 0$, as is also the case for the RPA and Hubbard approximations.

\section{Hyperuniformity of interacting electrons}
\label{res}

\subsection{Total and magnetic structure factors}
\label{summary_all_d}
\subsubsection{Unpolarized fermions: RPA and Hubbard approximations}
\label{unp}
In this section, we present the leading-order small-$k$ asymptotics of $S(k)$ and $\tilde{S}(k)$ for statistically homogeneous ground states of quasi-1D, 2D, and 3D unpolarized interacting fermions ($r_s > 0, \zeta = 0$) in the RPA and Hubbard approximations.
The detailed derivations are documented in Appendix~\ref{sec:appA}, and the results are summarized in Table \ref{tab:small-k} in terms of the dimensionless wavenumber $q = k/k_F$, where $k_F = k_{F\uparrow} = k_{F\downarrow}$ for $\zeta = 0$.
It is evident that regardless of the model used, the  total structure factor is hyperuniform, i.e., $S(k) \rightarrow 0$ as $k\rightarrow 0$. 

This observation can easily be understood in terms of the fluctuation-compressibility (FC) theorem applied at zero temperature~\cite{To18a}.
For single-component homogeneous equilibrium systems of point-like particles, the FC relation states that fluctuations in particle
number are proportional to the isothermal compressibility~\cite{goldstein1951theory,Ha86}:
\begin{equation}
\rho k^{}_B T \kappa^{}_T= S(0),
\label{FC}
\end{equation} 
Thus, in thermal equilibrium, any ground state ($T = 0$), ordered or disordered, for which the isothermal compressibility  $\kappa_T$ is bounded and positive must be hyperuniform~\cite{To18a}. 
More generally, the ground state of a single-component system is hyperuniform if the limit $\lim_{T\rightarrow 0}\kappa_T T$ vanishes~\cite{To08a, cherny2002sum}.
For quantum many-body systems, both ordered and disordered ground states are common occurrences, so it is not especially unusual to find disordered hyperuniform ground states of generic Hamiltonians, such as, e.g., superfluid helium-4 \cite{Fe56, Re67}.
In contrast, classical ground states almost invariably have high crystallographic symmetries \cite{We81, Fr96, Tr14}, except for some special interactions that result in disordered ground states~\cite{To18a}.
Note that while the FC relation holds for quantum electron liquids at all densities~\cite{cherny2002sum,dunkel2014consistent}, for certain quantum systems, its underlying assumptions may not be satisfied. Such violations of the FC relation occur, for instance, in systems with  inhomogeneous external potentials~\cite{Li09b}, strongly degenerate Bose gases such as liquid helium~\cite{Li14d} (due to nonextensive parts of thermodynamic potentials), and photon gases~\cite{Le15} (the compressibility of which does not exist).
Importantly, we recall that hyperuniformity is only defined for statistically homogeneous systems~\cite{To03a, To18a}.

\bgroup
\def\arraystretch{2}
\begin{table*}[htp]
\centering
{
\setlength{\tabcolsep}{10pt}
    \begin{tabular}{c l l l}
    \hline
    \hline
      System & $S_\text{RPA}(q),\, S_\text{Hubbard}(q)$ & $\tilde{S}_\text{RPA}(q)$ & $\tilde{S}_\text{Hubbard}(q)$\\
    \hline
    Quasi-1D & ${\displaystyle\frac{\pi  q}{8 \sqrt{-r_s \ln{q}}}}$ &  $q/2$ & ${\displaystyle\left[ \frac{1}{2} + \displaystyle\frac{e^{\,\ell^2/4} K_0\left(\ell^2/4\right) r_s}{\pi ^2}\right] q^{\phantom{1^{1^{1^{1^{1^1}}}}}}}$ \\
    2D & ${\displaystyle\frac{1}{2^{3/4} \sqrt{r_s}}q^{3/2}}$~\cite{Go93} &   $2q/\pi$ & ${\displaystyle\left[\frac{2}{\pi} + \frac{\left(4 \sqrt{2}-\sqrt{2} \pi \right) r_s}{2 \pi }\right]q}$ \\
    3D & ${\displaystyle\frac{3^{5/6} \pi ^{2/3} q^2}{4 \sqrt[3]{2} \sqrt{r_s}}}$~\cite{Go93, Wa91} &  $3q/4$~\cite{Wa91} & ${\displaystyle\left[\frac{3}{4} + \frac{2 \left(\frac{2}{3}\right)^{2/3} (1 - \ln 2)}{\pi ^{4/3}} r_s\right] q}$\smallskip \\
    \hline
    \hline
    \end{tabular}
        \caption{\label{tab:small-k}Leading-order small-$k$ expressions of the total structure factor $S(k)$ and the magnetic structure factors $\tilde{S}(k)$ for unpolarized interacting electron liquids ($r_s > 0, \zeta = 0$). As before, we define $q = k/k_F$ here. }
        }
\end{table*}
\egroup

For an equilibrium multicomponent system (as is the case for spin-$1/2$ particles), the FC relation generalizes to~\cite{Ki51, Ma13a}
\begin{equation}
    k^{}_B T \kappa^{}_T = \frac{|\mathbf{B}|}{\sum_{\alpha, \beta = 1}^M \sqrt{\rho_\alpha \rho_\beta}|\mathbf{B}_{\alpha\beta}|},
    \label{FC_multicompo}
\end{equation}
where $M$ is the number of components, the elements of the matrix $\mathbf{B}$ are $B_{\alpha\beta} = \lim_{k\rightarrow 0} S_{\alpha\beta}(k)$, and $\mathbf{B}_{\alpha\beta}$ is the $\alpha\beta$-cofactor of $\mathbf{B}$.
For the case of an unpolarized electron liquid, $|\mathbf{B}|$\,$=$\,$ S_{\uparrow\uparrow}^2(0) - S_{\uparrow\downarrow}^2(0)$ and $\sum_{\alpha, \beta = 1}^M |\mathbf{B}_{\alpha\beta}| = \rho[S_{\uparrow\uparrow}(0) - S_{\uparrow\downarrow}(0)]$.
We then have $k_B T \kappa_T = |\mathbf{B}|/\sum_{\alpha, \beta = 1}^M |\mathbf{B}_{\alpha\beta}| = S(0)/\rho$, i.e., Eq.~(\ref{FC}) is recovered.
Therefore, the FC relation implies that when summed over spin species, the ground-state configurations of unpolarized electron liquids are indeed hyperuniform.

We remark that hyperuniformity implies the direct-space sum rule  \cite{To18a}
\begin{equation}
    \rho \int_{\mathbb{R}^d}h(\boldsymbol{r})d\boldsymbol{r} = -1.
    \label{sum_rule}
\end{equation}
Importantly, while the analog of Eq.~(\ref{sum_rule}) holds for any \textit{finite} system with a fixed number of particles, number conservation does not imply hyperuniformity.
This is because hyperuniformity is a statistical structural property that concerns number density fluctuations in \textit{infinitely} large systems rather than finite systems; in other words, the Hubbard approximation $N\rightarrow\infty$ must be taken \textit{before} the limit $k\rightarrow 0$ or $r\rightarrow \infty$.

Table \ref{tab:small-k} demonstrates that, across dimensions, the total structure factors have leading-order terms that increase superlinearly (in contrast to the noninteracting electron gas, for which $S(k)\sim k$ for small $k/k_F$). 
In the language of hyperuniformity, for $d \geq 2$, $S(k)$ is class-I hyperuniform with $\alpha = (d+1)/2$. 
For $d = 1$, the increase of $S(k)$ at small $k$, given by $S(k)\propto k/\sqrt{-\ln (k/k_F)}$, is faster than any linear function in $k$ but slower than any power-law function with exponent $\alpha > 1$, i.e., $\lim_{k\rightarrow 0} S(k)/k = 0$ but $\lim_{k\rightarrow 0} S(k)/k^\alpha = \infty$ for any $\alpha > 1$. 
Therefore, one can regard the 1D $S(k)$ as being characterized by an exponent $\alpha = 1 + \epsilon$ with infinitesimal $\epsilon$, which still belongs to class I. 
This peculiar behavior is a consequence of the nature of the 1D potential, which scales as $\tilde{V}(k) \sim -2\ln(k\ell)$ as $k\rightarrow 0$ and is reflected in the dispersion of the collective charge-density excitations (the Tomonaga-Luttinger bosons)~\cite{li1992elementary,sarma1996dynamical}. We note, however, that a full description of realistic $1$D quantum fluids---in either solid-state materials or cold-atom systems---may require corrections beyond the Luttinger paradigm~\cite{imambekov2012one}.

The small-$k$ expressions for the total structure factors in $d$ dimensions are consistent with the well-known Bijl-Feynman formula~\cite{Si68, Go93}, i.e., the density excitation spectrum at long wavelengths---and consequently, the small-$k$ behavior of $S(k)$---is dominated by the plasmon contribution $S_p(k)$:
\begin{equation}
    S(q) \sim S_p(q) = \sqrt{\frac{q^2}{4 m \rho \tilde{V}(q)}}, \quad q\rightarrow 0.
    \label{plasmon}
\end{equation}
The plasmon is a basic density excitation of the electron liquid and can be regarded as the analog of longitudinal phonons in a solid.
Importantly, while $S(k)$ possesses the analytic leading-order term $\sim k^2$ in three dimensions, our derivation (see Appendix~\ref{sec:appA}) finds that it also contains a nonanalytic subleading $k^5$ term, implying that $h(r)$ decays as a power law given by $h(r)\sim r^{-(d + 5)} = r^{-8}$ at large $r$ \cite{To18a}. This $k^5$ term arises from the contribution of electron-hole pairs~\cite{Go91}, which are the other fundamental excitations of the electron liquid. 

Interestingly, the magnetic structure factors for unpolarized fermions are class-II hyperuniform with linear small-$k$ behaviors. 
In the RPA approximation, the coefficient of the leading linear term in $\tilde{S}(k)$ is identical to that of $S^0_{\sigma\sigma}(k)$ (\ref{S_free}) for unpolarized free fermions, i.e., $\tilde{S}(q)\sim c_dq$, where $c_1 = 1/2, c_2 = 2/\pi$, and $c_3 = 3/4$.
This is due to the fact that the RPA condition $I(q) = 0$ implies that $\chi^s(q, \omega) =g^2\mu_B^2\chi_0(q,\omega)$; see Eq.~(\ref{chis-I}).
Thus, Eq.~(\ref{tildeS-chis})---which amounts to an integration over the imaginary part of $\chi_0(q, \omega)$---yields the free-fermion structure factor in the respective dimensions~\cite{Li54}. 

On the other hand, in the Hubbard approximation, our analysis reveals that the magnetic structure factor increases with $r_s$ as $\tilde{S}(q)\sim (c_d + \lambda_d r_s)q$, where $\lambda_d>0$ depends on the dimensionality. Note that for quasi-1D systems, $\lambda_d$ also depends on the specific potential used.
We attribute this increase of $\tilde{S}(k)$ to the correlation hole effect discussed above~\cite{Br04b, Wa91}.
In the interacting electron liquid, the fermions are subject to an additional Coulomb repulsion compared to the noninteracting free Fermi gas.
The effect of this additional repulsion on the electron-electron correlations is experienced more strongly by oppositely aligned spins than by like spins, as the latter are already well-separated due to Pauli exclusion.
Thus, $S_{\uparrow\downarrow}(k)$ at small $k$ decreases significantly as $r_s$ increases, resulting in a magnetic structure factor (\ref{tildeS-spinspec}) that increases with $r_s$.
To the best of our knowledge, this increase of the linear term in $\tilde{S}(k)$ as a function of $r_s$ has not been identified previously.

\begin{figure*}[tb]
    \centering
    \includegraphics[width = \linewidth]{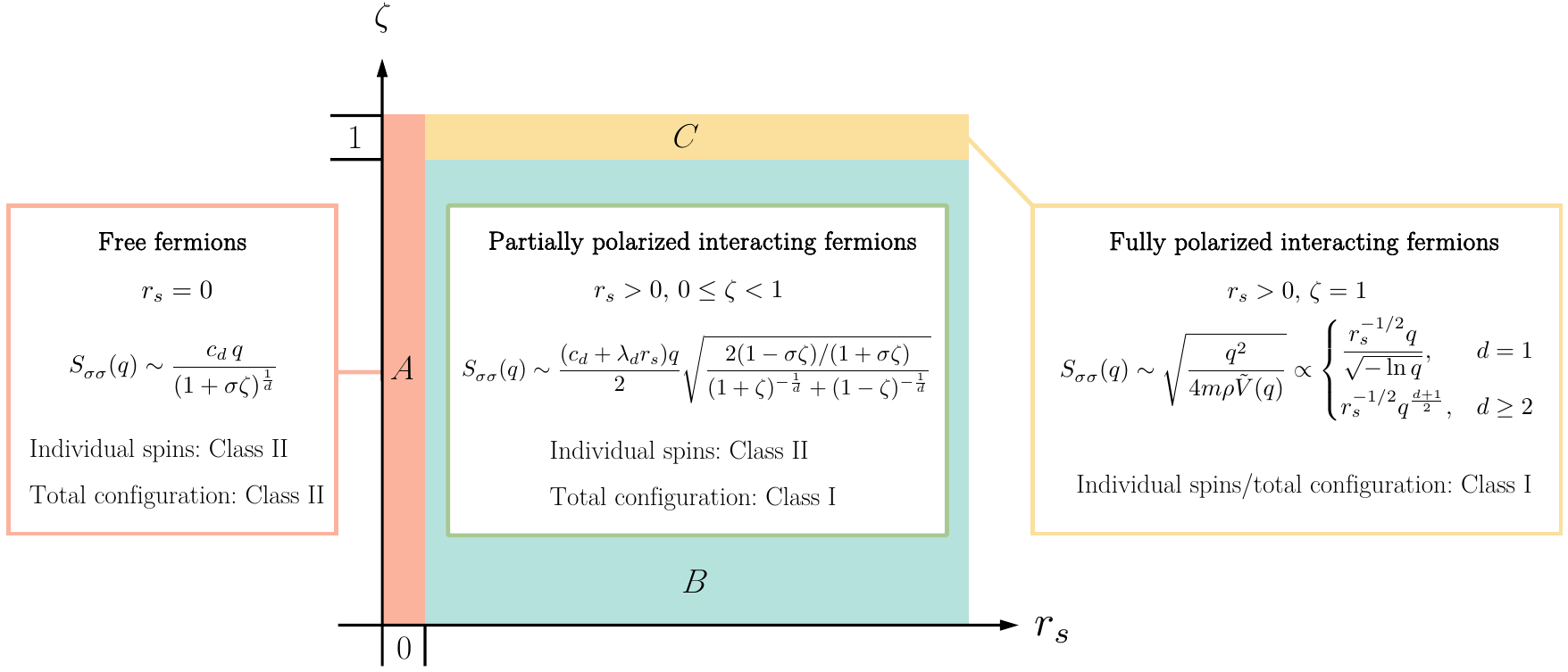}
    \caption{Regimes of free and interacting fermionic liquids---as characterized by the small-$k$ behaviors of parallel-spin structure factors $S_{\sigma\sigma}(k)$ in the Hubbard approximation---as a function of the interaction strength $r_s$ and the polarization $\zeta$.
    While regimes $A$ (free fermions) and $C$ (fully polarized interacting fermions) are technically lines on this diagram; they have been drawn as expanded regions for clarity. Regime $B$ corresponds to a partially polarized liquid of interacting fermions.}
    \label{phasediag}
\end{figure*}

\subsubsection{Generalization to arbitrary polarizations}
\label{smallk_analytic}
For interacting fermions ($r_s>0$) with arbitrary partial polarization, $0$\,$\leq$\,$ \zeta$\,$< 1$, we generalize the expressions obtained by \citet{Da03} for the spin-resolved 3D structure factors to all dimensions, yielding
\begin{widetext}
\begin{equation}
    S_{\sigma\sigma}(q; \zeta) \sim  \sqrt{\frac{2(1-\sigma\zeta)/(1+\sigma\zeta)}{(1+\zeta)^{-1/d} + (1-\zeta)^{-1/d}}}S_{\sigma\sigma}(q; \zeta = 0) + \frac{S_p(q)}{2}, \quad  q\rightarrow 0,
\end{equation}
\begin{equation}
    S_{\uparrow\downarrow}(q; \zeta) \sim -\sqrt{\frac{2}{(1+\zeta)^{-1/d} + (1-\zeta)^{-1/d}}} S_{\uparrow\downarrow}(q; \zeta = 0) + \frac{S_p(q)}{2}, \quad q \rightarrow 0,
\end{equation}
\end{widetext}
where $S_{\sigma\sigma'}(q; \zeta = 0)$ is the spin-resolved structure factor for unpolarized fermions, the small-$k$ behaviors of which are easily obtained from Table \ref{tab:small-k} and Eqs.~(\ref{Supup_unp}), (\ref{Supdown_unp}). 
Thus, $S_{\sigma\sigma}(k)$ for a liquid of partially polarized interacting fermions is class-II hyperuniform with a linear leading-order small-$k$ behavior.
On the other hand, fully polarized interacting fermions ($r_s > 0, \zeta = 1$) constitute a single-component system.
Assuming, without loss of generality, that the system consists of only up spins, we have $S_{\uparrow\uparrow}(k) = S(k; \zeta = 1) \sim S_p(k) = \sqrt{q_{\uparrow}^2/[4m\rho\tilde{V}(q_\uparrow)]}$, where $q_{\uparrow}$\,$ = $\,$k/k_{F\uparrow}$\,$ = $\,$(k/k_F)/2^{1/d}$.
To wit, the parallel-spin structure factor is now class-I hyperuniform in all dimensions.

Figure~\ref{phasediag} schematically summarizes the analysis of hyperuniformity in the Hubbard approximation, illustrating that the ground states of the electron liquid can be characterized into three regimes based on the parallel-spin structure factors $S_{\sigma\sigma}(k)$ at small $k$. 
Free fermions (regime $A$) form class-II hyperuniform states for both the individual spins and the total configuration. 
Partially polarized interacting fermions (regime $B$) form class-II hyperuniform states for the individual spins, but the total configuration as a whole is class-I hyperuniform.
This unusual form of multihyperuniformity will be further discussed in Sec.~\ref{multihu}.
Finally, fully polarized interacting fermions (regime $C$) are class-I hyperuniform.

\begin{figure*}[tb]
    \centering
\includegraphics[width=\linewidth]{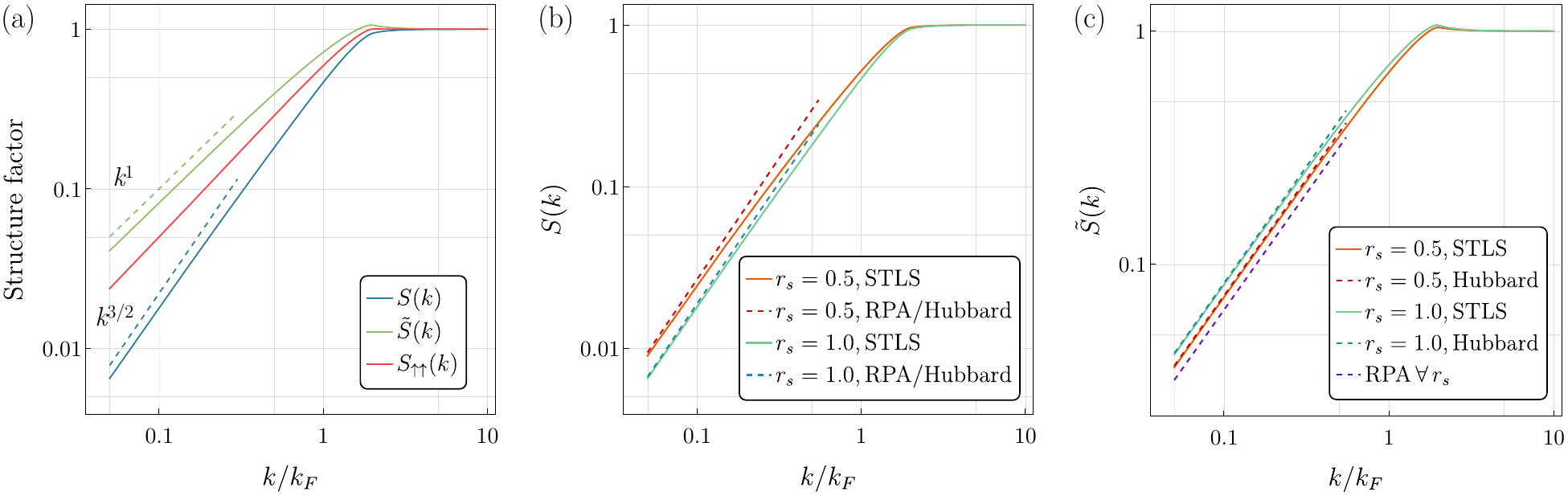}
    \caption{(a) The numerical structure factor computed using the STLS model shows unusual multihyperuniformity for a 2D unpolarized electron liquid with $r_s = 1$. 
    The small-$k$ scaling exponent $\alpha$ is $3/2$ for the total configuration and unity for the individual spin components. (b, c) Comparison of the analytical small-$k$ behaviors of (b) the total structure factor, and (c) the magnetic structure factor in Table \ref{tab:small-k} with the numerical STLS results, for 2D electron liquids with $r_s = 0.5$ and $r_s = 1$.
    }
    \label{fig:compare_stls}
\end{figure*}

\subsubsection{Comparison to the STLS scheme}
To verify whether the analytical small-$k$ expressions derived for the RPA and Hubbard approximations provide accurate representations of the electron liquid, we compare these expressions to the numerical structure factors obtained via the self-consistent STLS model described in Sec.~\ref{sec:models}, which is known to achieve excellent agreement with quantum Monte Carlo simulations for $r_s \leq 1$~\cite{Mo92, Go93} but can only be solved numerically~\cite{Go97}.
Figure~\ref{fig:compare_stls} presents such a comparison for  unpolarized electrons in 2D with $r_s = 0.5$ and 1, which are in the partially polarized interacting regime (regime $B$).
One can clearly observe from the STLS structure factors for the $r_s = 1$ state [Fig.~\ref{fig:compare_stls}(a)] that the system exhibits unusual multihyperuniform behavior.
The total structure factor $S(k)$ grows as $k^{3/2}$ at small $k$, whereas the parallel-spin structure factor $S_{\sigma\sigma}(k)$ is linear in $k$.
Figure~\ref{fig:compare_stls}(b) underscores that the leading-order term of the small-$k$ asymptotics for $S(k)$, given by the plasmon contribution~(\ref{plasmon}), agrees well with the STLS structure factors. 
Interestingly, this agreement is better for larger $r_s$ because the plasmon contribution to the longitudinal spectrum (which the RPA \textit{does} describe accurately for long wavelengths) becomes significant over a larger range of $k$ for stronger Coulomb interactions.
Figure~\ref{fig:compare_stls}(c) shows that the small-$k$ behaviors of the magnetic structure factor derived from the Hubbard approximation match the STLS results closely at different values of $r_s$. 
Contrarily, the RPA model does not capture the increase of $\tilde{S}(k)$ as $r_s$ increases.

Taken together, the results in Fig.~\ref{fig:compare_stls} convey that the Hubbard approximation provides a realistic description of the structure factors of ground-state electron liquids at small $k$, and thus our corresponding expressions in Table \ref{tab:small-k} can be used to analytically describe their behaviors.
Note that while the Hubbard approximation is often used as an initial guess for the STLS algorithm, and is therefore less accurate than the latter, our results show that their differences arise mainly at intermediate to large $k$ rather than at small $k$.

\subsection{Multihyperuniformity of partially polarized interacting fermions}
\label{multihu}
A multihyperuniform state is a multicomponent system that is simultaneously hyperuniform for both the individual components and the total configuration~\cite{Ji14}. 
They were first identified in avian cone photoreceptors consisting of five different cell types as an evolutionary strategy to sample light of different wavelengths~\cite{Ji14}.
A minimal statistical-mechanical model has been developed to describe such patterns in avian retina via long-range repulsive pair interactions between like cell types~\cite{Lo20}.
More recently, the concept of multihyperuniformity has been applied in digital cameras to create bionic birdlike imaging using LED arrays, achieving less chromatic moir{\'e} patterns and color misregistration artifacts~\cite{Zh21b}.
Multihyperuniformity has also been observed in the electron density distribution in a quasiperiodic potential~\cite{Sa22}.

In all of the aforementioned cases, both the components and the total configurations are class-II hyperuniform.
Thus, partially polarized interacting electron liquids (regime $B$ in Fig.~\ref{phasediag}) constitute examples of a highly unusual form of multihyperuniformity that has never been seen in any previous work on quantum and classical many-body systems. 
Here, each spin component is class-II hyperuniform, whereas the total configuration is of the strongest hyperuniformity class, namely, class I.
The higher hyperuniformity class for the total configuration suggest that the cross-correlations between opposite spin components must be nontrivial, i.e., $S_{\uparrow\downarrow}(k) \neq 0$, as theoretically expected.
If the opposite spins were uncorrelated, $S(k)$ would have been of the same hyperuniformity class as the individual spins per Eq.~(\ref{Stotal_from_spinspec}).

\begin{figure*}[tb]
    \centering
\includegraphics[width = \linewidth]{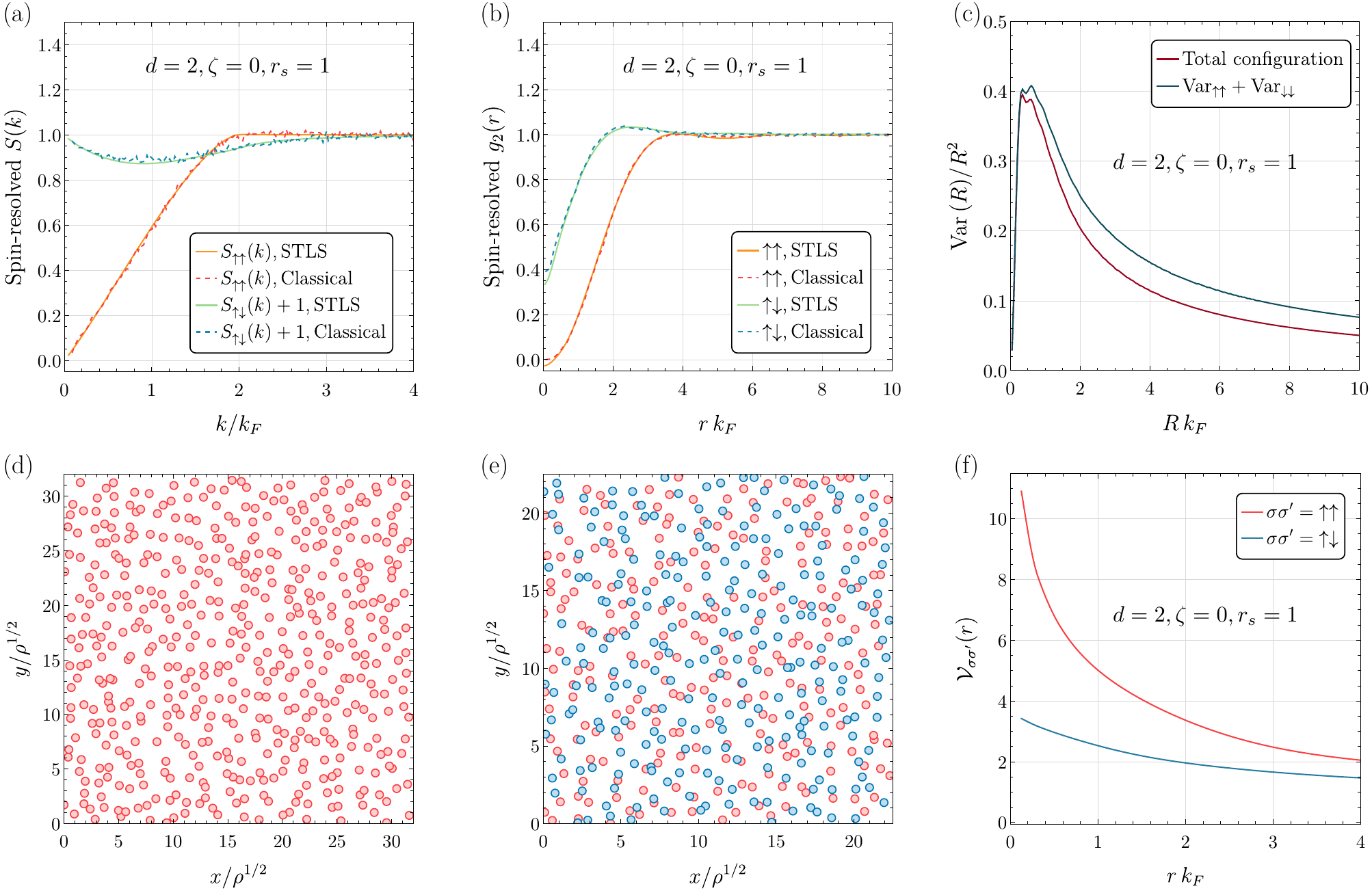}      
\caption{Unusual multihyperuniformity of the  2D unpolarized interacting electron liquid (for $r_s = 1$) and its equivalent classical system with effective one- and two-body interactions.
    (a) Spin-resolved structure factors, and (b) spin-resolved pair correlation functions  as computed for both the STLS model and the equivalent  classical system mimicking the electron liquid.
    (c) Scaled local number variance $\text{Var}(R)/R^2$ for the total configuration and the sum of $\text{Var}(R)/R^2$ for the individual spin components.
    (d, e) Snapshots of the equivalent classical system with a total number of 1000 particles for (d) an individual spin component ($\alpha = 1$), and (e) the total configuration ($\alpha = 3/2$). The red and blue points denote the up and down spins. 
    (f) Spin-specific effective potentials for the classical system obtained.
}
    \label{fig:multihu2D}
\end{figure*}

To more closely examine the effects of such cross-correlations, we plot the spin-resolved structure factors and pair correlation functions for the 2D unpolarized electron liquid with $r_s$\,$=$\,$1$, obtained from the STLS approximation [Figs.~\ref{fig:multihu2D}(a) and (b), solid curves].
Figure~\ref{fig:multihu2D}(a) shows that $S_{\uparrow\downarrow}(k)$ decreases linearly with $k$ at small $k$. 
This decreasing linear term exactly cancels out the linear terms in the parallel-spin structure factors, yielding a total structure factor without a linear term via Eq.~(\ref{Stotal_from_spinspec}); the resulting leading-order term is given by $k^{(d+1)/2} = k^{3/2}$.
The fact that $S_{\uparrow\downarrow}(k)$ displays a broad well with a minimum at $k \sim k_F$ indicates that, in real space, the oppositely aligned spins experience significant repulsion.
Indeed, the pair correlation functions in Fig.~\ref{fig:multihu2D}(b) highlight that while the same-spin $g_2(r)$ closely resembles that for the free fermions [see Fig.~\ref{pair_stat_free_pol}(b)], the opposite-spin $g_2(r)$ exhibits strong negative correlation at small to intermediate $r$, which is sharply different from the free-fermion opposite-spin correlation function, $g_{2,\uparrow\downarrow}(r) = 1$.
This high sensitivity of the opposite-spin correlation functions to $r_s$ is a result of the correlation hole effect discussed in Sec.~\ref{unp}.

In order to study the implications of this unusual multihyperuniformity for the number density variance, we numerically compute $\text{Var}(R)$ given by~\cite{To18a}
\begin{equation}
    \text{Var}(R) = \rho v_1(R)\left[\frac{s_1(1)}{(2\pi)^d}\int_0^\infty k^{d-1}S(k)\tilde{\alpha}_2(k, R) dk \right],
\label{sigmaSq_2DSk}
\end{equation}
where 
\begin{equation}
    \tilde{\alpha}_2(k, R) \equiv 2^d \pi^{d/2}\Gamma(1 + d/2)\frac{(J_{d/2}(kR))^2}{k^d}.
\end{equation}
We evaluate Eq.~(\ref{sigmaSq_2DSk}) for the total and spin-resolved structure factors for 2D unpolarized fermions (with $r_s = 1$) obtained from the STLS approximation.
Figure~\ref{fig:multihu2D}(c) plots the scaled local number variance $\text{Var}(R)/R^2$ for both the total configuration of such a system and the individual spins.
The fact that the scaled variance decreases at large $R$ clearly demonstrates hyperuniformity, as $\text{Var}(R)$ grows more slowly than the area of the observation window $v_1(R) = \pi R^2$ \cite{To18a}.
At large $R$, the total configuration has a significantly smaller local number variance than the sum of the number variances of individual spins, since the former grows as $R$ whereas the latter grows as $R\ln R$ according to Eq.~(\ref{alpha}).
One therefore has
\begin{equation}
\begin{split}
    \text{Var}_{\text{total}}(R) &\equiv \text{Var}_{\uparrow\uparrow}(R) + \text{Var}_{\downarrow\downarrow}(R) + 2\text{Cov}_{\uparrow\downarrow}(R) \\
    &< \text{Var}_{\uparrow\uparrow}(R) + \text{Var}_{\downarrow\downarrow}(R),
\end{split}
\end{equation}
which indicates that the covariance between the number densities of the opposite spins $\text{Cov}_{\uparrow\downarrow}(R)$ is negative due to their repulsive interaction, as noted above.

\section{Classical systems with pair statistics of electron liquids}
\label{class}
Recently, there has been considerable interest in the investigation of \textit{classical} systems that mimic pair statistics of quantum states~\cite{Ca94, Bu02, Cr04, Sm15, Zh20, Li14b}. 
These classical mappings enable one to efficiently predict thermodynamic properties, such as the exchange-correlation free energy~\cite{Li14b} and to simulate quantum systems via density functional theory (DFT)~\cite{Gu79, Pe81} or path-integral molecular dynamics (PIMD) approaches~\cite{Ca94, Cr04, Sm15}. 
Unfortunately, such mappings are usually limited to the hypernetted chain (HNC) approximation, which is known to be inaccurate for high-density systems~\cite{Wa20}.
However, to reliably simulate quantum states, one requires highly accurate formulations of the equivalent classical interactions~\cite{Da03}.
In this section, we describe how our analytical small-$k$ expressions for fermionic systems in conjunction with
efficient inverse algorithms \cite{To22, Wa24} enable us to construct corresponding classical systems, equilibrated at positive temperatures, that precisely mimic the zero-temperature fermionic pair statistics.

\subsection{Inverse algorithm}
\label{meth}
To determine equilibrium classical many-body systems with pair statistics that match those of interacting electron liquids, we use a recently devised precise inverse algorithm~\cite{To22, Wa24} that extracts effective one- and two-body interactions from targeted $g_2(r)$ and $S(k)$.
While the algorithm presented in Refs.~\cite{To22, Wa24} is for single-component particle systems, it can be easily generalized to multicomponent variants, including spins systems.
Here, we sketch this algorithm for a partially polarized electron liquid with both up and down spin constituents.

Our methodology uses parametrized spin-specific pair potentials $\mathcal{V}_{\sigma\sigma'}(r;\mathbf{a}_{\sigma\sigma'})$, the initial functional forms of which are informed by the small- and large-distance behaviors of the target pair statistics~\cite{To18a}. 
The scalar components $a_j$ of the potential parameter vector $\mathbf{a}_{\sigma\sigma'}$ are of three types: dimensionless energy scales $\varepsilon_j$, dimensionless distance scales $\sigma_j$, and dimensionless phases $\theta_j$~\cite{To22}.
To obtain an initial form of the small- and intermediate-$r$ behavior of $\mathcal{V}_{\sigma\sigma'}(r;\mathbf{a})$, we numerically fit the HNC approximation for the target spin-resolved pair statistics, given by 
\begin{equation}
    \beta \mathcal{V}_{\sigma\sigma'}(r) = h_{\sigma\sigma'}(r) - \ln(g_{2,\sigma\sigma'}(r)) - c_{\sigma\sigma'}(r),
\end{equation}
where $c_{\sigma\sigma'}(r)$ is the spin-resolved direct correlation function.
The large-$r$ behavior of $\mathcal{V}_{\sigma\sigma'}(r;\mathbf{a}_{\sigma\sigma'})$ is given by the target direct correlation function $c_{\sigma\sigma'}(r)$ under mild conditions~\cite{Ha86, To22}.
The direct correlation functions are related to the targeted pair statistics via the multicomponent Ornstein-Zernike equation~\cite{Ha73, Ma13a, Fi23}
\begin{equation}
    \tilde{h}_{\nu\mu}(k) = \tilde{c}_{\nu\mu}(k) + \sum_{\lambda = 1}^n \rho_{\lambda}\tilde{c}_{\nu\lambda}(k)\tilde{h}_{\lambda\mu}(k),
    \label{multicompo_oz}
\end{equation}
where $n$ is the number of components, $\nu, \mu$, and $\lambda$ are indices of the components, and $\tilde{h}(k)$ and $\tilde{c}(k)$ are the Fourier transforms of $h(r)$ and $c(r)$, respectively.

In the specific case of single-component hyperuniform targets (i.e., fully polarized fermionic liquids), the large-$r$ asymptotic form of $\beta \mathcal{V}(r) \sim -c(r)$ is given by~\cite{To18a}
\begin{equation}
    \mathcal{V}(r)\sim 
    \begin{cases}
    r^{-(d-\alpha)}, \quad &d \ne \alpha\\
    -\ln(r), \quad &d = \alpha
    \end{cases}.
    \label{v_long_hu}
\end{equation}
That is, the potential is long-ranged in the sense that its volume integral is unbounded. 
Thus, one requires a neutralizing background one-body potential to maintain stability~\cite{To18a, Ha73, Ga79, Dy62a}, and the Ewald summation technique~\cite{Ew21} is used to compute the total potential energy.
For nonhyperuniform targets, the volume integral of $c(r)$ is bounded and one only requires two-body interactions to maintain stability.

Once the initial form of $\mathcal{V}_{\sigma\sigma'}(r;\mathbf{a}_{\sigma\sigma'})$ is chosen, the low-storage Broyden–Fletcher–Goldfarb–Shanno (BFGS) algorithm~\cite{Liu89} is used to minimize an objective function $\Psi({\bf a})$, where $\mathbf{a} = [\mathbf{a}_{\uparrow\uparrow}, \mathbf{a}_{\downarrow\downarrow}, \mathbf{a}_{\uparrow\downarrow}]$, based on the distance between target and trial pair statistics in both direct and Fourier spaces:
\begin{equation}
\begin{split}
\Psi(\mathbf{a}) = \int_{\mathbb{R}^d}\hspace*{-0.2cm}
 w^{}_{g_2}({\boldsymbol{r}})\sum_{\sigma,\sigma'}\frac{\rho_\sigma\rho_{\sigma'}}{\rho}\left(g^{\textsc{t}}_{2,\sigma\sigma'}({\boldsymbol{r}})-g^{}_{2,\sigma\sigma'}({\boldsymbol{r}};\mathbf{a})\right)^2
d{\boldsymbol{r}} \\
+ \frac{1}{(2\pi)^d}\int_{\mathbb{R}^d}\hspace*{-0.2cm} w^{}_{S}({\bf k})\sum_{\sigma,\sigma'}\frac{\sqrt{\rho_\sigma\rho_{\sigma'}}}{\rho^2}\left(S^{\textsc{t}}_{\sigma\sigma'}({\bf k})-S^{}_{\sigma\sigma'}(k;\mathbf{a})\right)^2 d\boldsymbol{k},
\label{Psi}
\end{split}
\end{equation}
where $g^{\textsc{t}}_{2,\sigma\sigma'}(\boldsymbol{r})$ and $S^{\textsc{t}}_{\sigma\sigma'}(\boldsymbol{k})$ are the targeted spin-resolved pair statistics, $w_{g_2}({\boldsymbol{r}})$ and $w_{S}({\bf k})$ are weight functions, and $g_{2,\sigma\sigma'}({\boldsymbol{r}};\mathbf{a})$ and $S_{\sigma\sigma'}(\boldsymbol{k};\mathbf{a})$ are the equilibrated spin-resolved pair statistics under $\mathcal{V}_{\sigma\sigma'}(r;\mathbf{a}_{\sigma\sigma'})$ obtained from Monte Carlo simulations.
As in our previous works~\cite{To22, Wa22b, Wa23, Wa24}, here, we use Gaussian weight functions $w_{g_2}(r) = \exp[-(r\rho^{1/d}/4)^2]$ and $w_S(k) = \exp[-(k/(2\rho^{1/d}))^2]$. 

The BFGS algorithm requires the gradient of the objective function $\Psi(\mathbf{a})$.
Akin to Ref.~\cite{Wa24}, we estimate the gradient of $\Psi(\mathbf{a}_i)$ in iteration $i$ by performing automatic differentiation (AD) on an approximant function $\psi$ of $\Psi$. 
For a simulated canonical ensemble $\mathcal{C}_i$ in iteration $i$, $\psi$ is defined by reweighing the pair statistics of each configuration in $\mathcal{C}_i$ by a Boltzmann factor involving the difference in total configurational energies $\Phi(\boldsymbol{r}^N) = \sum_{j<k}^N v_{\sigma_j\sigma_k}(|\boldsymbol{r}_j - \boldsymbol{r}_k|)$ under $\mathcal{V}_{\sigma\sigma'}(r;\mathbf{a}_i)$ and $\mathcal{V}_{\sigma\sigma'}(r;\mathbf{a}')$~\cite{No08b}, i.e.,
\begin{equation}
\begin{split}
    &\varphi(x; \mathbf{a}') = \\
    &\frac{\sum_{\boldsymbol{r}^N \in \mathcal{C}_i} \varphi(x; \boldsymbol{r}^N) \exp[-(\Phi(\boldsymbol{r}^N; \mathbf{a}') - \Phi(\boldsymbol{r}^N; \mathbf{a}_i))/T] }{\sum_{\boldsymbol{r}^N \in \mathcal{C}_i} \exp[-(\Phi(\boldsymbol{r}^N; \mathbf{a}') - \Phi(\boldsymbol{r}^N; \mathbf{a}_i))/T]},
\end{split}
    \label{reweigh}
\end{equation}
where $\varphi(x; \boldsymbol{r}^N)$ is either $g_{2,\sigma\sigma'}(r)$ or $S_{\sigma\sigma'}(k)$ computed from each configuration $\boldsymbol{r}^N$ in $\mathcal{C}_i$, and $T$ denotes absolute temperature. 
The approximate objective function $\psi(\mathbf{a}')$ at the perturbed parameters $\mathbf{a}'$ in the vicinity of $\mathbf{a}_i$ can be computed by inserting Eq.~(\ref{reweigh}) into Eq.~(\ref{Psi}).
One can then apply BFGS optimization with AD-generated gradients to find the optimal parameters $\mathbf{a}_{i+1}$ that minimize $\psi(\mathbf{a}')$.
In the next iteration, simulations are performed with the potential $\mathcal{V}(r;\mathbf{a}_{i+1})$ to obtain a new set of configurations $\mathcal{C}_{i+1}$. 
The iterations repeat until $\Psi(\mathbf{a}_i)$ computed from $\mathcal{C}_i$ is smaller than some given convergence threshold $\epsilon$, set to be $10^{-3}$ in this study.
If convergence is not achieved, a different set of basis functions is chosen and the optimization process is repeated. 
Note that this inverse methodology is highly efficient as it requires just one simulation per iteration, regardless of the number of potential parameters $|\mathbf{a}|$.

\begin{figure*}[htp]
\includegraphics[width=\linewidth]{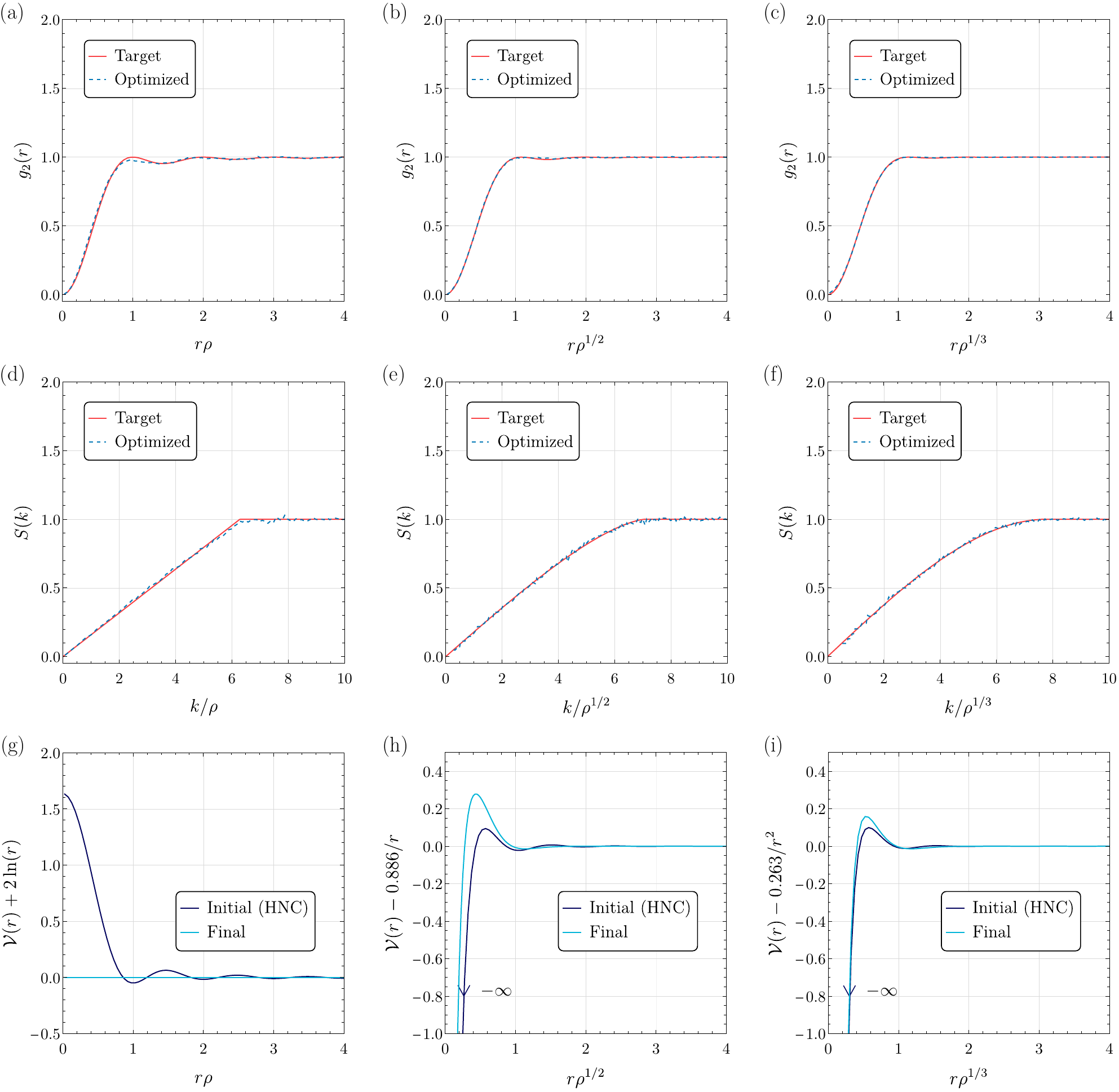}
    \caption{(a)--(c) Target and optimized pair correlation functions, and (d)--(f) target and optimized static structure factors  for polarized free fermions in one, two and three dimensions.
    (g)--(i) The short-ranged parts of the optimized potentials in one, two and three dimensions, respectively, as well as the corresponding HNC approximations used as the initial guess for the potentials.}
    \label{pair_stat_free_pol}
\end{figure*}

\begin{figure*}[htp]
\includegraphics[width = \linewidth]{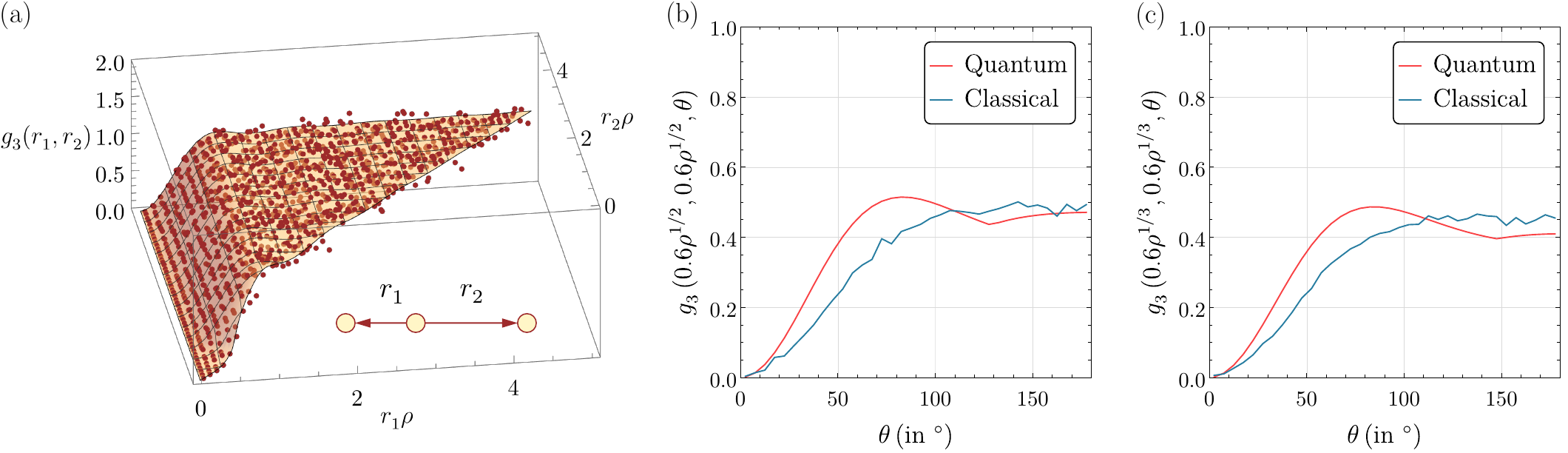}
    \caption{Three-body correlation function $g_3$ for polarized free fermions and their corresponding classical two-component system with effective one- and two-body potentials. (a) For 1D free fermions, the quantum and classical systems are exactly mappable to each other, and thus, their $g_3$ functions agree. The surface corresponds to the $g_3$ function for the quantum state, and the points plot the classical $g_3$ data.
    (b, c) Three-body correlation function $g_3$ for 2D and 3D free fermions, respectively, for points on small isosceles triangles with side lengths of 0.6$\rho^{1/d}$. 
    The quantum state displays stronger $g_3$ correlations for acute triangles compared to the classical state.}
    \label{g3_free_pol}
\end{figure*}

\subsection{Results for classical equilibrium systems}
Our analytical small-$k$ expressions for the spin-resolved structure factors enable us to precisely determine the spin-specific effective pair potentials that yield classical equilibrium states with pair statistics identical to interacting electron liquids.
Here, we first carry out such an analysis for the 2D unpolarized interacting electron liquid with $r_s = 1$ as an illustrative example. 
We then study the structural degeneracy problems for 2D and 3D fully polarized free Fermi gases, in light of their important connections to random matrix theory.

The leading terms of the spin-specific structure factors at small $k$ for 2D unpolarized interacting fermions are given by Eqs.~(\ref{2DSupup}) and (\ref{2DSupdown}) in Appendix~\ref{sec:appA}.
To derive the large-$r$ behaviors of the effective potentials, we note that the multicomponent Ornstein-Zernike equation (\ref{multicompo_oz}) for this two-component system is given by
\begin{alignat}{1}
\nonumber
    \tilde{h}_{\uparrow\uparrow}(k) &= \tilde{c}_{\uparrow\uparrow}(k) + \rho_{\uparrow}\tilde{c}_{\uparrow\uparrow}(k)\tilde{h}_{\uparrow\uparrow}(k) 
    + \rho_{\downarrow}\tilde{c}_{\uparrow\downarrow}(k)\tilde{h}_{\uparrow\downarrow}(k),\\
    \tilde{h}_{\uparrow\downarrow}(k) &= \tilde{c}_{\uparrow\downarrow}(k) + \rho_{\uparrow}\tilde{c}_{\uparrow\uparrow}(k)\tilde{h}_{\uparrow\downarrow}(k) 
    + \rho_{\downarrow}\tilde{c}_{\uparrow\downarrow}(k)\tilde{h}_{\downarrow\downarrow}(k),
\label{oz_2Dunp}
\end{alignat}
where $\rho_\uparrow = \rho_\downarrow = k_F^2/(4\pi)$.
Using Eqs.~(\ref{g2-S}), (\ref{2DSupup}), and (\ref{2DSupdown}) and solving Eq.~(\ref{oz_2Dunp}), one arrives at the large-$r$ behavior of the effective potentials $\beta \mathcal{V}_{\sigma\sigma'}(r) \sim - c_{\sigma\sigma'}(r)$. For $r_s = 1$, we find that as $r \rightarrow \infty$,
\begin{alignat}{1}
\nonumber
    \mathcal{V}_{\sigma\sigma}(r) &\sim 
    \frac{1}{2 \left({\displaystyle\frac{4 \sqrt{2}-\sqrt{2} \pi }{4 \pi }+\frac{1}{\pi }}\right) rk_F}+\frac{\sqrt[4]{2} \Gamma \left(\frac{1}{4}\right)}{\sqrt{rk_F} \Gamma \left(\frac{3}{4}\right)}, \\
    \mathcal{V}_{\uparrow\downarrow}(r) &\sim -\frac{1}{2 \left({\displaystyle\frac{4 \sqrt{2}-\sqrt{2} \pi }{4 \pi }+\frac{1}{\pi }}\right) rk_F}+\frac{\sqrt[4]{2} \Gamma \left(\frac{1}{4}\right)}{\sqrt{rk_F} \Gamma \left(\frac{3}{4}\right)}.
    \label{pot_2Dunp}
\end{alignat}

Equation~(\ref{pot_2Dunp}) now provides important insights into the emergence of multihyperuniformity in the classical system. 
Between any pair of particles---irrespective of whether they are of the same or the opposite spin species---there is always a long-ranged repulsive interaction term proportional to $r^{-1/2} = r^{-(d-1)/2} = r^{-(d-3/2)}$, which gives rise to the $k^{3/2}$ behavior at small $k$ in the total structure factor. 
Interestingly, for two electrons with the same spin, there is also a weaker repulsive term proportional to $r^{-1}=r^{-(d-1)}$, corresponding to the linear rise in  $S_{\uparrow\uparrow}(k)$ with $k$.
Between opposite spins, however, this $r^{-1}$ term has the same magnitude but is attractive, corresponding to the linear decrease in $S_{\uparrow\downarrow}(k)$.
Thus, we infer that the spin-independent repulsive term $r^{-(d-1)/2}$ describes correlation-hole effects due to the Coulomb interactions, whereas the spin-dependent term $r^{-(d-1)}$ accounts for the exchange hole.

Using our inverse algorithm described in Sec.~\ref{meth}, we determine a classical two-component system with pair statistics that match those for the 2D unpolarized electron liquid with $r_s = 1$.
The corresponding classical pair statistics are shown as dashed curves in Figs.~\ref{fig:multihu2D}(a) and (b).
Figure~\ref{fig:multihu2D}(f) plots the spin-specific effective potentials for the equivalent classical  system, illustrating that antiparallel spins experience a weaker effective classical repulsion than parallel ones (the functional forms of the potentials are given in Appendix~\ref{sec:appB}).
Figures~\ref{fig:multihu2D}(d) and (e) show snapshots of this equivalent classical system for one of the spin components and for the total configuration, respectively.

Such a quantum-classical correspondence between systems with matching pair statistics also allows one to probe the so-called structural degeneracy problem~\cite{Le75a, Cr03, Cos04, To06b, St19}.
It is known that for a homogeneous many-body system, one- and two-body correlations are insufficient to uniquely determine the higher-body correlation functions $g_3, g_4, ...$~\cite{To06b}.
Thus, we expect that the differences in the higher-order correlation functions reflect crucial quantum effects.

To facilitate the investigation of this degeneracy problem, we consider fully polarized free fermions, because their $n$-body statistics are exactly known via the corresponding determinantal point process~\cite{To08b}. 
For ground states of free fermions, each of the $n$-body correlation functions can be expressed completely in terms of $g_2(r)$~\cite{To08b}:
\begin{equation}
    g_n(\boldsymbol{r}_{12}, \boldsymbol{r}_{13}, \dots) = \det\left[ \sqrt{1 - g_2(r_{ij})}\right]_{i, j = 1, \dots, n}.
\end{equation}

Figure~\ref{pair_stat_free_pol} demonstrates that our inverse algorithm can successfully determine classical 1D, 2D, and 3D many-body systems interacting with one- and two-body potentials that closely match the pair correlation functions [Figs.~\ref{pair_stat_free_pol}(a)--(c)] and structure factors [Figs.~\ref{pair_stat_free_pol}(d)--(f)] of spin-polarized free fermions.
The short-ranged parts of the final optimized pair potentials, as well as the HNC potentials used as initial guesses, are shown in Figs.~\ref{pair_stat_free_pol}(g)--(i). 
In the 1D case, it is known~\cite{Dy62a} that the pair potential corresponding to the fermionic pair statistics (\ref{g2_1D_pol_free}) and (\ref{s_1D_pol_free}) is given by $\mathcal{V}(r) = -2\ln(r)$.
This potential is unique up to an additive constant due to Henderson's theorem~\cite{He74}. 
Indeed, starting from the HNC initial guess that has a nonzero short-ranged part, i.e., $\mathcal{V}_{\text{HNC}}(r) + 2\ln(r) \ne 0$, our inverse algorithm yields a final potential whose short-ranged part vanishes, recovering the pure logarithmic potential.
This highlights the accuracy of the algorithm in finding the unique pair potential corresponding to target pair statistics, as noted in our previous works~\cite{To22, Wa24}.
As $d$ increases, the differences between the HNC potential and the final optimized potential are reduced. This reflects the decorrelation principle in higher dimensions, which asserts that a particular functional form for a hypothetical correlation function corresponding to a disordered system becomes easier to be realized by many-body systems with increasing spatial dimensionality~\cite{To06b}.

Seeking to study the structural degeneracy of the polarized free fermionic quantum system and its effective classical equivalent, we compute their respective three-body correlation functions, as shown in Fig.~\ref{g3_free_pol}.
The classical and quantum 1D systems are exactly mappable to each other, as shown by \citet{Dy62a}. 
Thus, all their $n$-body correlation functions must agree.
Accordingly, Fig.~\ref{g3_free_pol}(a) shows that $g_3(r_1, r_2)$ for the free Fermi gas and that for the corresponding classical system match each other very well at all values of $r_1$ and $r_2$.

On the other hand, for $d \geq 2$, we observe distinct three-body statistics between the classical and quantum states.
For example, Figs.~\ref{g3_free_pol}(b) and (c) display $g_3$ for points positioned on small isosceles triangles with side lengths $0.6\rho^{1/d}$, as a function of the vertex angle $\theta$ for $d=2$ and 3, respectively.
In both the 2D and 3D cases, the quantum states exhibit the strongest three-body correlations for nearly equilateral triangles with $\theta \approx 60^\circ$, whereas the classical states do so for triplets of particles that are chain-like, with $\theta \approx 180^\circ$.
The prevalence of nearly equilateral triangles is associated with the clustering of particles~\cite{Wa23}.
To the contrary, particles in the classical systems favor the formation of linear chains and fewer clusters.
These differences in classical and quantum higher-body correlations confirms the prediction by \citet{To08b} that ground states of free Fermi gases in two and higher dimensions cannot be described as classical particles interacting through one- and two-body potentials at a finite temperature, i.e., intrinsic $n$-body interactions with $n \geq 3$ are generally needed.

\section{Discussion and conclusions}
\label{conc}

In this work, we systematically studied the many-body statistics of ground states of interacting electron liquids through the lens of disordered hyperuniformity, which is
characterized by the simultaneous statistical isotropy typical of liquids and suppression of density fluctuations on large scales in the manner of crystals. 
Classical ground states almost invariably have high crystallographic symmetries \cite{We81, Fr96, Tr14}, and one requires highly unusual ``stealthy'' potentials to create disordered hyperuniform ground states \cite{Ba08, Zh17b}.  
In contrast, disordered hyperuniformity is widely prevalent in statistically homogeneous fermionic quantum ground states.
Studying quantum ground states with diverse parent Hamiltonians, boundary conditions, and degrees of freedom (spins, fermions, bosons, etc.) can provide an important and largely unexplored arena to study the formation and design of hyperuniform states of matter with tunable long-wavelength behaviors of their associated structure factors. Moreover, the bipartite fluctuations of particle number and spin densities---as probed by the structure factors studied here---can often be directly related to the entanglement entropy~\cite{PhysRevB.85.035409}, thus providing an experimentally accessible route for its measurement.

In this regard, we obtained several analytical small-$k$ results for statistically homogeneous ground states of interacting fermions and probed the equivalent ``designer'' classical systems that achieve such fermionic pair correlations.
We showed that all such states are hyperuniform, as must be the case due to the fluctuation-compressibility relation. 
Moreover, we find that the small-$k$ scaling exponent $\alpha$ and the coefficient $\mathcal{B}$ in the hyperuniform scaling relation (\ref{scaling}) exhibit rich behaviors as functions of the dimensionality, the interaction parameter, and the polarization.
Specifically, the total structure factor in two and three dimensions is characterized by an exponent $\alpha = (d + 1)/2$ as long as $r_s > 0$---as opposed to the case of the free Fermi gas with $\alpha = 1$---and the origin of this relation can be traced back to the plasmon excitations of the electron liquid.
On the other hand, the individual spin components are always class-II hyperuniform with $\alpha = 1$.
Consequently, partially polarized interacting electron liquids exhibit a highly unusual form of multihyperuniformity, in which the net configuration belongs to a stronger hyperuniformity class than each spin component individually due to the negative density correlations between oppositely aligned spins up to large length scales.
This novel type of multihyperuniform patterns presents abundant opportunities for applications, such as the design of optical arrays~\cite{Zh21b} for computer vision, or in other contexts where accurate sampling of disordered signals on large length scales is desired.
The fact that one can achieve these structures in both quantum and classical systems suggests that the fabrication of such multihyperuniform materials is highly promising.
For example, it has been shown that various functional forms of effective interparticle interactions can be achieved in classical polymer systems \cite{Ya04, Ca12}, the interaction parameters of which are tunable via the chemical composition of the polymer particles, thereby realizing a plethora of prescribed pair statistics \cite{Wa24}. 
Thus, to achieve the aforementioned unusual multihyperuniform states, one can design two types of polymers interacting via some component-specific pair potentials, the large-$r$ asymptotic forms of which are given by Eq.~(\ref{pot_2Dunp}).

Our work also has important implications for the development of density functional theories (DFT) for fermionic states.
Previous DFT formulations have often introduced  functionals using the HNC approximation for pair functions within the RPA framework ~\cite{Da03}, which is exact in the limit $r_s\rightarrow 0$ but does not capture the increase of $\tilde{S}(k)$ at fixed small $k$ as $r_s$ increases, as identified in this work and shown in Fig.~\ref{fig:compare_stls}(c). 
Proceeding beyond the RPA scheme, here, we have derived small-$k$ expressions for the structure factor and pair correlations via the Hubbard approximation that precisely capture this increase.
These expressions, together with our inverse algorithm that determines effective one- and two-body potentials beyond the HNC approximation (Sec. \ref{meth}), can thus be used as constraints for the development of more accurate DFT functionals.

Moreover, our study of the structural degeneracy of pair statistics for free Fermi gases confirms the existence of intrinsic three- and higher-body interactions in fermionic systems in two and higher dimensions, as predicted in Ref.~\cite{To08b}.
Determining classical systems with pair statistics that match those of target quantum states provides an efficient tool to investigate intrinsically quantum effects, such as the inherent wavefunction antisymmetry for fermions, or the ``full counting statistics'' of particle correlations beyond the two-body level. 
A natural extension of our work would be to determine classical $2, \dots, n$-body interactions that generate equilibrium systems matching the corresponding $g_2, \dots, g_n$ correlations for quantum systems.
One could then directly compare the $(n+1)$-body correlations between the classical and the quantum states to find the intrinsically quantum effects on the $(n+1)$-body level. This analytical framework can of course be applied to many other quantum systems, including fermionic gases subject to non-Coulombic interactions as well as bosonic systems.\\

\begin{acknowledgments}
 H. W. and S. T. are supported by the Army Research Office under Cooperative Agreement No. W911NF-22-2-0103. R.S. is supported by the Princeton Quantum Initiative Fellowship.\\
\end{acknowledgments}

\appendix

\section{Small-$k$ behaviors of structure factors}
\label{sec:appA}

Here, we present a detailed derivation of the analytical long-wavelength behaviors of the total and spin-resolved structure factors of  electron liquids described with the RPA and Hubbard approximations. 
We focus here on unpolarized electron liquids, and results for arbitrary polarizations are documented in Sec.~\ref{summary_all_d} in the main text.
Note that for unpolarized fermions, the spin-resolved structure factors $S_{\sigma\sigma'}(k)$  can be expressed by the total and the magnetic structure factors via Eqs.~(\ref{Supup_unp}) and (\ref{Supdown_unp}).

\subsection{Quasi-1D systems}
Here, we show the derivation of the small-$q$ behaviors of $S(q)$ and $\tilde{S}(q)$ for the quasi-1D case, for which the potential is given by Eq.~(\ref{Vq1D}).
The real and imaginary parts of the 1D Lindhard function $\chi_0(q, \omega; r_s)$ are~\cite{Mi11}
\begin{widetext}
\begin{alignat}{1}
    \label{Rechi01D}
    \text{Re}[\chi_0(q, \omega; r_s)] &= \frac{4 r_s} {\pi ^2 q}\left[\ln \left(\left| \frac{2q -q^2 +\nu}{2q +q^2-\nu}\right| \right)-\ln \left(\left| \frac{{2q+q^2+\nu }}{2q-{q^2-\nu }}\right| \right)\right],\\
    \text{Im}[\chi_0(q, \omega; r_s)] &= \begin{cases}
 0, & \quad \nu <2 q-q^2 \text{ or } \nu \geq 2 q + q^2 \\
 -{\displaystyle\frac{4r_s}{\pi  q}}, & \quad 2 q-q^2\leq \nu <2 q + q^2 \end{cases},
 \label{chi01D}
\end{alignat}  
where $\nu = 2m\omega = 8\omega r_s/\pi$.\\

\citet{Si68} showed that Eq.~(\ref{S-chid}) can be written as 
\begin{equation}
\begin{split}
    S(q) &= -\frac{1}{\rho \pi} \frac{\omega}{\nu} \int_{0}^{2q+q^2} \text{Im}[\chi^d(q, \omega; r_s)] d\nu + \frac{1}{\rho} \left(\frac{\partial \text{Re}\frac{1}{\chi^d(q, \omega; r_s)}}{\partial \omega}\right)_{\omega = \omega_p(q)}^{-1},
\end{split}
    \label{S3DSingwi}
\end{equation}
where the second term comes from the singularity of the integrand $\chi^d(q, \omega; r_s)$ at the so-called plasmon frequency $\omega_p(q)$.
The integral along the imaginary axis in Eq.~(\ref{S-chid}) must pass around this singularity by a semicircular path.\\

We now proceed to evaluate both terms of Eq.~\eqref{S3DSingwi}. 
In the RPA scheme, we set $G(q) = 0$ and use Eqs.~(\ref{chid-G}) and (\ref{chi01D}) to obtain 
\begin{alignat}{1}
    \text{Im}[\chi^d(q, \omega; r_s)] &= 
    -4 \pi ^3 q r_s\left[{\displaystyle{\left(\pi ^2 q-4 r_s e^{\frac{\ell^2 q^2}{4}} K_0\left(\frac{\ell^2 q^2}{4}\right) \ln \left(\frac{(q^2-2q+\nu) |2 q-q^2+\nu| }{(q^2+2q-\nu ) | 2q+q^2+\nu | }\right)\right)^2 \hspace*{-0.1cm}+ 16 \pi ^2 r_s^2 e^{\frac{\ell^2 q^2}{2}} K_0\left(\frac{\ell^2 q^2}{4}\right)^2}}\right]^{-1}\\
    \nonumber
    &\mbox{ if }\,\, 2q - q^2 \leq \nu \leq 2q + q^2,
\end{alignat}
and $0$ otherwise.
For fixed $q$, this is a symmetric function about $\nu = 2q$, at which its magnitude achieves a maximum
\begin{equation}
\begin{split}
    \max_{\omega}\left|\text{Im}[\chi^d(q, \omega; r_s)]\right|
    &= \frac{4 \pi ^3 q r_s}{\left(\pi ^2 q-4 r_s e^{\frac{\ell^2 q^2}{4}} \ln \left({\displaystyle\frac{\left| 4 q-q^2\right| }{\left| 4 q+q^2\right| }}\right) K_0\left({\displaystyle\frac{\ell^2 q^2}{4}}\right)\right)^2+16 \pi ^2 r_s^2 e^{\frac{\ell^2 q^2}{2}} K_0\left({\displaystyle\frac{\ell^2 q^2}{4}}\right)^2} \\
    &\sim \frac{\pi }{16 r_s}\frac{q}{(\ln{q})^2}, \quad q \rightarrow 0.
\end{split}
\end{equation}
Thus, we obtain an upper bound for the the first term of Eq.~(\ref{S3DSingwi}):
\begin{equation}
\begin{split}
-\frac{1}{\rho \pi} \frac{\omega}{\nu} \int_{0}^{2q+q^2} \text{Im}[\chi^d(q, \omega; r_s)] d\nu &= -\frac{1}{\rho \pi} \frac{\omega}{\nu} \int_{2q - q^2}^{2q+q^2} \text{Im}[\chi^d(q, \omega; r_s)] d\nu \leq 2q^2 \frac{\pi }{16 r_s}\frac{q}{(\ln{q})^2}
\propto \frac{q^3}{\ln(q)^2}, \quad q \rightarrow 0.
\end{split} 
\end{equation}
\end{widetext}
To evaluate the second term in Eq.~(\ref{S3DSingwi}), we again set $G(q) = 0$ and use Eqs.~(\ref{chid-G}) and (\ref{chi01D}) to find
\begin{equation}
\begin{split}
    &\text{Re}\left[\frac{1}{\chi^d(q, \omega; r_s)}\right] \\
    &=\frac{8 r_s \ln \left(\frac{\ell^2 q^2}{8}\right)+8 \gamma  r_s-\pi ^2}{8 r_s}+\frac{2 r_s \omega ^2}{q^2} + \mathcal{O}(q).
\end{split}
    \label{ReChid1D}
\end{equation}
The plasmon frequency $\omega_p(q)$ is defined by the relation $\text{Re}\left[1/\chi^d(q, \omega_p; r_s)\right] = 0$.
Solving this with Eq.~(\ref{ReChid1D}) gives
\begin{equation}
    \omega_p(q) = \frac{q \sqrt{-8 r_s \ln \left(\frac{\ell^2 q^2}{8}\right)-8 \gamma  r_s+\pi ^2}}{4 r_s},
\end{equation}
which exhibits a characteristic $\vert q^2 \ln q\vert^{1/2}$ form~\cite{sarma1985screening,gold1990analytical,li1992elementary,schulz1993wigner,chou2020nonmonotonic}.
Inserting $\omega_p(q)$ into the second term of Eq.~(\ref{S3DSingwi}) results in an expression proportional to $q/\sqrt{-\ln q}$ at small $q$ (see below), dominating the $q^3/(\ln q)^2$ behavior from the first term. 
Thus, one has
\begin{equation}
\begin{split}
    S_{\text{RPA}}(q) &\sim \frac{1}{\rho}\frac{q}{\sqrt{-8 r_s \ln \left(\frac{\ell^2 q^2}{8}\right)-8 \gamma  r_s + \pi ^2}} \\
    &\sim {\frac{\pi  q}{8 \sqrt{-r_s \ln{q}}}}, \quad q \rightarrow 0.
\end{split}
    \label{SksmallkRPA1D}
\end{equation}

For the magnetic structure factor, we set $I(q) = 0$ in Eq.~(\ref{chis-I}). 
Equations (\ref{tildeS-chis}) and (\ref{chi01D}) yield
\begin{equation}
\begin{split}
    {\tilde{S}_{\text{RPA}}(q)} &= -\frac{1}{\rho\pi}\int_0^\infty \text{Im}[\chi_0(q, \omega; r_s)] \\
    &\sim -\frac{1}{\rho\pi}\frac{\omega}{\nu}\int_{2q-q^2}^{2q+q^2}\frac{-4\pi r_s}{q} d\nu {\sim \frac{q}{2}}, \quad q \rightarrow 0.
\end{split}
\end{equation}
Thus, the spin-resolved structure factors have the small-$k$ behaviors 
\begin{alignat}{2}
    S_{\sigma\sigma, \text{RPA}}(q) &\sim \phantom{-} \frac{q}{4} + \frac{\pi  q}{16 \sqrt{-r_s \ln{q}}}, \quad &&q \rightarrow 0,\\
    S_{\uparrow\downarrow, \text{RPA}}(q) &\sim -\frac{q}{4} + \frac{\pi  q}{16 \sqrt{-r_s \ln{q}}}, \quad &&q \rightarrow 0,
\end{alignat}

For the Hubbard approximation, one can again show that the second term of Eq.~(\ref{S3DSingwi}) dominates the first at small $q$.
Using Eqs.~(\ref{Vq1D}), (\ref{chid-G}), (\ref{HubbardG}),  and (\ref{S3DSingwi}), the small-$q$ behavior of $S(q)$ in the Hubbard approximation is given by
\begin{equation}
    \frac{q}{\sqrt{-8 r_s \ln \left(\frac{\ell^2 q^2}{8}\right)-4 e^{\frac{\ell^2}{4}} r_s K_0\left(\frac{\ell^2}{4}\right)-8 \gamma  r_s+\pi ^2}},
\end{equation}
and consequently,
\begin{alignat}{1}
    \label{S1Dhubbard}
    S_{\text{Hubbard}}(q) &\sim   {\frac{\pi  q}{8 \sqrt{-r_s \ln{q}}}}, \quad q \rightarrow 0.
\end{alignat}

For the magnetic structure factor, one can first obtain the small-$q$ behavior of $\chi^s(q, \omega; r_s)$ from Eqs.~(\ref{chis-I}) and (\ref{Vq1D}), i.e,
\begin{alignat}{1}
    \chi^s(q, \omega; r_s) &= \frac{\chi_0(q, \omega; r_s)}{1 + \frac{1}{2}\tilde{V}(\sqrt{q+1})\chi_0(q, \omega; r_s)} \\
    \nonumber
    &\sim \frac{\chi_0(q, \omega; r_s)}{1 + \frac{1}{2} e^{\frac{\ell^2}{4}} K_0\left(\frac{\ell^2}{4}\right)\chi_0(q, \omega; r_s)}, \quad q\rightarrow 0.
\end{alignat}
Plotting $\chi^s(q, \omega; r_s)$ against $\omega$ at fixed $q$ shows that this function contains a singularity on the interval $0\leq \nu \leq 2q - q^2$ for small $q$.
Thus, we have an expression for $\tilde{S}(q)$ analogous to Eq.~(\ref{S3DSingwi}):
\begin{equation}
\begin{split}
    \tilde{S}(q) &= \frac{1}{\rho \pi} \frac{\omega}{\nu} \int_{0}^{2q+q^2} \text{Im}[\chi^s(q, \omega; r_s)] d\nu\\
    &+ \frac{1}{\rho} \left(\frac{\partial \text{Re}\frac{1}{\chi^s(q, \omega; r_s)}}{\partial \omega}\right)_{\omega = \omega_s(q)}^{-1},
\end{split}
\label{Stilde1D}
\end{equation}
where $\omega_s(q)$ is the frequency of the singularity of $\chi^s(q, \omega; r_s)$.
Solving $1/\chi^s(q, \omega; r_s) = 0$ at small $k$ gives 
\begin{equation}
    \omega_s(q) \sim \frac{\sqrt{q^2 \left[\pi ^2-4 e^{\frac{\ell^2}{4}} r_s K_0\left(\frac{\ell^2}{4}\right)\right]}}{4 r_s}, \quad q\rightarrow 0.
    \label{omegas}
\end{equation}
Equations (\ref{Stilde1D}) and (\ref{omegas}) then result in 
\begin{equation}
\begin{split}
    \tilde{S}_{\text{Hubbard}}(q) &\sim \frac{q}{\sqrt{\pi ^2-4 e^{\frac{\ell^2}{4}} r_s K_0\left(\frac{\ell^2}{4}\right)}}\\
    &\sim {\left[ \frac{1}{2} + \frac{e^{\frac{\ell^2}{4}} K_0\left(\frac{\ell^2}{4}\right) r_s}{\pi ^2}\right]q}, \quad q \rightarrow 0.
\end{split}
\label{Stilde1DHubbard}
\end{equation}
To validate Eq.~(\ref{Stilde1DHubbard}), we first set $r_s = 1, \ell = 2/m = (2\pi)/(4r_s) = \pi/2$, as done by \citet{Ta96}, and find that $\tilde{S}_{\text{Hubbard}}(q) \sim 0.642 q$.
This is consistent with the numerical calculations for the STLS model in Ref.~\cite{Ta96}, which show that $S(q)$ is linear in $q$ up to $q = 1.5$ and that $\tilde{S}(q = 1) \approx 0.64$.

Equations (\ref{S1Dhubbard}) and (\ref{Stilde1DHubbard}) yield the small-$k$ behaviors of the spin-resolved structure factors as $q \rightarrow 0$:
\begin{alignat}{1}
    S_{\sigma\sigma, \text{Hubbard}}(q) &\sim \left[ \frac{1}{4} + \frac{e^{\frac{\ell^2}{4}} K_0\left(\frac{\ell^2}{4}\right) r_s}{2\pi ^2}\right]q + \frac{\pi  q}{16 \sqrt{-r_s \ln{q}}},\\
    S_{\uparrow\downarrow, \text{Hubbard}}(q) &\sim -\left[ \frac{1}{4} + \frac{e^{\frac{\ell^2}{4}} K_0\left(\frac{\ell^2}{4}\right) r_s}{2\pi ^2}\right]q + \frac{\pi  q}{16 \sqrt{-r_s \ln{q}}}. 
\end{alignat}\\

\subsection{Two dimensions}
To find $S(k)$ and $\tilde{S}(k)$ for 2D interacting unpolarized fermions with the RPA and Hubbard approximations, we follow the procedure outlined in Ref.~\cite{Mo95}, which simplifies Eqs.~(\ref{S-chid})--(\ref{chis-I}) to 
\begin{widetext}
\begin{alignat}{1}
    S(q) &= \frac{2q^2}{\pi} \int_0^{\alpha(q)}\left(\sqrt{1-\frac{q^2 \sin^2(\beta)}{4}}
    +\frac{\cot^2(\beta)}{\sqrt{1-\frac{q^2 \sin^2(\beta)}{4}}}\right)\frac{1-\cos(\beta)}{q+\sqrt{2}r_s[1-\cos(\beta)][1-G(q)]} d\beta,
    \label{G-S}\\
    \tilde{S}(q) &= \frac{2q^2}{\pi} \int_0^{\alpha(q)}\left(\sqrt{1-\frac{q^2 \sin^2(\beta)}{4}}
    +\frac{\cot^2(\beta)}{\sqrt{1-\frac{q^2 \sin^2(\beta)}{4}}}\right)\frac{1-\cos(\beta)}{q+\sqrt{2}r_s[1-\cos(\beta)]I(q)} d\beta,
    \label{I-STilde}
\end{alignat}
\end{widetext}
where
\begin{equation}
    \alpha(q) = 
    \begin{cases}
        \pi/2, \quad &q < 2\\
        \arcsin(2/q), \quad &q \geq 2.
    \end{cases}
\end{equation}
To derive the small-$q$ behavior of $S(q)$ and $\tilde{S}(q)$ for $r_s>0$ within the RPA model, we set the corresponding LFCs $G(q) = I(q) = 0$ in Eqs.~(\ref{G-S}) and (\ref{I-STilde}). 
The integrals now have closed-form solutions in the limit of small $q$:
\begin{widetext}
\begin{alignat}{1}
\nonumber
    {S_{\text{RPA}}(q) \sim}& \frac{2q^2}{\pi} \int_{0}^{\pi/2}\frac{[\csc ^2(\beta )+\frac{1}{8} q^2 \cos (2 \beta ) + \mathcal{O}(q^4)](1-\cos(\beta))}{q+\sqrt{2}r_s[1-\cos(\beta)]}d\beta 
     = \frac{2q^2}{\pi} \int_{0}^{\pi/2}\hspace*{-0.3cm}\frac{1}{q[1 + \cos(\beta)]+\sqrt{2}r_s\sin^2(\beta)}d\beta + \mathcal{O}(q^{9/2})\\
     \label{StotalRPA}
     =& \frac{2 q^2}{\pi} \left(\frac{1}{q+2 \sqrt{2} r_s}+\frac{2 r_s \left(\sqrt{2} q+4 r_s\right) \tan ^{-1}\left(\frac{\sqrt{q+2 \sqrt{2} r_s}}{\sqrt{q}}\right)}{\sqrt{q} \left(q+2 \sqrt{2} r_s\right)^{5/2}}\right) + \mathcal{O}(q^{9/2})={\frac{1}{2^{3/4} \sqrt{r_s}}q^{3/2}}  + \mathcal{O}(q^{5/2}), \quad q \rightarrow 0,
\\
\nonumber
    \tilde{S}_{\text{RPA}}(q) &\sim \frac{2q^2}{\pi} \int_{0}^{\pi/2}\left[\frac{(1 + \cot^2(\beta))(1-\cos(\beta))}{q} + \mathcal{O}(q) \right]d\beta  = \frac{2q^2}{\pi} \int_{0}^{\pi/2}\frac{1}{q[1 + \cos(\beta)]}d\beta  + \mathcal{O}(q^3) \\
    \label{RPAS2D}
    &= {\frac{2 q}{\pi}} + \mathcal{O}(q^3), \quad q \rightarrow 0.
\end{alignat}
\end{widetext}
Using Eqs.~(\ref{Supup_unp}) and (\ref{Supdown_unp}), we find that the same- and opposite-spin small-$q$ expressions for RPA are given by
\begin{alignat}{2}
    S_{\sigma\sigma, \text{RPA}}(q) &\sim \phantom{-}\frac{q}{\pi} + \frac{1}{2^{7/4}\sqrt{r_s}}q^{3/2}, \quad &&q \rightarrow 0,\\
    S_{\uparrow\downarrow, \text{RPA}}(q) &= - \frac{q}{\pi} + \frac{1}{2^{7/4}\sqrt{r_s}}q^{3/2}, \quad &&q \rightarrow 0.
\end{alignat}

In the Hubbard approximation, the LFC (\ref{HubbardG}) for 2D electron liquids is given by $G_{\text{Hubbard}}(q) = -I_{\text{Hubbard}}(q) = \frac{1}{2}\sqrt{q^2/(q^2 + 1)}$. 
Thus, at small $q$, we have $G_{\text{Hubbard}}(q) = -I_{\text{Hubbard}}(q) \sim q/2$.
Inserting these expressions into Eqs.~(\ref{G-S}) and (\ref{I-STilde}) yields the small-$q$ behaviors of the total and the magnetic structure factors:
\begin{widetext}
\begin{alignat}{1}
\nonumber
    S_{\text{Hubbard}}(q) =& \frac{2q^2}{\pi} \int_{0}^{\frac{\pi}{2}}\frac{[\csc ^2(\beta )+\frac{1}{8} q^2 \cos (2 \beta ) + \mathcal{O}(q^4)](1-\cos(\beta))}{q+\sqrt{2}r_s[1-\cos(\beta)](1 - \frac{q}{2})}d\beta \\
\nonumber
    =& \frac{2q^2}{\pi} \int_{0}^{\frac{\pi}{2}}\frac{1}{q[1 + \cos(\beta)]+\sqrt{2}r_s\sin^2(\beta)(1 - \frac{q}{2})}d\beta + \mathcal{O}(q^{9/2})\\
\nonumber
    =& \frac{2 q^2}{\pi} \left(\frac{1}{-\sqrt{2} q r_s+q+2 \sqrt{2} r_s}+\frac{\sqrt{2} (q-2) r_s \tanh ^{-1}\left(\frac{\sqrt{\sqrt{2} q r_s-q-2 \sqrt{2} r_s}}{\sqrt{q}}\right)}{\sqrt{q} \left(q \left(\sqrt{2} r_s-1\right)-2 \sqrt{2} r_s\right)^{3/2}}\right) + \mathcal{O}(q^{9/2})\\
    \label{hubbardStotal}
    =& \frac{1}{2^{3/4} \sqrt{r_s}}q^{3/2} + \mathcal{O}(q^{5/2}), \quad q \rightarrow 0\\
\nonumber
    \tilde{S}_{\text{Hubbard}}(q) & =  \frac{2q^2}{\pi} \int_{0}^{\frac{\pi}{2}}\left[\frac{\sec ^2\left(\frac{\beta }{2}\right)}{q \left[\sqrt{2} r_s \cos (\beta ) - \sqrt{2}r_s + 2\right]} + \mathcal{O}(q) \right]d\beta = \frac{2 q \left(\frac{1}{1-\sqrt{2} \text{rs}}+\frac{\sqrt{2} r_s \tanh ^{-1}\left(\sqrt{\sqrt{2} r_s - 1}\right)}{\left(\sqrt{2} r_s - 1\right)^{3/2}}\right)}{\pi } + \mathcal{O}(q^3)\\
    & = \left[\frac{2}{\pi} + \frac{\left(4 \sqrt{2}-\sqrt{2} \pi \right) r_s}{2 \pi } + \mathcal{O}(r_s^2)\right]q + \mathcal{O}(q^3), \quad q \rightarrow 0.
\label{hubbardS2D}
\end{alignat}
\end{widetext}
Using Eqs.~(\ref{Supup_unp}) and (\ref{Supdown_unp}), we find that the same- and opposite-spin small-$q$ structure factors for  the Hubbard approximation are given by
\begin{widetext}
\begin{alignat}{2}
\label{2DSupup}
   S_{\sigma\sigma, \text{Hubbard}}(q) &\sim\phantom{-} \left[\frac{1}{\pi} + \frac{\left(4 \sqrt{2}-\sqrt{2} \pi \right) r_s}{4 \pi }\right]q + \frac{1}{2^{7/4}\sqrt{r_s}}q^{3/2}, \quad &&q \rightarrow 0,\\
   \label{2DSupdown}
    S_{\uparrow\downarrow, \text{Hubbard}}(q) &\sim - \left[\frac{1}{\pi} + \frac{\left(4 \sqrt{2}-\sqrt{2} \pi \right) r_s}{4 \pi }\right]q + \frac{1}{2^{7/4}\sqrt{r_s}}q^{3/2}, \quad &&q \rightarrow 0.
\end{alignat}
\end{widetext}
\subsection{Three dimensions}
The small-$k$ behaviors of the spin-resolved structure factors for 3D unpolarized interacting Fermi gases within the RPA model have been derived by Wang and Perdew~\cite{Wa91}.
However, we include the derivation here for completeness.
The real and imaginary parts of the 3D Lindhard function are given by~\cite{Mi11}
\begin{widetext}
\begin{subequations}
\begin{equation}
    \text{Re}[\chi_0(q, \omega; r_s)] = \frac{ r_s}{(18\pi^7)^{1/3}} \left( \frac{\left(4 q^2-\left(\nu -q^2\right)^2\right) \ln \left| {\displaystyle\frac{2q -q^2 +\nu}{2q +q^2-\nu}}\right| - \left(4 q^2-\left(\nu +q^2\right)^2\right) \ln \left| {\displaystyle\frac{{2q+q^2+\nu }}{2q-{q^2-\nu }}}\right| }{8 q^3}-1\right)
    \label{rechi03d}
\end{equation}
\begin{equation}
    \text{Im}[\chi_0(q, \omega; r_s)] =
    \begin{cases}
        -{\displaystyle\frac{\nu r_s }{2^{4/3} 3^{2/3} \pi ^{4/3} q}}, \quad &\nu < 2q - q^2\\[1ex]
        {\displaystyle\frac{[\nu ^2+q^4-2 (\nu +2) q^2]r_s}{2^{10/3}\, 3^{2/3}\, \pi ^{4/3}\, q^3}}, \quad &|2q - q^2| \leq \nu < 2q + q^2\\
        0, \quad &\nu \geq 2q + q^2,
    \end{cases}
    \label{imchi03d}
\end{equation}
\label{chi03d}
\end{subequations}
\end{widetext}
where $\nu = 2m\omega = (2 \left(2/3\right)^{2/3} \omega r_s )/(\sqrt[3]{\pi })$ is a scaled frequency.

We proceed to evaluate both terms of Eq.~(\ref{S3DSingwi}) in the RPA framework. 
Setting $G(q) = 0$ in Eq.~(\ref{chid-G}), $\text{Im}[\chi^d(q, \omega; r_s)]$ can be derived from Eqs.~(\ref{chid-G}) and (\ref{imchi03d}).
As is the case for Eq.~(\ref{chi03d}), $\text{Im}[\chi^d(q, \omega; r_s)]$ is also continuous and piecewise smooth.
At small $q$, the portion of $\text{Im}[\chi^d(q, \omega; r_s)]$ on the interval $0 \leq \nu < 2q - q^2$ has the largest contribution to the integral in the first term of Eq.~(\ref{S3DSingwi}).
It is given by 
\begin{widetext}
\begin{equation}
\begin{split}
\text{Im}[\chi^d(q, \omega; r_s)] 
&= r_s\frac{\text{Im}[\chi_0(q, \omega; r_s)]\, q^4}{\left(q^2 -4 \pi  \text{Re}[\chi_0(q, \omega; r_s)]\right)^2 + \left(4 \pi  \text{Im}[\chi_0(q, \omega; r_s)]\right)^2}\\[0.25ex]
&= \mathcal{O}(q^3)\nu + \mathcal{O}(q)\nu^3 + \mathcal{O}(q^{-1})\nu^5 + \dots + \mathcal{O}(q^4)\nu^2 + \mathcal{O}(q^2)\nu^4 + \mathcal{O}(1)\nu^6 + \dots,
\quad \nu < 2q - q^2.
\end{split}
\end{equation}
The first term of Eq.~(\ref{S3DSingwi}) is approximated as
\begin{equation}
    -\frac{1}{\rho \pi} \frac{\omega}{\nu} \int_{0}^{2q+q^2} \text{Im}[\chi^d(q, \omega; r_s)]\, d\nu \sim -\frac{1}{\rho \pi} \frac{\omega}{\nu} \int_{0}^{2q-q^2} \text{Im}[\chi^d(q, \omega; r_s)]\, d\nu \propto q^5, \quad q\rightarrow 0.
\end{equation}  
\end{widetext}
Numerical integration confirms the $q^5$ behavior of the first term of Eq.~(\ref{S3DSingwi}).

To evaluate the second term in Eq.~(\ref{S3DSingwi}), we note that the small-$q$ RPA behavior of $\text{Re}[\chi^d(q, \omega; r_s)]$  can be derived from Eqs.~(\ref{chid-G}) and (\ref{rechi03d}), setting $G(q) = 0$:
\begin{widetext}
\begin{equation}
    \text{Re}[\chi^d(q, \omega; r_s)] \sim \frac{\left(\frac{3}{2}\right)^{2/3} q^2}{\pi  \left(3 \pi ^{2/3} \omega^2 r_s-2^{4/3} 3^{2/3}\right)} + \mathcal{O}\left(q^3\right), \quad q \rightarrow 0.
    \label{Rechidsmallq}
\end{equation}
\end{widetext}
The singularity of $\chi^d(q, \omega; r_s)$ occurs where the denominator in Eq.~(\ref{Rechidsmallq}) vanishes, i.e.,
\begin{equation}
    \omega_p(q) = \frac{2^{2/3}}{3^{1/6} \pi^{1/3}  \sqrt{r_s}}, \quad q \rightarrow 0.
    \label{omegap0}
\end{equation}
Inserting Eqs.~(\ref{Rechidsmallq}) and (\ref{omegap0}) into the second term of Eq.~(\ref{S3DSingwi}) yields an expression quadratic in $q$, which dominates the $q^5$ behavior due to the first term. 
Thus, we have
\begin{equation}
    {S_{\text{RPA}}(q) \sim \frac{3^{5/6} \pi ^{2/3} q^2}{4 \sqrt[3]{2} \sqrt{r_s}}}, \quad q \rightarrow 0,
    \label{SRPA3D}
\end{equation}
which has exactly the same form as the known plasmon contribution (\ref{plasmon}), with $\rho = 1/(3\pi^2), m = r_s[4\pi\rho/3]^{1/3} = \left(2/3\right)^{2/3} r_s/\pi^{1/3}$ and $\tilde{V}(q) = 4\pi/q^2$. 

For the magnetic structure factor $\tilde{S}(q)$, one can show from Eqs.~(\ref{tildeS-chis}) and (\ref{chis-I}) with $I(q) = 0$ that
\begin{equation}
\begin{split}
    {\tilde{S}_{\text{RPA}}(q)} &= -\frac{1}{\rho\pi}\int_0^\infty \text{Im}[\chi_0(q, \omega; r_s)] d\omega \\
    &= -\frac{1}{\rho\pi}\frac{\omega}{\nu} \left(\int_0^{2q - q^2} -\frac{\nu r_s}{2^{4/3} 3^{2/3} \pi ^{4/3} q} d\nu \right. \\
    &\left. \quad+ \int_{2q - q^2}^{2q + q^2} \frac{[\nu ^2+q^4-2 (\nu +2) q^2]r_s}{2^{10/3} 3^{2/3} \pi ^{4/3} q^3} d\nu \right) \\
    &{\sim \frac{3q}{4} - \frac{q^3}{16}}, \quad q \rightarrow 0,
\end{split}
    \label{StildeRPA3D}
\end{equation}
where we have used the fact that in the small-$q$ limit, only the interval $0 < \nu < 2q - q^2$ in Eq.~(\ref{imchi03d}) contributes to the linear term in $\tilde{S}(k)$.
Combining Eqs.~(\ref{SRPA3D}) and (\ref{StildeRPA3D}) yields the spin-resolved structure factors
\begin{subequations}
\begin{equation}
    {S_{\sigma\sigma, \text{RPA}}(q) \sim \phantom{-} \frac{3q}{8} + \frac{3^{5/6} \pi ^{2/3} q^2}{8 \sqrt[3]{2} \sqrt{r_s}}}, \quad q \rightarrow 0,
\end{equation}
\begin{equation}
    {S_{\uparrow\downarrow, \text{RPA}}(q) \sim -\frac{3q}{8} + \frac{3^{5/6} \pi ^{2/3} q^2}{8 \sqrt[3]{2} \sqrt{r_s}}}, \quad q \rightarrow 0.
\end{equation}
\label{RPASresolved3D}
\end{subequations}

For the 3D Hubbard approximation, one has $G_{\text{Hubbard}}(q) = -I_{\text{Hubbard}}(q) = (q^2 + 1)/(2q^2)$.
Inserting $G_{\text{Hubbard}}(k)$ into Eq.~(\ref{chid-G}) and using Eq.~(\ref{rechi03d}) leads to the same form of $\text{Re}[\chi^d(q, \omega; r_s)]$ as Eq.~(\ref{Rechidsmallq}). 
Thus, $S(q)$ in the Hubbard approximation has the same leading quadratic term as in the RPA:
\begin{equation}
    {S_{\text{Hubbard}}(q) \sim \frac{3^{5/6} \pi ^{2/3} q^2}{4 \sqrt[3]{2} \sqrt{r_s}}}, \quad q \rightarrow 0.
    \label{StotalHubbard3D}
\end{equation}

To study the magnetic structure factor in the Hubbard approximation, we use the fact that $I_{\text{Hubbard}}(q) \sim q^2/2$ at small $q$, and rewrite Eq.~(\ref{chis-I}) as
\begin{alignat}{1}
\nonumber
    \frac{\chi^s(q, \omega; r_s)}{-g^2\mu_B^2} &\sim \frac{\chi_0(q, \omega; r_s)}{1 - \frac{4\pi}{q^2}\chi_0(q, \omega; r_s)\frac{q^2}{2}}\\
    &= \frac{\chi_0(q, \omega; r_s)}{1 - 2\pi \chi_0(q, \omega; r_s)}, \quad q \rightarrow 0.
\end{alignat}
Equation (\ref{chi03d}) indicates that $\chi_0(q, \omega; r_s)/r_s$ only depends on $q$ and $\nu$.
We set $\alpha_0(q, \nu)$ and $\beta_0(q, \nu)$ to be the real and imaginary parts of $\chi_0(q, \omega; r_s)/r_s$, respectively.
This gives a small-$r_s$ expansion of $\chi^s(q, \omega; r_s)$:
\begin{equation}
\begin{split}
     \frac{\text{Im}[\chi^s(q, \omega; r_s)]}{-g^2\mu_B^2} &= \frac{\beta_0 r_s}{1 + 4 \pi r_s \alpha_0 + 4\pi^2 r_s^2(\alpha_0^2 + \beta_0^2)} \\
     &\sim r_s[\beta_0 - 4\pi r_s \alpha_0\beta_0 + \mathcal{O}(r_s^2)], \\
     &\quad\,\, q\rightarrow 0, r_s \rightarrow 0,
\end{split}
\label{chis3d_smallrS}
\end{equation}
where we have omitted the arguments of $\alpha_0(q, \nu)$ and $\beta_0(q, \nu)$ for clarity. 
From Eq.~(\ref{tildeS-chis}), we then find
\begin{widetext}
\begin{equation}
\begin{split}
    {\tilde{S}_{\text{Hubbard}}(q) \sim} & \frac{1}{\rho\pi}\frac{\omega}{\nu}r_s\int_0^{2q - q^2} \beta_0(q, \nu) d\nu - \left(\frac{4}{\rho}\frac{\omega}{\nu}r_s\right) r_s \int_0^{2q - q^2} \alpha_0(q, \nu)\beta_0(q, \nu) d\nu \\
    \sim& \frac{3q}{4} + \left(\frac{3}{1024\pi^4}\right)^{\frac{1}{3}}r_s \times \\
    &\int_0^{2q - q^2} \left(-\frac{\left(\nu ^2+q^4-2 (\nu +2) q^2\right) \tanh ^{-1}\left(\frac{\nu -q^2}{2 q}\right)}{q^3}+\frac{\left(\nu ^2+q^4+2 (\nu -2) q^2\right) \tanh ^{-1}\left(\frac{\nu +q^2}{2 q}\right)}{q^3}-4\right)\frac{\nu}{q} d\nu\\
    \sim & {\left[\frac{3}{4} + \frac{2 \left(\frac{2}{3}\right)^{2/3} (1 - \ln 2)}{\pi ^{4/3}} r_s\right] q}, \quad q \rightarrow 0.
\end{split}
\end{equation}
\end{widetext}
Thus, the leading-order terms of the spin-resolved structure factors are given by 
\begin{widetext}
\begin{subequations}
\begin{equation}
    {S_{\sigma\sigma, \text{Hubbard}}(q)  \sim  \phantom{-}
    \left(\frac{3}{8} + \frac{\left(\frac{2}{3}\right)^{2/3} (1 - \ln 2)}{\pi ^{4/3}} r_s \right)q + \frac{3^{5/6} \pi ^{2/3} q^2}{8 \sqrt[3]{2} \sqrt{r_s}}}, \quad q \rightarrow 0, 
    \label{hubbardSupup3D}
\end{equation}
\begin{equation}
    {S_{\uparrow\downarrow, \text{Hubbard}}(q) \sim 
    -\left(\frac{3}{8} + \frac{\left(\frac{2}{3}\right)^{2/3} (1 - \ln 2)}{\pi ^{4/3}} r_s \right)q + \frac{3^{5/6} \pi ^{2/3} q^2}{8 \sqrt[3]{2} \sqrt{r_s}}}, \quad q \rightarrow 0.
\end{equation}
\end{subequations}
\end{widetext}

\section{Effective classical potentials}
\label{sec:appB}
Here, we list the explicit forms of the effective classical potentials for the fermionic systems considered in Sec.~\ref{class}. 
Note that the pair distances $r$ are measured in units of $k_F^{-1}$.

\paragraph{1D free fermions:}
\begin{equation}
    \mathcal{V}(r) = -2\ln(r)
\end{equation}

\paragraph{2D free fermions:}
\begin{equation}
    \mathcal{V}(r) = -\varepsilon_1\exp\left(-\frac{r^{*2}}{\sigma_1^2}\right)\ln(r^*) + \frac{\sqrt{\pi}}{2r^*}\left[1 - \exp\left(-\frac{r^{*2}}{\sigma_2^2}\right)\right],
\end{equation}
where $r^* = r/\sqrt{2\pi}$, $\varepsilon_1 = 2.099, \sigma_1 = 0.6666$, and $\sigma_2 = 0.4739$.

\paragraph{3D free fermions:}
\begin{equation}
\begin{split}
    \mathcal{V}(r) &= -\varepsilon_1\exp\left(-\frac{r^{*2}}{\sigma_1^2}\right)\ln(r^*) \\
    &+ \frac{2 \sqrt[3]{2}}{3^{2/3} \pi ^{4/3} r^{*2}}\left[1 - \exp\left(-\frac{r^{*2}}{\sigma_2^2}\right)\right],
\end{split}
\end{equation}
where $r^* = r/(3\pi^2)^{1/3}$, $\varepsilon_1 = 1.303$, $\sigma_1 = 0.7198$, and $\sigma_2 = 0.5175$.

\paragraph{2D unpolarized electron liquid with $r_s = 1$:}
\begin{alignat}{1}
\nonumber
    \mathcal{V}_{\sigma\sigma}(r) &= \varepsilon_1^{(1)}\exp\left(-\frac{r^2}{\sigma_1^2}\right)\left(-\ln(r) + \varepsilon_1^{(2)}\right) \\
    &+ \left[1 - \exp\left(-\frac{r^2}{\sigma_2^2}\right)\right]\times \\
    \nonumber
    &\phantom{+}\,\left[\frac{1}{2 \left(\frac{4 \sqrt{2}-\sqrt{2} \pi } {4 \pi}+\frac{1}{\pi }\right) r} +\frac{\sqrt[4]{2} \Gamma \left(\frac{1}{4}\right)}{\sqrt{r} \Gamma \left(\frac{3}{4}\right)}\right],\\
    \nonumber
    \mathcal{V}_{\uparrow\downarrow}(r) &= 
    \varepsilon_3\exp\left(-\frac{r}{\sigma_3}\right) + \varepsilon_4\exp\left(-\frac{r^2}{\sigma_4^2}\right) \\
    &+ \left[1 - \exp\left(-\frac{r^2}{\sigma_5^2}\right)\right]\times\\
    \nonumber
    &\phantom{+}\,\left[-\frac{1}{2 \left(\frac{4 \sqrt{2}-\sqrt{2} \pi } {4 \pi}+\frac{1}{\pi }\right) r} +\frac{\sqrt[4]{2} \Gamma \left(\frac{1}{4}\right)}{\sqrt{r} \Gamma \left(\frac{3}{4}\right)}\right],
\end{alignat}
where $\varepsilon_1^{(1)} = 3.192$, $\sigma_1 = 2.172$, $\varepsilon_1^{(2)} = 1.309$, $\sigma_2 = 1.537$, $\varepsilon_3 = 1.486$, $\sigma_3 = 0.7748$, $\varepsilon_4 = 2.180$, $\sigma_4 = 1.141$, and $\sigma_5 = 1.238$.\\


\begin{thebibliography}{120}%
\makeatletter
\providecommand \@ifxundefined [1]{%
 \@ifx{#1\undefined}
}%
\providecommand \@ifnum [1]{%
 \ifnum #1\expandafter \@firstoftwo
 \else \expandafter \@secondoftwo
 \fi
}%
\providecommand \@ifx [1]{%
 \ifx #1\expandafter \@firstoftwo
 \else \expandafter \@secondoftwo
 \fi
}%
\providecommand \natexlab [1]{#1}%
\providecommand \enquote  [1]{``#1''}%
\providecommand \bibnamefont  [1]{#1}%
\providecommand \bibfnamefont [1]{#1}%
\providecommand \citenamefont [1]{#1}%
\providecommand \href@noop [0]{\@secondoftwo}%
\providecommand \href [0]{\begingroup \@sanitize@url \@href}%
\providecommand \@href[1]{\@@startlink{#1}\@@href}%
\providecommand \@@href[1]{\endgroup#1\@@endlink}%
\providecommand \@sanitize@url [0]{\catcode `\\12\catcode `\$12\catcode `\&12\catcode `\#12\catcode `\^12\catcode `\_12\catcode `\%12\relax}%
\providecommand \@@startlink[1]{}%
\providecommand \@@endlink[0]{}%
\providecommand \url  [0]{\begingroup\@sanitize@url \@url }%
\providecommand \@url [1]{\endgroup\@href {#1}{\urlprefix }}%
\providecommand \urlprefix  [0]{URL }%
\providecommand \Eprint [0]{\href }%
\providecommand \doibase [0]{https://doi.org/}%
\providecommand \selectlanguage [0]{\@gobble}%
\providecommand \bibinfo  [0]{\@secondoftwo}%
\providecommand \bibfield  [0]{\@secondoftwo}%
\providecommand \translation [1]{[#1]}%
\providecommand \BibitemOpen [0]{}%
\providecommand \bibitemStop [0]{}%
\providecommand \bibitemNoStop [0]{.\EOS\space}%
\providecommand \EOS [0]{\spacefactor3000\relax}%
\providecommand \BibitemShut  [1]{\csname bibitem#1\endcsname}%
\let\auto@bib@innerbib\@empty
\bibitem [{\citenamefont {Fisher}(1967)}]{Fi67}%
  \BibitemOpen
  \bibfield  {author} {\bibinfo {author} {\bibfnamefont {M.~E.}\ \bibnamefont {Fisher}},\ }\bibfield  {title} {\bibinfo {title} {The theory of equilibrium critical phenomena},\ }\href {https://doi.org/10.1088/0034-4885/30/2/306} {\bibfield  {journal} {\bibinfo  {journal} {Rep. Prog. Phys.}\ }\textbf {\bibinfo {volume} {30}},\ \bibinfo {pages} {615} (\bibinfo {year} {1967})}\BibitemShut {NoStop}%
\bibitem [{\citenamefont {Hansen}(1973)}]{Ha73}%
  \BibitemOpen
  \bibfield  {author} {\bibinfo {author} {\bibfnamefont {J.~P.}\ \bibnamefont {Hansen}},\ }\bibfield  {title} {\bibinfo {title} {{Statistical Mechanics of Dense Ionized Matter. I. Equilibrium Properties of the Classical One-Component Plasma}},\ }\href {https://doi.org/10.1103/PhysRevA.8.3096} {\bibfield  {journal} {\bibinfo  {journal} {Phys. Rev. A}\ }\textbf {\bibinfo {volume} {8}},\ \bibinfo {pages} {3096} (\bibinfo {year} {1973})}\BibitemShut {NoStop}%
\bibitem [{\citenamefont {Ziff}(1977)}]{Zi77}%
  \BibitemOpen
  \bibfield  {author} {\bibinfo {author} {\bibfnamefont {R.~M.}\ \bibnamefont {Ziff}},\ }\bibfield  {title} {\bibinfo {title} {On the bulk distribution functions and fluctuation theorems},\ }\href {https://doi.org/10.1063/1.523496} {\bibfield  {journal} {\bibinfo  {journal} {J. Math. Phys.}\ }\textbf {\bibinfo {volume} {18}},\ \bibinfo {pages} {1825} (\bibinfo {year} {1977})}\BibitemShut {NoStop}%
\bibitem [{\citenamefont {Landau}\ and\ \citenamefont {Lifshitz}(1980)}]{La80}%
  \BibitemOpen
  \bibfield  {author} {\bibinfo {author} {\bibfnamefont {L.~D.}\ \bibnamefont {Landau}}\ and\ \bibinfo {author} {\bibfnamefont {E.~M.}\ \bibnamefont {Lifshitz}},\ }\href@noop {} {\emph {\bibinfo {title} {Statistical Physics}}}\ (\bibinfo  {publisher} {Pergamon Press},\ \bibinfo {address} {New York},\ \bibinfo {year} {1980})\BibitemShut {NoStop}%
\bibitem [{\citenamefont {Torquato}\ and\ \citenamefont {Stillinger}(2003)}]{To03a}%
  \BibitemOpen
  \bibfield  {author} {\bibinfo {author} {\bibfnamefont {S.}~\bibnamefont {Torquato}}\ and\ \bibinfo {author} {\bibfnamefont {F.~H.}\ \bibnamefont {Stillinger}},\ }\bibfield  {title} {\bibinfo {title} {Local density fluctuations, hyperuniform systems, and order metrics},\ }\href {https://doi.org/10.1103/PhysRevE.68.041113} {\bibfield  {journal} {\bibinfo  {journal} {Phys. Rev. E}\ }\textbf {\bibinfo {volume} {68}},\ \bibinfo {pages} {041113} (\bibinfo {year} {2003})}\BibitemShut {NoStop}%
\bibitem [{\citenamefont {Torquato}(2018)}]{To18a}%
  \BibitemOpen
  \bibfield  {author} {\bibinfo {author} {\bibfnamefont {S.}~\bibnamefont {Torquato}},\ }\bibfield  {title} {\bibinfo {title} {Hyperuniform states of matter},\ }\href {https://doi.org/10.1016/j.physrep.2018.03.001} {\bibfield  {journal} {\bibinfo  {journal} {Physics Reports}\ }\textbf {\bibinfo {volume} {745}},\ \bibinfo {pages} {1} (\bibinfo {year} {2018})}\BibitemShut {NoStop}%
\bibitem [{\citenamefont {Florescu}\ \emph {et~al.}(2009)\citenamefont {Florescu}, \citenamefont {Torquato},\ and\ \citenamefont {Steinhardt}}]{Fl09b}%
  \BibitemOpen
  \bibfield  {author} {\bibinfo {author} {\bibfnamefont {M.}~\bibnamefont {Florescu}}, \bibinfo {author} {\bibfnamefont {S.}~\bibnamefont {Torquato}},\ and\ \bibinfo {author} {\bibfnamefont {P.~J.}\ \bibnamefont {Steinhardt}},\ }\bibfield  {title} {\bibinfo {title} {Designer disordered materials with large complete photonic band gaps},\ }\href {https://doi.org/10.1073/pnas.0907744106} {\bibfield  {journal} {\bibinfo  {journal} {Proc. Nat. Acad. Sci. U. S. A.}\ }\textbf {\bibinfo {volume} {106}},\ \bibinfo {pages} {20658} (\bibinfo {year} {2009})}\BibitemShut {NoStop}%
\bibitem [{\citenamefont {Marcotte}\ \emph {et~al.}(2013)\citenamefont {Marcotte}, \citenamefont {Stillinger},\ and\ \citenamefont {Torquato}}]{Ma13a}%
  \BibitemOpen
  \bibfield  {author} {\bibinfo {author} {\bibfnamefont {{\' E}.}~\bibnamefont {Marcotte}}, \bibinfo {author} {\bibfnamefont {F.~H.}\ \bibnamefont {Stillinger}},\ and\ \bibinfo {author} {\bibfnamefont {S.}~\bibnamefont {Torquato}},\ }\bibfield  {title} {\bibinfo {title} {Nonequilibrium static growing length scales in supercooled liquids on approaching the glass transition},\ }\href {https://doi.org/10.1063/1.4769422} {\bibfield  {journal} {\bibinfo  {journal} {J. Chem. Phys.}\ }\textbf {\bibinfo {volume} {138}},\ \bibinfo {pages} {12A508} (\bibinfo {year} {2013})}\BibitemShut {NoStop}%
\bibitem [{\citenamefont {Klatt}\ \emph {et~al.}(2022)\citenamefont {Klatt}, \citenamefont {Steinhardt},\ and\ \citenamefont {Torquato}}]{Kl22}%
  \BibitemOpen
  \bibfield  {author} {\bibinfo {author} {\bibfnamefont {M.~A.}\ \bibnamefont {Klatt}}, \bibinfo {author} {\bibfnamefont {P.~J.}\ \bibnamefont {Steinhardt}},\ and\ \bibinfo {author} {\bibfnamefont {S.}~\bibnamefont {Torquato}},\ }\bibfield  {title} {\bibinfo {title} {Wave propagation and band tails of two-dimensional disordered systems in the thermodynamic limit},\ }\href {https://doi.org/10.1073/pnas.2213633119} {\bibfield  {journal} {\bibinfo  {journal} {Proc. Nat. Acad. Sci. U. S. A.}\ }\textbf {\bibinfo {volume} {119}},\ \bibinfo {pages} {e2213633119} (\bibinfo {year} {2022})}\BibitemShut {NoStop}%
\bibitem [{\citenamefont {Torquato}(2021)}]{To21d}%
  \BibitemOpen
  \bibfield  {author} {\bibinfo {author} {\bibfnamefont {S.}~\bibnamefont {Torquato}},\ }\bibfield  {title} {\bibinfo {title} {Diffusion spreadability as a probe of the microstructure of complex media across length scales},\ }\href {https://doi.org/10.1103/PhysRevE.104.054102} {\bibfield  {journal} {\bibinfo  {journal} {Phys. Rev. E}\ }\textbf {\bibinfo {volume} {104}},\ \bibinfo {pages} {054102} (\bibinfo {year} {2021})}\BibitemShut {NoStop}%
\bibitem [{\citenamefont {Donev}\ \emph {et~al.}(2005)\citenamefont {Donev}, \citenamefont {Stillinger},\ and\ \citenamefont {Torquato}}]{Do05d}%
  \BibitemOpen
  \bibfield  {author} {\bibinfo {author} {\bibfnamefont {A.}~\bibnamefont {Donev}}, \bibinfo {author} {\bibfnamefont {F.~H.}\ \bibnamefont {Stillinger}},\ and\ \bibinfo {author} {\bibfnamefont {S.}~\bibnamefont {Torquato}},\ }\bibfield  {title} {\bibinfo {title} {Unexpected density fluctuations in disordered jammed hard-sphere packings},\ }\href {https://doi.org/10.1103/PhysRevLett.95.090604} {\bibfield  {journal} {\bibinfo  {journal} {Phys. Rev. Lett.}\ }\textbf {\bibinfo {volume} {95}},\ \bibinfo {pages} {090604} (\bibinfo {year} {2005})}\BibitemShut {NoStop}%
\bibitem [{\citenamefont {Zachary}\ and\ \citenamefont {Torquato}(2009)}]{Za09}%
  \BibitemOpen
  \bibfield  {author} {\bibinfo {author} {\bibfnamefont {C.~E.}\ \bibnamefont {Zachary}}\ and\ \bibinfo {author} {\bibfnamefont {S.}~\bibnamefont {Torquato}},\ }\bibfield  {title} {\bibinfo {title} {Hyperuniformity in point patterns and two-phase heterogeneous media},\ }\href {https://doi.org/10.1088/1742-5468/2009/12/P12015} {\bibfield  {journal} {\bibinfo  {journal} {J. Stat. Mech.: Theory \& Exp.}\ }\textbf {\bibinfo {volume} {2009}},\ \bibinfo {pages} {P12015} (\bibinfo {year} {2009})}\BibitemShut {NoStop}%
\bibitem [{\citenamefont {Levesque}\ \emph {et~al.}(2000)\citenamefont {Levesque}, \citenamefont {Weis},\ and\ \citenamefont {Lebowitz}}]{Le00}%
  \BibitemOpen
  \bibfield  {author} {\bibinfo {author} {\bibfnamefont {D.}~\bibnamefont {Levesque}}, \bibinfo {author} {\bibfnamefont {J.-J.}\ \bibnamefont {Weis}},\ and\ \bibinfo {author} {\bibfnamefont {J.}~\bibnamefont {Lebowitz}},\ }\bibfield  {title} {\bibinfo {title} {Charge fluctuations in the two-dimensional one-component plasma},\ }\href {https://doi.org/10.1023/A:1018643829340} {\bibfield  {journal} {\bibinfo  {journal} {J. Stat. Phys.}\ }\textbf {\bibinfo {volume} {100}},\ \bibinfo {pages} {209} (\bibinfo {year} {2000})}\BibitemShut {NoStop}%
\bibitem [{\citenamefont {Zhang}\ \emph {et~al.}(2016)\citenamefont {Zhang}, \citenamefont {Stillinger},\ and\ \citenamefont {Torquato}}]{Zh16a}%
  \BibitemOpen
  \bibfield  {author} {\bibinfo {author} {\bibfnamefont {G.}~\bibnamefont {Zhang}}, \bibinfo {author} {\bibfnamefont {F.~H.}\ \bibnamefont {Stillinger}},\ and\ \bibinfo {author} {\bibfnamefont {S.}~\bibnamefont {Torquato}},\ }\bibfield  {title} {\bibinfo {title} {The perfect glass paradigm: Disordered hyperuniform glasses down to absolute zero},\ }\href {https://doi.org/10.1038/srep36963} {\bibfield  {journal} {\bibinfo  {journal} {Sci. Rep.}\ }\textbf {\bibinfo {volume} {6}},\ \bibinfo {pages} {36963} (\bibinfo {year} {2016})}\BibitemShut {NoStop}%
\bibitem [{\citenamefont {{Hexner}}\ and\ \citenamefont {{Levine}}(2015)}]{He15}%
  \BibitemOpen
  \bibfield  {author} {\bibinfo {author} {\bibfnamefont {D.}~\bibnamefont {{Hexner}}}\ and\ \bibinfo {author} {\bibfnamefont {D.}~\bibnamefont {{Levine}}},\ }\bibfield  {title} {\bibinfo {title} {{Hyperuniformity of critical absorbing states}},\ }\href {https://doi.org/10.1103/PhysRevLett.114.110602} {\bibfield  {journal} {\bibinfo  {journal} {Phys. Rev. Lett.}\ }\textbf {\bibinfo {volume} {114}},\ \bibinfo {pages} {110602} (\bibinfo {year} {2015})}\BibitemShut {NoStop}%
\bibitem [{\citenamefont {Cort{\'{e}}}\ \emph {et~al.}(2008)\citenamefont {Cort{\'{e}}}, \citenamefont {Chaikin}, \citenamefont {Gollub},\ and\ \citenamefont {Pine}}]{Co08}%
  \BibitemOpen
  \bibfield  {author} {\bibinfo {author} {\bibfnamefont {L.}~\bibnamefont {Cort{\'{e}}}}, \bibinfo {author} {\bibfnamefont {P.~M.}\ \bibnamefont {Chaikin}}, \bibinfo {author} {\bibfnamefont {J.~P.}\ \bibnamefont {Gollub}},\ and\ \bibinfo {author} {\bibfnamefont {D.~J.}\ \bibnamefont {Pine}},\ }\bibfield  {title} {\bibinfo {title} {{Random organization in periodically driven systems}},\ }\href {https://doi.org/10.1038/nphys891} {\bibfield  {journal} {\bibinfo  {journal} {Nat. Phys.}\ }\textbf {\bibinfo {volume} {4}},\ \bibinfo {pages} {420} (\bibinfo {year} {2008})}\BibitemShut {NoStop}%
\bibitem [{\citenamefont {Torquato}\ \emph {et~al.}(2000)\citenamefont {Torquato}, \citenamefont {Truskett},\ and\ \citenamefont {Debenedetti}}]{To00b}%
  \BibitemOpen
  \bibfield  {author} {\bibinfo {author} {\bibfnamefont {S.}~\bibnamefont {Torquato}}, \bibinfo {author} {\bibfnamefont {T.~M.}\ \bibnamefont {Truskett}},\ and\ \bibinfo {author} {\bibfnamefont {P.~G.}\ \bibnamefont {Debenedetti}},\ }\bibfield  {title} {\bibinfo {title} {Is random close packing of spheres well defined?},\ }\href {https://doi.org/10.1103/PhysRevLett.84.2064} {\bibfield  {journal} {\bibinfo  {journal} {Phys. Rev. Lett.}\ }\textbf {\bibinfo {volume} {84}},\ \bibinfo {pages} {2064} (\bibinfo {year} {2000})}\BibitemShut {NoStop}%
\bibitem [{\citenamefont {Jiao}\ \emph {et~al.}(2014)\citenamefont {Jiao}, \citenamefont {Lau}, \citenamefont {Hatzikirou}, \citenamefont {Meyer-Hermann}, \citenamefont {Corbo},\ and\ \citenamefont {Torquato}}]{Ji14}%
  \BibitemOpen
  \bibfield  {author} {\bibinfo {author} {\bibfnamefont {Y.}~\bibnamefont {Jiao}}, \bibinfo {author} {\bibfnamefont {T.}~\bibnamefont {Lau}}, \bibinfo {author} {\bibfnamefont {H.}~\bibnamefont {Hatzikirou}}, \bibinfo {author} {\bibfnamefont {M.}~\bibnamefont {Meyer-Hermann}}, \bibinfo {author} {\bibfnamefont {J.~C.}\ \bibnamefont {Corbo}},\ and\ \bibinfo {author} {\bibfnamefont {S.}~\bibnamefont {Torquato}},\ }\bibfield  {title} {\bibinfo {title} {Avian photoreceptor patterns represent a disordered hyperuniform solution to a multiscale packing problem},\ }\href {https://doi.org/10.1103/PhysRevE.89.022721} {\bibfield  {journal} {\bibinfo  {journal} {Phys. Rev. E}\ }\textbf {\bibinfo {volume} {89}},\ \bibinfo {pages} {022721} (\bibinfo {year} {2014})}\BibitemShut {NoStop}%
\bibitem [{\citenamefont {Huang}\ \emph {et~al.}(2021)\citenamefont {Huang}, \citenamefont {Hu}, \citenamefont {Yang}, \citenamefont {Liu},\ and\ \citenamefont {Zhang}}]{Hu21}%
  \BibitemOpen
  \bibfield  {author} {\bibinfo {author} {\bibfnamefont {M.}~\bibnamefont {Huang}}, \bibinfo {author} {\bibfnamefont {W.}~\bibnamefont {Hu}}, \bibinfo {author} {\bibfnamefont {S.}~\bibnamefont {Yang}}, \bibinfo {author} {\bibfnamefont {Q.~X.}\ \bibnamefont {Liu}},\ and\ \bibinfo {author} {\bibfnamefont {H.~P.}\ \bibnamefont {Zhang}},\ }\bibfield  {title} {\bibinfo {title} {{Circular swimming motility and disordered hyperuniform state in an algae system}},\ }\href {https://doi.org/10.1073/pnas.2100493118} {\bibfield  {journal} {\bibinfo  {journal} {Proc. Natl. Acad. Sci. U.S.A.}\ }\textbf {\bibinfo {volume} {118}},\ \bibinfo {pages} {e2100493118} (\bibinfo {year} {2021})}\BibitemShut {NoStop}%
\bibitem [{\citenamefont {Wigner}(1934)}]{Wi34}%
  \BibitemOpen
  \bibfield  {author} {\bibinfo {author} {\bibfnamefont {E.}~\bibnamefont {Wigner}},\ }\bibfield  {title} {\bibinfo {title} {On the interaction of electrons in metals},\ }\href {https://doi.org/10.1103/PhysRev.46.1002} {\bibfield  {journal} {\bibinfo  {journal} {Phys. Rev.}\ }\textbf {\bibinfo {volume} {46}},\ \bibinfo {pages} {1002} (\bibinfo {year} {1934})}\BibitemShut {NoStop}%
\bibitem [{\citenamefont {LeRoy}\ \emph {et~al.}(2002)\citenamefont {LeRoy}, \citenamefont {Topinka}, \citenamefont {Westervelt}, \citenamefont {Maranowski},\ and\ \citenamefont {Gossard}}]{Le02}%
  \BibitemOpen
  \bibfield  {author} {\bibinfo {author} {\bibfnamefont {B.~J.}\ \bibnamefont {LeRoy}}, \bibinfo {author} {\bibfnamefont {M.~A.}\ \bibnamefont {Topinka}}, \bibinfo {author} {\bibfnamefont {R.~M.}\ \bibnamefont {Westervelt}}, \bibinfo {author} {\bibfnamefont {K.~D.}\ \bibnamefont {Maranowski}},\ and\ \bibinfo {author} {\bibfnamefont {A.~C.}\ \bibnamefont {Gossard}},\ }\bibfield  {title} {\bibinfo {title} {{Imaging electron density in a two-dimensional electron gas}},\ }\href {https://doi.org/10.1063/1.1484548} {\bibfield  {journal} {\bibinfo  {journal} {Appl. Phys. Lett.}\ }\textbf {\bibinfo {volume} {80}},\ \bibinfo {pages} {4431} (\bibinfo {year} {2002})}\BibitemShut {NoStop}%
\bibitem [{\citenamefont {Go\~ni}\ \emph {et~al.}(1991)\citenamefont {Go\~ni}, \citenamefont {Pinczuk}, \citenamefont {Weiner}, \citenamefont {Calleja}, \citenamefont {Dennis}, \citenamefont {Pfeiffer},\ and\ \citenamefont {West}}]{Go91}%
  \BibitemOpen
  \bibfield  {author} {\bibinfo {author} {\bibfnamefont {A.~R.}\ \bibnamefont {Go\~ni}}, \bibinfo {author} {\bibfnamefont {A.}~\bibnamefont {Pinczuk}}, \bibinfo {author} {\bibfnamefont {J.~S.}\ \bibnamefont {Weiner}}, \bibinfo {author} {\bibfnamefont {J.~M.}\ \bibnamefont {Calleja}}, \bibinfo {author} {\bibfnamefont {B.~S.}\ \bibnamefont {Dennis}}, \bibinfo {author} {\bibfnamefont {L.~N.}\ \bibnamefont {Pfeiffer}},\ and\ \bibinfo {author} {\bibfnamefont {K.~W.}\ \bibnamefont {West}},\ }\bibfield  {title} {\bibinfo {title} {{One-dimensional plasmon dispersion and dispersionless intersubband excitations in GaAs quantum wires}},\ }\href {https://doi.org/10.1103/PhysRevLett.67.3298} {\bibfield  {journal} {\bibinfo  {journal} {Phys. Rev. Lett.}\ }\textbf {\bibinfo {volume} {67}},\ \bibinfo {pages} {3298} (\bibinfo {year} {1991})}\BibitemShut {NoStop}%
\bibitem [{\citenamefont {Andrei}(1997)}]{An97}%
  \BibitemOpen
  \bibfield  {author} {\bibinfo {author} {\bibfnamefont {E.~Y.}\ \bibnamefont {Andrei}},\ }\href@noop {} {\emph {\bibinfo {title} {Two-Dimensional Electron Systems on Helium and Other Cryogenic Substrates}}}\ (\bibinfo  {publisher} {Kluwer},\ \bibinfo {address} {Boston},\ \bibinfo {year} {1997})\BibitemShut {NoStop}%
\bibitem [{\citenamefont {Feynman}(1998)}]{Fe98}%
  \BibitemOpen
  \bibfield  {author} {\bibinfo {author} {\bibfnamefont {R.~P.}\ \bibnamefont {Feynman}},\ }\href@noop {} {\emph {\bibinfo {title} {Statistical Mechanics}}}\ (\bibinfo  {publisher} {Westview Press},\ \bibinfo {address} {Boulder, Colorado},\ \bibinfo {year} {1998})\BibitemShut {NoStop}%
\bibitem [{\citenamefont {Feynman}\ and\ \citenamefont {Cohen}(1956)}]{Fe56}%
  \BibitemOpen
  \bibfield  {author} {\bibinfo {author} {\bibfnamefont {R.~P.}\ \bibnamefont {Feynman}}\ and\ \bibinfo {author} {\bibfnamefont {M.}~\bibnamefont {Cohen}},\ }\bibfield  {title} {\bibinfo {title} {Energy spectrum of the excitations in liquid helium},\ }\href {https://doi.org/10.1103/PhysRev.102.1189} {\bibfield  {journal} {\bibinfo  {journal} {Phys. Rev.}\ }\textbf {\bibinfo {volume} {102}},\ \bibinfo {pages} {1189} (\bibinfo {year} {1956})}\BibitemShut {NoStop}%
\bibitem [{\citenamefont {Ginibre}(1965)}]{Gi65}%
  \BibitemOpen
  \bibfield  {author} {\bibinfo {author} {\bibfnamefont {J.}~\bibnamefont {Ginibre}},\ }\bibfield  {title} {\bibinfo {title} {Statistical ensembles of complex, quaternion, and real matrices},\ }\href {https://doi.org/10.1063/1.1704292} {\bibfield  {journal} {\bibinfo  {journal} {J. Math. Phys.}\ }\textbf {\bibinfo {volume} {6}},\ \bibinfo {pages} {440} (\bibinfo {year} {1965})}\BibitemShut {NoStop}%
\bibitem [{\citenamefont {Rougerie}(2019)}]{Ro18}%
  \BibitemOpen
  \bibfield  {author} {\bibinfo {author} {\bibfnamefont {N.}~\bibnamefont {Rougerie}},\ }\bibfield  {title} {\bibinfo {title} {On the {Laughlin} function and its perturbations},\ }\href {https://doi.org/10.5802/slsedp.131} {\bibfield  {journal} {\bibinfo  {journal} {S{\'e}minaire Laurent Schwartz -- EDP et applications}\ ,\ \bibinfo {pages} {1}} (\bibinfo {year} {2018-2019})},\ \bibinfo {note} {talk:2}\BibitemShut {NoStop}%
\bibitem [{\citenamefont {Abreu}\ \emph {et~al.}(2017)\citenamefont {Abreu}, \citenamefont {Pereira}, \citenamefont {Romero},\ and\ \citenamefont {Torquato}}]{Ab17}%
  \BibitemOpen
  \bibfield  {author} {\bibinfo {author} {\bibfnamefont {L.~D.}\ \bibnamefont {Abreu}}, \bibinfo {author} {\bibfnamefont {J.~M.}\ \bibnamefont {Pereira}}, \bibinfo {author} {\bibfnamefont {J.~L.}\ \bibnamefont {Romero}},\ and\ \bibinfo {author} {\bibfnamefont {S.}~\bibnamefont {Torquato}},\ }\bibfield  {title} {\bibinfo {title} {The {W}eyl-{H}eisenberg ensemble: {H}yperuniformity and higher {L}andau levels},\ }\href {https://doi.org/10.1088/1742-5468/aa68a7} {\bibfield  {journal} {\bibinfo  {journal} {J. Stat. Mech.: Th. and Exper.}\ }\textbf {\bibinfo {volume} {2017}},\ \bibinfo {pages} {043103} (\bibinfo {year} {2017})}\BibitemShut {NoStop}%
\bibitem [{\citenamefont {Dyson}(1962)}]{Dy62a}%
  \BibitemOpen
  \bibfield  {author} {\bibinfo {author} {\bibfnamefont {F.~J.}\ \bibnamefont {Dyson}},\ }\bibfield  {title} {\bibinfo {title} {Statistical theory of the energy levels of complex systems. {I}},\ }\href {https://doi.org/10.1063/1.1703773} {\bibfield  {journal} {\bibinfo  {journal} {J. Math. Phys.}\ }\textbf {\bibinfo {volume} {3}},\ \bibinfo {pages} {140} (\bibinfo {year} {1962})}\BibitemShut {NoStop}%
\bibitem [{\citenamefont {Montgomery}(1973)}]{Mon73}%
  \BibitemOpen
  \bibfield  {author} {\bibinfo {author} {\bibfnamefont {H.~L.}\ \bibnamefont {Montgomery}},\ }\bibfield  {title} {\bibinfo {title} {The pair correlation of zeros of the zeta function},\ }\href {https://websites.umich.edu/~hlm/paircor1.pdf} {\bibfield  {journal} {\bibinfo  {journal} {Proc. Symp. Pure Math}\ ,\ \bibinfo {pages} {181}} (\bibinfo {year} {1973})}\BibitemShut {NoStop}%
\bibitem [{\citenamefont {Odlyzko}(1987)}]{Od87}%
  \BibitemOpen
  \bibfield  {author} {\bibinfo {author} {\bibfnamefont {A.~M.}\ \bibnamefont {Odlyzko}},\ }\bibfield  {title} {\bibinfo {title} {On the distribution of spacings between zeros of the zeta function},\ }\href {https://doi.org/10.1090/S0025-5718-1987-0866115-0} {\bibfield  {journal} {\bibinfo  {journal} {Math. Comp.}\ }\textbf {\bibinfo {volume} {48}},\ \bibinfo {pages} {273} (\bibinfo {year} {1987})}\BibitemShut {NoStop}%
\bibitem [{\citenamefont {Metha}(1991)}]{Me91}%
  \BibitemOpen
  \bibfield  {author} {\bibinfo {author} {\bibfnamefont {M.~L.}\ \bibnamefont {Metha}},\ }\href@noop {} {\emph {\bibinfo {title} {Random Matrices}}}\ (\bibinfo  {publisher} {Academic Press},\ \bibinfo {address} {New York},\ \bibinfo {year} {1991})\BibitemShut {NoStop}%
\bibitem [{\citenamefont {Torquato}\ \emph {et~al.}(2008)\citenamefont {Torquato}, \citenamefont {Scardicchio},\ and\ \citenamefont {Zachary}}]{To08b}%
  \BibitemOpen
  \bibfield  {author} {\bibinfo {author} {\bibfnamefont {S.}~\bibnamefont {Torquato}}, \bibinfo {author} {\bibfnamefont {A.}~\bibnamefont {Scardicchio}},\ and\ \bibinfo {author} {\bibfnamefont {C.~E.}\ \bibnamefont {Zachary}},\ }\bibfield  {title} {\bibinfo {title} {Point processes in arbitrary dimension from {F}ermionic gases, random matrix theory, and number theory},\ }\href {https://doi.org/10.1088/1742-5468/2008/11/P11019} {\bibfield  {journal} {\bibinfo  {journal} {J. Stat. Mech.: Theory Exp.}\ }\textbf {\bibinfo {volume} {2008}},\ \bibinfo {pages} {P11019}}\BibitemShut {NoStop}%
\bibitem [{Note8()}]{note8}%
  \BibitemOpen
  \bibinfo {note} {Dyson showed that the pair correlation function of the eigenvalues of the Gaussian Unitary Ensemble (GUE) is given by Eq.~(\ref {g2_1D_pol_free}). The GUE eigenvalues are mappable to a classical many-body system at positive temperature $T>0$ in which particles on a unit circle interact under a logarithmic repulsive potential~\cite {Dy62a}.}\BibitemShut {Stop}%
\bibitem [{\citenamefont {Ashcroft}\ and\ \citenamefont {Mermin}(1976)}]{As76}%
  \BibitemOpen
  \bibfield  {author} {\bibinfo {author} {\bibfnamefont {N.~W.}\ \bibnamefont {Ashcroft}}\ and\ \bibinfo {author} {\bibfnamefont {D.~N.}\ \bibnamefont {Mermin}},\ }\href@noop {} {\emph {\bibinfo {title} {Solid State Physics}}}\ (\bibinfo  {publisher} {Thomson Learning},\ \bibinfo {address} {Toronto},\ \bibinfo {year} {1976})\BibitemShut {NoStop}%
\bibitem [{\citenamefont {Pines}\ and\ \citenamefont {Nozi\`eres}(1966)}]{Pi66}%
  \BibitemOpen
  \bibfield  {author} {\bibinfo {author} {\bibfnamefont {D.}~\bibnamefont {Pines}}\ and\ \bibinfo {author} {\bibfnamefont {P.}~\bibnamefont {Nozi\`eres}},\ }\href@noop {} {\emph {\bibinfo {title} {The Theory of Quantum Liquids}}}\ (\bibinfo  {publisher} {Benjamin},\ \bibinfo {address} {New York},\ \bibinfo {year} {1966})\ \bibinfo {note} {vol. I}\BibitemShut {NoStop}%
\bibitem [{\citenamefont {Bohm}\ and\ \citenamefont {Pines}(1951)}]{Bo51}%
  \BibitemOpen
  \bibfield  {author} {\bibinfo {author} {\bibfnamefont {D.}~\bibnamefont {Bohm}}\ and\ \bibinfo {author} {\bibfnamefont {D.}~\bibnamefont {Pines}},\ }\bibfield  {title} {\bibinfo {title} {{A Collective Description of Electron Interactions. I. Magnetic Interactions}},\ }\href {https://doi.org/10.1103/PhysRev.82.625} {\bibfield  {journal} {\bibinfo  {journal} {Phys. Rev.}\ }\textbf {\bibinfo {volume} {82}},\ \bibinfo {pages} {625} (\bibinfo {year} {1951})}\BibitemShut {NoStop}%
\bibitem [{\citenamefont {Hubbard}\ and\ \citenamefont {Peierls}(1957)}]{Hu57}%
  \BibitemOpen
  \bibfield  {author} {\bibinfo {author} {\bibfnamefont {J.}~\bibnamefont {Hubbard}}\ and\ \bibinfo {author} {\bibfnamefont {R.~E.}\ \bibnamefont {Peierls}},\ }\bibfield  {title} {\bibinfo {title} {The description of collective motions in terms of many-body perturbation theory},\ }\href {https://doi.org/10.1098/rspa.1957.0106} {\bibfield  {journal} {\bibinfo  {journal} {Proc. R. Soc. Lond. A}\ }\textbf {\bibinfo {volume} {240}},\ \bibinfo {pages} {539} (\bibinfo {year} {1957})}\BibitemShut {NoStop}%
\bibitem [{\citenamefont {Singwi}\ \emph {et~al.}(1968)\citenamefont {Singwi}, \citenamefont {Tosi}, \citenamefont {Land},\ and\ \citenamefont {Sj\"olander}}]{Si68}%
  \BibitemOpen
  \bibfield  {author} {\bibinfo {author} {\bibfnamefont {K.~S.}\ \bibnamefont {Singwi}}, \bibinfo {author} {\bibfnamefont {M.~P.}\ \bibnamefont {Tosi}}, \bibinfo {author} {\bibfnamefont {R.~H.}\ \bibnamefont {Land}},\ and\ \bibinfo {author} {\bibfnamefont {A.}~\bibnamefont {Sj\"olander}},\ }\bibfield  {title} {\bibinfo {title} {Electron correlations at metallic densities},\ }\href {https://doi.org/10.1103/PhysRev.176.589} {\bibfield  {journal} {\bibinfo  {journal} {Phys. Rev.}\ }\textbf {\bibinfo {volume} {176}},\ \bibinfo {pages} {589} (\bibinfo {year} {1968})}\BibitemShut {NoStop}%
\bibitem [{\citenamefont {Gold}\ and\ \citenamefont {Calmels}(1993)}]{Go93}%
  \BibitemOpen
  \bibfield  {author} {\bibinfo {author} {\bibfnamefont {A.}~\bibnamefont {Gold}}\ and\ \bibinfo {author} {\bibfnamefont {L.}~\bibnamefont {Calmels}},\ }\bibfield  {title} {\bibinfo {title} {{Correlation in Fermi liquids: Analytical results for the local-field correction in two and three dimensions}},\ }\href {https://doi.org/10.1103/PhysRevB.48.11622} {\bibfield  {journal} {\bibinfo  {journal} {Phys. Rev. B}\ }\textbf {\bibinfo {volume} {48}},\ \bibinfo {pages} {11622} (\bibinfo {year} {1993})}\BibitemShut {NoStop}%
\bibitem [{\citenamefont {Wang}\ and\ \citenamefont {Perdew}(1991)}]{Wa91}%
  \BibitemOpen
  \bibfield  {author} {\bibinfo {author} {\bibfnamefont {Y.}~\bibnamefont {Wang}}\ and\ \bibinfo {author} {\bibfnamefont {J.~P.}\ \bibnamefont {Perdew}},\ }\bibfield  {title} {\bibinfo {title} {Correlation hole of the spin-polarized electron gas, with exact small-wave-vector and high-density scaling},\ }\href {https://doi.org/10.1103/PhysRevB.44.13298} {\bibfield  {journal} {\bibinfo  {journal} {Phys. Rev. B}\ }\textbf {\bibinfo {volume} {44}},\ \bibinfo {pages} {13298} (\bibinfo {year} {1991})}\BibitemShut {NoStop}%
\bibitem [{\citenamefont {Davoudi}\ \emph {et~al.}(2003)\citenamefont {Davoudi}, \citenamefont {Asgari}, \citenamefont {Polini},\ and\ \citenamefont {Tosi}}]{Da03}%
  \BibitemOpen
  \bibfield  {author} {\bibinfo {author} {\bibfnamefont {B.}~\bibnamefont {Davoudi}}, \bibinfo {author} {\bibfnamefont {R.}~\bibnamefont {Asgari}}, \bibinfo {author} {\bibfnamefont {M.}~\bibnamefont {Polini}},\ and\ \bibinfo {author} {\bibfnamefont {M.~P.}\ \bibnamefont {Tosi}},\ }\bibfield  {title} {\bibinfo {title} {Analytic theory of ground-state properties of a three-dimensional electron gas with arbitrary spin polarization},\ }\href {https://doi.org/10.1103/PhysRevB.68.155112} {\bibfield  {journal} {\bibinfo  {journal} {Phys. Rev. B}\ }\textbf {\bibinfo {volume} {68}},\ \bibinfo {pages} {155112} (\bibinfo {year} {2003})}\BibitemShut {NoStop}%
\bibitem [{\citenamefont {Tanatar}(1996)}]{Ta96}%
  \BibitemOpen
  \bibfield  {author} {\bibinfo {author} {\bibfnamefont {B.}~\bibnamefont {Tanatar}},\ }\bibfield  {title} {\bibinfo {title} {Spin correlations in a quasi-one-dimensional electron gas},\ }\href {https://doi.org/https://doi.org/10.1016/S0921-4526(96)00477-2} {\bibfield  {journal} {\bibinfo  {journal} {Physica B: Condens. Matter}\ }\textbf {\bibinfo {volume} {228}},\ \bibinfo {pages} {329} (\bibinfo {year} {1996})}\BibitemShut {NoStop}%
\bibitem [{\citenamefont {Gabrielli}\ \emph {et~al.}(2008)\citenamefont {Gabrielli}, \citenamefont {Joyce},\ and\ \citenamefont {Torquato}}]{Ga08}%
  \BibitemOpen
  \bibfield  {author} {\bibinfo {author} {\bibfnamefont {A.}~\bibnamefont {Gabrielli}}, \bibinfo {author} {\bibfnamefont {M.}~\bibnamefont {Joyce}},\ and\ \bibinfo {author} {\bibfnamefont {S.}~\bibnamefont {Torquato}},\ }\bibfield  {title} {\bibinfo {title} {Tilings of space and superhomogeneous point processes},\ }\href {https://doi.org/10.1103/PhysRevE.77.031125} {\bibfield  {journal} {\bibinfo  {journal} {Phys. Rev. E}\ }\textbf {\bibinfo {volume} {77}},\ \bibinfo {pages} {031125} (\bibinfo {year} {2008})}\BibitemShut {NoStop}%
\bibitem [{\citenamefont {Lobo}\ \emph {et~al.}(1969)\citenamefont {Lobo}, \citenamefont {Singwi},\ and\ \citenamefont {Tosi}}]{Lo69}%
  \BibitemOpen
  \bibfield  {author} {\bibinfo {author} {\bibfnamefont {R.}~\bibnamefont {Lobo}}, \bibinfo {author} {\bibfnamefont {K.~S.}\ \bibnamefont {Singwi}},\ and\ \bibinfo {author} {\bibfnamefont {M.~P.}\ \bibnamefont {Tosi}},\ }\bibfield  {title} {\bibinfo {title} {Spin correlations in the electron liquid},\ }\href {https://doi.org/10.1103/PhysRev.186.470} {\bibfield  {journal} {\bibinfo  {journal} {Phys. Rev.}\ }\textbf {\bibinfo {volume} {186}},\ \bibinfo {pages} {470} (\bibinfo {year} {1969})}\BibitemShut {NoStop}%
\bibitem [{\citenamefont {Torquato}\ and\ \citenamefont {Wang}(2022)}]{To22}%
  \BibitemOpen
  \bibfield  {author} {\bibinfo {author} {\bibfnamefont {S.}~\bibnamefont {Torquato}}\ and\ \bibinfo {author} {\bibfnamefont {H.}~\bibnamefont {Wang}},\ }\bibfield  {title} {\bibinfo {title} {Precise determination of pair interactions from pair statistics of many-body systems in and out of equilibrium},\ }\href {https://doi.org/10.1103/PhysRevE.106.044122} {\bibfield  {journal} {\bibinfo  {journal} {Phys. Rev. E}\ }\textbf {\bibinfo {volume} {106}},\ \bibinfo {pages} {044122} (\bibinfo {year} {2022})}\BibitemShut {NoStop}%
\bibitem [{\citenamefont {Wang}\ and\ \citenamefont {Torquato}(2024)}]{Wa24}%
  \BibitemOpen
  \bibfield  {author} {\bibinfo {author} {\bibfnamefont {H.}~\bibnamefont {Wang}}\ and\ \bibinfo {author} {\bibfnamefont {S.}~\bibnamefont {Torquato}},\ }\bibfield  {title} {\bibinfo {title} {{Designer pair statistics of disordered many-particle systems with novel properties}},\ }\href {https://doi.org/10.1063/5.0189769} {\bibfield  {journal} {\bibinfo  {journal} {J. Chem. Phys.}\ }\textbf {\bibinfo {volume} {160}},\ \bibinfo {pages} {044911} (\bibinfo {year} {2024})}\BibitemShut {NoStop}%
\bibitem [{\citenamefont {Stillinger}\ and\ \citenamefont {Torquato}(2019)}]{St19}%
  \BibitemOpen
  \bibfield  {author} {\bibinfo {author} {\bibfnamefont {F.~H.}\ \bibnamefont {Stillinger}}\ and\ \bibinfo {author} {\bibfnamefont {S.}~\bibnamefont {Torquato}},\ }\bibfield  {title} {\bibinfo {title} {Structural degeneracy in pair distance distributions},\ }\href {https://doi.org/10.1063/1.5096894} {\bibfield  {journal} {\bibinfo  {journal} {J. Chem. Phys.}\ }\textbf {\bibinfo {volume} {150}},\ \bibinfo {pages} {204125} (\bibinfo {year} {2019})}\BibitemShut {NoStop}%
\bibitem [{\citenamefont {Torquato}\ and\ \citenamefont {Stillinger}(2008)}]{To08a}%
  \BibitemOpen
  \bibfield  {author} {\bibinfo {author} {\bibfnamefont {S.}~\bibnamefont {Torquato}}\ and\ \bibinfo {author} {\bibfnamefont {F.~H.}\ \bibnamefont {Stillinger}},\ }\bibfield  {title} {\bibinfo {title} {New duality relations for classical ground states},\ }\href {https://doi.org/10.1103/PhysRevLett.100.020602} {\bibfield  {journal} {\bibinfo  {journal} {Phys. Rev. Lett.}\ }\textbf {\bibinfo {volume} {100}},\ \bibinfo {pages} {020602} (\bibinfo {year} {2008})}\BibitemShut {NoStop}%
\bibitem [{\citenamefont {Tomonaga}(1950)}]{tomonaga1950remarks}%
  \BibitemOpen
  \bibfield  {author} {\bibinfo {author} {\bibfnamefont {S.-i.}\ \bibnamefont {Tomonaga}},\ }\bibfield  {title} {\bibinfo {title} {{Remarks on Bloch's method of sound waves applied to many-fermion problems}},\ }\href {https://doi.org/10.1143/ptp/5.4.544} {\bibfield  {journal} {\bibinfo  {journal} {Prog. Theor. Phys.}\ }\textbf {\bibinfo {volume} {5}},\ \bibinfo {pages} {544} (\bibinfo {year} {1950})}\BibitemShut {NoStop}%
\bibitem [{\citenamefont {Luttinger}(1963)}]{luttinger1963exactly}%
  \BibitemOpen
  \bibfield  {author} {\bibinfo {author} {\bibfnamefont {J.~M.}\ \bibnamefont {Luttinger}},\ }\bibfield  {title} {\bibinfo {title} {{An Exactly Soluble Model of a Many‐Fermion System}},\ }\href {https://doi.org/10.1063/1.1704046} {\bibfield  {journal} {\bibinfo  {journal} {J. Math. Phys.}\ }\textbf {\bibinfo {volume} {4}},\ \bibinfo {pages} {1154} (\bibinfo {year} {1963})}\BibitemShut {NoStop}%
\bibitem [{\citenamefont {Das~Sarma}\ and\ \citenamefont {Lai}(1985)}]{sarma1985screening}%
  \BibitemOpen
  \bibfield  {author} {\bibinfo {author} {\bibfnamefont {S.}~\bibnamefont {Das~Sarma}}\ and\ \bibinfo {author} {\bibfnamefont {W.-y.}\ \bibnamefont {Lai}},\ }\bibfield  {title} {\bibinfo {title} {Screening and elementary excitations in narrow-channel semiconductor microstructures},\ }\href {https://doi.org/10.1103/PhysRevB.32.1401} {\bibfield  {journal} {\bibinfo  {journal} {Phys. Rev. B}\ }\textbf {\bibinfo {volume} {32}},\ \bibinfo {pages} {1401} (\bibinfo {year} {1985})}\BibitemShut {NoStop}%
\bibitem [{\citenamefont {Li}\ \emph {et~al.}(1992)\citenamefont {Li}, \citenamefont {Das~Sarma},\ and\ \citenamefont {Joynt}}]{li1992elementary}%
  \BibitemOpen
  \bibfield  {author} {\bibinfo {author} {\bibfnamefont {Q.~P.}\ \bibnamefont {Li}}, \bibinfo {author} {\bibfnamefont {S.}~\bibnamefont {Das~Sarma}},\ and\ \bibinfo {author} {\bibfnamefont {R.}~\bibnamefont {Joynt}},\ }\bibfield  {title} {\bibinfo {title} {{Elementary excitations in one-dimensional quantum wires: Exact equivalence between the random-phase approximation and the Tomonaga-Luttinger model}},\ }\href {https://doi.org/10.1103/PhysRevB.45.13713} {\bibfield  {journal} {\bibinfo  {journal} {Phys. Rev. B}\ }\textbf {\bibinfo {volume} {45}},\ \bibinfo {pages} {13713} (\bibinfo {year} {1992})}\BibitemShut {NoStop}%
\bibitem [{\citenamefont {Schulz}(1993)}]{schulz1993wigner}%
  \BibitemOpen
  \bibfield  {author} {\bibinfo {author} {\bibfnamefont {H.~J.}\ \bibnamefont {Schulz}},\ }\bibfield  {title} {\bibinfo {title} {Wigner crystal in one dimension},\ }\href {https://doi.org/10.1103/PhysRevLett.71.1864} {\bibfield  {journal} {\bibinfo  {journal} {Phys. Rev. Lett.}\ }\textbf {\bibinfo {volume} {71}},\ \bibinfo {pages} {1864} (\bibinfo {year} {1993})}\BibitemShut {NoStop}%
\bibitem [{\citenamefont {Iucci}\ and\ \citenamefont {Na\'{o}n}(2000)}]{iucci2000exact}%
  \BibitemOpen
  \bibfield  {author} {\bibinfo {author} {\bibfnamefont {A.}~\bibnamefont {Iucci}}\ and\ \bibinfo {author} {\bibfnamefont {C.}~\bibnamefont {Na\'{o}n}},\ }\bibfield  {title} {\bibinfo {title} {{Exact electronic Green functions in a Luttinger liquid with long-range interactions}},\ }\href {https://doi.org/10.1103/PhysRevB.61.15530} {\bibfield  {journal} {\bibinfo  {journal} {Phys. Rev. B}\ }\textbf {\bibinfo {volume} {61}},\ \bibinfo {pages} {15530} (\bibinfo {year} {2000})}\BibitemShut {NoStop}%
\bibitem [{\citenamefont {Wang}\ \emph {et~al.}(2001)\citenamefont {Wang}, \citenamefont {Millis},\ and\ \citenamefont {Das~Sarma}}]{wang2001coulomb}%
  \BibitemOpen
  \bibfield  {author} {\bibinfo {author} {\bibfnamefont {D.~W.}\ \bibnamefont {Wang}}, \bibinfo {author} {\bibfnamefont {A.~J.}\ \bibnamefont {Millis}},\ and\ \bibinfo {author} {\bibfnamefont {S.}~\bibnamefont {Das~Sarma}},\ }\bibfield  {title} {\bibinfo {title} {{Coulomb Luttinger liquid}},\ }\href {https://doi.org/10.1103/PhysRevB.64.193307} {\bibfield  {journal} {\bibinfo  {journal} {Phys. Rev. B}\ }\textbf {\bibinfo {volume} {64}},\ \bibinfo {pages} {193307} (\bibinfo {year} {2001})}\BibitemShut {NoStop}%
\bibitem [{\citenamefont {Creffield}\ \emph {et~al.}(2001)\citenamefont {Creffield}, \citenamefont {H{\"a}usler},\ and\ \citenamefont {MacDonald}}]{creffield2001spin}%
  \BibitemOpen
  \bibfield  {author} {\bibinfo {author} {\bibfnamefont {C.~E.}\ \bibnamefont {Creffield}}, \bibinfo {author} {\bibfnamefont {W.}~\bibnamefont {H{\"a}usler}},\ and\ \bibinfo {author} {\bibfnamefont {A.~H.}\ \bibnamefont {MacDonald}},\ }\bibfield  {title} {\bibinfo {title} {{Spin and charge Tomonaga-Luttinger parameters in quantum wires}},\ }\href {https://doi.org/10.1209/epl/i2001-00140-7} {\bibfield  {journal} {\bibinfo  {journal} {EPL}\ }\textbf {\bibinfo {volume} {53}},\ \bibinfo {pages} {221} (\bibinfo {year} {2001})}\BibitemShut {NoStop}%
\bibitem [{\citenamefont {Fogler}(2005)}]{fogler2005ground}%
  \BibitemOpen
  \bibfield  {author} {\bibinfo {author} {\bibfnamefont {M.~M.}\ \bibnamefont {Fogler}},\ }\bibfield  {title} {\bibinfo {title} {{Ground-State Energy of the Electron Liquid in Ultrathin Wires}},\ }\href {https://doi.org/10.1103/PhysRevLett.94.056405} {\bibfield  {journal} {\bibinfo  {journal} {Phys. Rev. Lett.}\ }\textbf {\bibinfo {volume} {94}},\ \bibinfo {pages} {056405} (\bibinfo {year} {2005})}\BibitemShut {NoStop}%
\bibitem [{\citenamefont {Casula}\ \emph {et~al.}(2006)\citenamefont {Casula}, \citenamefont {Sorella},\ and\ \citenamefont {Senatore}}]{casula2006ground}%
  \BibitemOpen
  \bibfield  {author} {\bibinfo {author} {\bibfnamefont {M.}~\bibnamefont {Casula}}, \bibinfo {author} {\bibfnamefont {S.}~\bibnamefont {Sorella}},\ and\ \bibinfo {author} {\bibfnamefont {G.}~\bibnamefont {Senatore}},\ }\bibfield  {title} {\bibinfo {title} {{Ground state properties of the one-dimensional Coulomb gas using the lattice regularized diffusion Monte Carlo method}},\ }\href {https://doi.org/10.1103/PhysRevB.74.245427} {\bibfield  {journal} {\bibinfo  {journal} {Phys. Rev. B}\ }\textbf {\bibinfo {volume} {74}},\ \bibinfo {pages} {245427} (\bibinfo {year} {2006})}\BibitemShut {NoStop}%
\bibitem [{\citenamefont {Shulenburger}\ \emph {et~al.}(2008)\citenamefont {Shulenburger}, \citenamefont {Casula}, \citenamefont {Senatore},\ and\ \citenamefont {Martin}}]{shulenburger2008correlation}%
  \BibitemOpen
  \bibfield  {author} {\bibinfo {author} {\bibfnamefont {L.}~\bibnamefont {Shulenburger}}, \bibinfo {author} {\bibfnamefont {M.}~\bibnamefont {Casula}}, \bibinfo {author} {\bibfnamefont {G.}~\bibnamefont {Senatore}},\ and\ \bibinfo {author} {\bibfnamefont {R.~M.}\ \bibnamefont {Martin}},\ }\bibfield  {title} {\bibinfo {title} {{Correlation effects in quasi-one-dimensional quantum wires}},\ }\href {https://doi.org/10.1103/PhysRevB.78.165303} {\bibfield  {journal} {\bibinfo  {journal} {Phys. Rev. B}\ }\textbf {\bibinfo {volume} {78}},\ \bibinfo {pages} {165303} (\bibinfo {year} {2008})}\BibitemShut {NoStop}%
\bibitem [{\citenamefont {Gold}\ and\ \citenamefont {Ghazali}(1990)}]{gold1990analytical}%
  \BibitemOpen
  \bibfield  {author} {\bibinfo {author} {\bibfnamefont {A.}~\bibnamefont {Gold}}\ and\ \bibinfo {author} {\bibfnamefont {A.}~\bibnamefont {Ghazali}},\ }\bibfield  {title} {\bibinfo {title} {{Analytical results for semiconductor quantum-well wire: Plasmons, shallow impurity states, and mobility}},\ }\href {https://doi.org/10.1103/PhysRevB.41.7626} {\bibfield  {journal} {\bibinfo  {journal} {Phys. Rev. B}\ }\textbf {\bibinfo {volume} {41}},\ \bibinfo {pages} {7626} (\bibinfo {year} {1990})}\BibitemShut {NoStop}%
\bibitem [{\citenamefont {Hu}\ and\ \citenamefont {Das~Sarma}(1993)}]{hu1993many}%
  \BibitemOpen
  \bibfield  {author} {\bibinfo {author} {\bibfnamefont {B.~Y.-K.}\ \bibnamefont {Hu}}\ and\ \bibinfo {author} {\bibfnamefont {S.}~\bibnamefont {Das~Sarma}},\ }\bibfield  {title} {\bibinfo {title} {Many-body exchange-correlation effects in the lowest subband of semiconductor quantum wires},\ }\href {https://doi.org/10.1103/PhysRevB.48.5469} {\bibfield  {journal} {\bibinfo  {journal} {Phys. Rev. B}\ }\textbf {\bibinfo {volume} {48}},\ \bibinfo {pages} {5469} (\bibinfo {year} {1993})}\BibitemShut {NoStop}%
\bibitem [{\citenamefont {Sun}\ and\ \citenamefont {Kirczenow}(1993)}]{sun1993density}%
  \BibitemOpen
  \bibfield  {author} {\bibinfo {author} {\bibfnamefont {Y.}~\bibnamefont {Sun}}\ and\ \bibinfo {author} {\bibfnamefont {G.}~\bibnamefont {Kirczenow}},\ }\bibfield  {title} {\bibinfo {title} {{Density-functional theory of the electronic structure of Coulomb-confined quantum wires}},\ }\href {https://doi.org/10.1103/PhysRevB.47.4413} {\bibfield  {journal} {\bibinfo  {journal} {Phys. Rev. B}\ }\textbf {\bibinfo {volume} {47}},\ \bibinfo {pages} {4413} (\bibinfo {year} {1993})}\BibitemShut {NoStop}%
\bibitem [{\citenamefont {Friesen}\ and\ \citenamefont {Bergersen}(1980)}]{friesen1980dielectric}%
  \BibitemOpen
  \bibfield  {author} {\bibinfo {author} {\bibfnamefont {W.~I.}\ \bibnamefont {Friesen}}\ and\ \bibinfo {author} {\bibfnamefont {B.}~\bibnamefont {Bergersen}},\ }\bibfield  {title} {\bibinfo {title} {Dielectric response of a one-dimensional electron gas},\ }\href {https://doi.org/10.1088/0022-3719/13/36/016} {\bibfield  {journal} {\bibinfo  {journal} {J. Phys. C: Solid State Phys.}\ }\textbf {\bibinfo {volume} {13}},\ \bibinfo {pages} {6627} (\bibinfo {year} {1980})}\BibitemShut {NoStop}%
\bibitem [{\citenamefont {Hu}\ and\ \citenamefont {O'Connell}(1990)}]{Hu90}%
  \BibitemOpen
  \bibfield  {author} {\bibinfo {author} {\bibfnamefont {G.~Y.}\ \bibnamefont {Hu}}\ and\ \bibinfo {author} {\bibfnamefont {R.~F.}\ \bibnamefont {O'Connell}},\ }\bibfield  {title} {\bibinfo {title} {Electron-electron interactions in quasi-one-dimensional electron systems},\ }\href {https://doi.org/10.1103/PhysRevB.42.1290} {\bibfield  {journal} {\bibinfo  {journal} {Phys. Rev. B}\ }\textbf {\bibinfo {volume} {42}},\ \bibinfo {pages} {1290} (\bibinfo {year} {1990})}\BibitemShut {NoStop}%
\bibitem [{\citenamefont {Garg}\ \emph {et~al.}(2008)\citenamefont {Garg}, \citenamefont {Moudgil}, \citenamefont {Kumar},\ and\ \citenamefont {Ahluwalia}}]{Ga08b}%
  \BibitemOpen
  \bibfield  {author} {\bibinfo {author} {\bibfnamefont {V.}~\bibnamefont {Garg}}, \bibinfo {author} {\bibfnamefont {R.~K.}\ \bibnamefont {Moudgil}}, \bibinfo {author} {\bibfnamefont {K.}~\bibnamefont {Kumar}},\ and\ \bibinfo {author} {\bibfnamefont {P.~K.}\ \bibnamefont {Ahluwalia}},\ }\bibfield  {title} {\bibinfo {title} {Ground-state properties of a quasi-one-dimensional electron gas within a dynamical self-consistent mean-field approximation},\ }\href {https://doi.org/10.1103/PhysRevB.78.045406} {\bibfield  {journal} {\bibinfo  {journal} {Phys. Rev. B}\ }\textbf {\bibinfo {volume} {78}},\ \bibinfo {pages} {045406} (\bibinfo {year} {2008})}\BibitemShut {NoStop}%
\bibitem [{\citenamefont {Li}\ and\ \citenamefont {Das~Sarma}(1991)}]{li1991elementary}%
  \BibitemOpen
  \bibfield  {author} {\bibinfo {author} {\bibfnamefont {Q.~P.}\ \bibnamefont {Li}}\ and\ \bibinfo {author} {\bibfnamefont {S.}~\bibnamefont {Das~Sarma}},\ }\bibfield  {title} {\bibinfo {title} {{Elementary excitation spectrum of one-dimensional electron systems in confined semiconductor structures: Zero magnetic field}},\ }\href {https://doi.org/10.1103/PhysRevB.43.11768} {\bibfield  {journal} {\bibinfo  {journal} {Phys. Rev. B}\ }\textbf {\bibinfo {volume} {43}},\ \bibinfo {pages} {11768} (\bibinfo {year} {1991})}\BibitemShut {NoStop}%
\bibitem [{\citenamefont {Hansen}\ and\ \citenamefont {McDonald}(1986)}]{Ha86}%
  \BibitemOpen
  \bibfield  {author} {\bibinfo {author} {\bibfnamefont {J.~P.}\ \bibnamefont {Hansen}}\ and\ \bibinfo {author} {\bibfnamefont {I.~R.}\ \bibnamefont {McDonald}},\ }\href@noop {} {\emph {\bibinfo {title} {Theory of Simple Liquids}}}\ (\bibinfo  {publisher} {Academic Press},\ \bibinfo {address} {New York},\ \bibinfo {year} {1986})\BibitemShut {NoStop}%
\bibitem [{\citenamefont {Ng}\ and\ \citenamefont {Singwi}(1987)}]{Ng87}%
  \BibitemOpen
  \bibfield  {author} {\bibinfo {author} {\bibfnamefont {T.~K.}\ \bibnamefont {Ng}}\ and\ \bibinfo {author} {\bibfnamefont {K.~S.}\ \bibnamefont {Singwi}},\ }\bibfield  {title} {\bibinfo {title} {{Arbitrarily polarized model Fermi liquid}},\ }\href {https://doi.org/10.1103/PhysRevB.35.6683} {\bibfield  {journal} {\bibinfo  {journal} {Phys. Rev. B}\ }\textbf {\bibinfo {volume} {35}},\ \bibinfo {pages} {6683} (\bibinfo {year} {1987})}\BibitemShut {NoStop}%
\bibitem [{\citenamefont {Kumar}\ \emph {et~al.}(2009)\citenamefont {Kumar}, \citenamefont {Garg},\ and\ \citenamefont {Moudgil}}]{Ku09}%
  \BibitemOpen
  \bibfield  {author} {\bibinfo {author} {\bibfnamefont {K.}~\bibnamefont {Kumar}}, \bibinfo {author} {\bibfnamefont {V.}~\bibnamefont {Garg}},\ and\ \bibinfo {author} {\bibfnamefont {R.~K.}\ \bibnamefont {Moudgil}},\ }\bibfield  {title} {\bibinfo {title} {Spin-resolved correlations and ground state of a three-dimensional electron gas: Spin-polarization effects},\ }\href {https://doi.org/10.1103/PhysRevB.79.115304} {\bibfield  {journal} {\bibinfo  {journal} {Phys. Rev. B}\ }\textbf {\bibinfo {volume} {79}},\ \bibinfo {pages} {115304} (\bibinfo {year} {2009})}\BibitemShut {NoStop}%
\bibitem [{\citenamefont {Giuliani}\ and\ \citenamefont {Vignale}(2005)}]{giuliani2008quantum}%
  \BibitemOpen
  \bibfield  {author} {\bibinfo {author} {\bibfnamefont {G.}~\bibnamefont {Giuliani}}\ and\ \bibinfo {author} {\bibfnamefont {G.}~\bibnamefont {Vignale}},\ }\href {https://doi.org/10.1017/CBO9780511619915} {\emph {\bibinfo {title} {{Quantum Theory of the Electron Liquid}}}}\ (\bibinfo  {publisher} {Cambridge University Press},\ \bibinfo {address} {New York},\ \bibinfo {year} {2005})\BibitemShut {NoStop}%
\bibitem [{\citenamefont {Lindhard}(1954)}]{Li54}%
  \BibitemOpen
  \bibfield  {author} {\bibinfo {author} {\bibfnamefont {J.}~\bibnamefont {Lindhard}},\ }\bibfield  {title} {\bibinfo {title} {On the properties of a gas of charged particles},\ }\href {https://www.osti.gov/biblio/4405425} {\bibfield  {journal} {\bibinfo  {journal} {Kgl. Danske Videnskab. Selskab Mat.-fys. Medd.}\ }\textbf {\bibinfo {volume} {28}},\ \bibinfo {pages} {8} (\bibinfo {year} {1954})}\BibitemShut {NoStop}%
\bibitem [{\citenamefont {Mihaila}(2011)}]{Mi11}%
  \BibitemOpen
  \bibfield  {author} {\bibinfo {author} {\bibfnamefont {B.}~\bibnamefont {Mihaila}},\ }\href@noop {} {\bibinfo {title} {{Lindhard function of a $d$-dimensional Fermi gas}}} (\bibinfo {year} {2011}),\ \Eprint {https://arxiv.org/abs/1111.5337} {arXiv:1111.5337 [cond-mat.quant-gas]} \BibitemShut {NoStop}%
\bibitem [{\citenamefont {Goldstein}(1951)}]{goldstein1951theory}%
  \BibitemOpen
  \bibfield  {author} {\bibinfo {author} {\bibfnamefont {L.}~\bibnamefont {Goldstein}},\ }\bibfield  {title} {\bibinfo {title} {{On the Theory of Coherent Scattering Processes in Liquids}},\ }\href {https://doi.org/10.1103/PhysRev.84.466} {\bibfield  {journal} {\bibinfo  {journal} {Phys. Rev.}\ }\textbf {\bibinfo {volume} {84}},\ \bibinfo {pages} {466} (\bibinfo {year} {1951})}\BibitemShut {NoStop}%
\bibitem [{\citenamefont {Cherny}(2002)}]{cherny2002sum}%
  \BibitemOpen
  \bibfield  {author} {\bibinfo {author} {\bibfnamefont {A.~Y.}\ \bibnamefont {Cherny}},\ }\href@noop {} {\bibinfo {title} {Sum rule for the pair correlation function}} (\bibinfo {year} {2002}),\ \Eprint {https://arxiv.org/abs/cond-mat/0209378} {arXiv:cond-mat/0209378 [cond-mat.stat-mech]} \BibitemShut {NoStop}%
\bibitem [{\citenamefont {Reatto}\ and\ \citenamefont {Chester}(1967)}]{Re67}%
  \BibitemOpen
  \bibfield  {author} {\bibinfo {author} {\bibfnamefont {L.}~\bibnamefont {Reatto}}\ and\ \bibinfo {author} {\bibfnamefont {G.~V.}\ \bibnamefont {Chester}},\ }\bibfield  {title} {\bibinfo {title} {{Phonons and the Properties of a Bose System}},\ }\href {https://doi.org/10.1103/PhysRev.155.88} {\bibfield  {journal} {\bibinfo  {journal} {Phys. Rev.}\ }\textbf {\bibinfo {volume} {155}},\ \bibinfo {pages} {88} (\bibinfo {year} {1967})}\BibitemShut {NoStop}%
\bibitem [{\citenamefont {Weber}\ and\ \citenamefont {Stillinger}(1981)}]{We81}%
  \BibitemOpen
  \bibfield  {author} {\bibinfo {author} {\bibfnamefont {T.~A.}\ \bibnamefont {Weber}}\ and\ \bibinfo {author} {\bibfnamefont {F.~H.}\ \bibnamefont {Stillinger}},\ }\bibfield  {title} {\bibinfo {title} {{Gaussian core model in two dimensions. II. Solid and fluid phase topological distribution functions}},\ }\href {https://doi.org/10.1063/1.441582} {\bibfield  {journal} {\bibinfo  {journal} {J. Chem. Phys.}\ }\textbf {\bibinfo {volume} {74}},\ \bibinfo {pages} {4020} (\bibinfo {year} {1981})}\BibitemShut {NoStop}%
\bibitem [{\citenamefont {Frenkel}\ and\ \citenamefont {Smit}(1996)}]{Fr96}%
  \BibitemOpen
  \bibfield  {author} {\bibinfo {author} {\bibfnamefont {D.}~\bibnamefont {Frenkel}}\ and\ \bibinfo {author} {\bibfnamefont {B.}~\bibnamefont {Smit}},\ }\href@noop {} {\emph {\bibinfo {title} {Understanding Molecular Simulation}}}\ (\bibinfo  {publisher} {Academic Press},\ \bibinfo {address} {New York},\ \bibinfo {year} {1996})\BibitemShut {NoStop}%
\bibitem [{\citenamefont {Travesset}(2014)}]{Tr14}%
  \BibitemOpen
  \bibfield  {author} {\bibinfo {author} {\bibfnamefont {A.}~\bibnamefont {Travesset}},\ }\bibfield  {title} {\bibinfo {title} {{Phase diagram of power law and Lennard-Jones systems: Crystal phases}},\ }\href {https://doi.org/10.1063/1.4898371} {\bibfield  {journal} {\bibinfo  {journal} {J. Chem. Phys.}\ }\textbf {\bibinfo {volume} {141}},\ \bibinfo {pages} {164501} (\bibinfo {year} {2014})}\BibitemShut {NoStop}%
\bibitem [{\citenamefont {Dunkel}\ and\ \citenamefont {Hilbert}(2014)}]{dunkel2014consistent}%
  \BibitemOpen
  \bibfield  {author} {\bibinfo {author} {\bibfnamefont {J.}~\bibnamefont {Dunkel}}\ and\ \bibinfo {author} {\bibfnamefont {S.}~\bibnamefont {Hilbert}},\ }\bibfield  {title} {\bibinfo {title} {Consistent thermostatistics forbids negative absolute temperatures},\ }\href {https://doi.org/10.1038/nphys2815} {\bibfield  {journal} {\bibinfo  {journal} {Nature Phys.}\ }\textbf {\bibinfo {volume} {10}},\ \bibinfo {pages} {67} (\bibinfo {year} {2014})}\BibitemShut {NoStop}%
\bibitem [{\citenamefont {Lim}\ \emph {et~al.}(2009)\citenamefont {Lim}, \citenamefont {Jung}, \citenamefont {Bae}, \citenamefont {Park}, \citenamefont {Jang},\ and\ \citenamefont {Sung}}]{Li09b}%
  \BibitemOpen
  \bibfield  {author} {\bibinfo {author} {\bibfnamefont {Y.~R.}\ \bibnamefont {Lim}}, \bibinfo {author} {\bibfnamefont {W.}~\bibnamefont {Jung}}, \bibinfo {author} {\bibfnamefont {J.~H.}\ \bibnamefont {Bae}}, \bibinfo {author} {\bibfnamefont {B.~J.}\ \bibnamefont {Park}}, \bibinfo {author} {\bibfnamefont {J.}~\bibnamefont {Jang}},\ and\ \bibinfo {author} {\bibfnamefont {J.}~\bibnamefont {Sung}},\ }\bibfield  {title} {\bibinfo {title} {Excess grand potential for a system under an external field: Effects of external field driven nonextensivity},\ }\href {https://doi.org/10.1021/jp900629d} {\bibfield  {journal} {\bibinfo  {journal} {J. Phys. Chem. B}\ }\textbf {\bibinfo {volume} {113}},\ \bibinfo {pages} {7982} (\bibinfo {year} {2009})}\BibitemShut {NoStop}%
\bibitem [{\citenamefont {Lim}\ \emph {et~al.}(2014)\citenamefont {Lim}, \citenamefont {Park}, \citenamefont {Song}, \citenamefont {Yang}, \citenamefont {Yoon}, \citenamefont {Kim},\ and\ \citenamefont {Sung}}]{Li14d}%
  \BibitemOpen
  \bibfield  {author} {\bibinfo {author} {\bibfnamefont {Y.~R.}\ \bibnamefont {Lim}}, \bibinfo {author} {\bibfnamefont {S.~J.}\ \bibnamefont {Park}}, \bibinfo {author} {\bibfnamefont {S.}~\bibnamefont {Song}}, \bibinfo {author} {\bibfnamefont {G.-S.}\ \bibnamefont {Yang}}, \bibinfo {author} {\bibfnamefont {Y.-G.}\ \bibnamefont {Yoon}}, \bibinfo {author} {\bibfnamefont {J.-H.}\ \bibnamefont {Kim}},\ and\ \bibinfo {author} {\bibfnamefont {J.}~\bibnamefont {Sung}},\ }\bibfield  {title} {\bibinfo {title} {Exchange symmetry, fluctuation-compressibility relation, and thermodynamic potentials of quantum liquids},\ }\href {https://doi.org/10.1103/PhysRevE.89.062131} {\bibfield  {journal} {\bibinfo  {journal} {Phys. Rev. E}\ }\textbf {\bibinfo {volume} {89}},\ \bibinfo {pages} {062131} (\bibinfo {year} {2014})}\BibitemShut {NoStop}%
\bibitem [{\citenamefont {Leff}(2015)}]{Le15}%
  \BibitemOpen
  \bibfield  {author} {\bibinfo {author} {\bibfnamefont {H.~S.}\ \bibnamefont {Leff}},\ }\bibfield  {title} {\bibinfo {title} {{Fluctuations in particle number for a photon gas}},\ }\href {https://doi.org/10.1119/1.4904322} {\bibfield  {journal} {\bibinfo  {journal} {Am. J. Phys.}\ }\textbf {\bibinfo {volume} {83}},\ \bibinfo {pages} {362} (\bibinfo {year} {2015})}\BibitemShut {NoStop}%
\bibitem [{\citenamefont {Kirkwood}\ and\ \citenamefont {Buff}(1951)}]{Ki51}%
  \BibitemOpen
  \bibfield  {author} {\bibinfo {author} {\bibfnamefont {J.~G.}\ \bibnamefont {Kirkwood}}\ and\ \bibinfo {author} {\bibfnamefont {F.~P.}\ \bibnamefont {Buff}},\ }\bibfield  {title} {\bibinfo {title} {{The Statistical Mechanical Theory of Solutions. I}},\ }\href {https://doi.org/10.1063/1.1748352} {\bibfield  {journal} {\bibinfo  {journal} {J. Chem. Phys.}\ }\textbf {\bibinfo {volume} {19}},\ \bibinfo {pages} {774} (\bibinfo {year} {1951})}\BibitemShut {NoStop}%
\bibitem [{\citenamefont {Das~Sarma}\ and\ \citenamefont {Hwang}(1996)}]{sarma1996dynamical}%
  \BibitemOpen
  \bibfield  {author} {\bibinfo {author} {\bibfnamefont {S.}~\bibnamefont {Das~Sarma}}\ and\ \bibinfo {author} {\bibfnamefont {E.~H.}\ \bibnamefont {Hwang}},\ }\bibfield  {title} {\bibinfo {title} {Dynamical response of a one-dimensional quantum-wire electron system},\ }\href {https://doi.org/10.1103/PhysRevB.54.1936} {\bibfield  {journal} {\bibinfo  {journal} {Phys. Rev. B}\ }\textbf {\bibinfo {volume} {54}},\ \bibinfo {pages} {1936} (\bibinfo {year} {1996})}\BibitemShut {NoStop}%
\bibitem [{\citenamefont {Imambekov}\ \emph {et~al.}(2012)\citenamefont {Imambekov}, \citenamefont {Schmidt},\ and\ \citenamefont {Glazman}}]{imambekov2012one}%
  \BibitemOpen
  \bibfield  {author} {\bibinfo {author} {\bibfnamefont {A.}~\bibnamefont {Imambekov}}, \bibinfo {author} {\bibfnamefont {T.~L.}\ \bibnamefont {Schmidt}},\ and\ \bibinfo {author} {\bibfnamefont {L.~I.}\ \bibnamefont {Glazman}},\ }\bibfield  {title} {\bibinfo {title} {{One-dimensional quantum liquids: Beyond the Luttinger liquid paradigm}},\ }\href {https://doi.org/10.1103/RevModPhys.84.1253} {\bibfield  {journal} {\bibinfo  {journal} {Rev. Mod. Phys.}\ }\textbf {\bibinfo {volume} {84}},\ \bibinfo {pages} {1253} (\bibinfo {year} {2012})}\BibitemShut {NoStop}%
\bibitem [{\citenamefont {Bruus}\ and\ \citenamefont {Flensberg}(2004)}]{Br04b}%
  \BibitemOpen
  \bibfield  {author} {\bibinfo {author} {\bibfnamefont {H.}~\bibnamefont {Bruus}}\ and\ \bibinfo {author} {\bibfnamefont {K.}~\bibnamefont {Flensberg}},\ }\href {https://books.google.com/books?id=zeaMBAAAQBAJ} {\emph {\bibinfo {title} {Many-Body Quantum Theory in Condensed Matter Physics: An Introduction}}},\ Oxford Graduate Texts\ (\bibinfo  {publisher} {OUP Oxford},\ \bibinfo {year} {2004})\BibitemShut {NoStop}%
\bibitem [{\citenamefont {Moroni}\ \emph {et~al.}(1992)\citenamefont {Moroni}, \citenamefont {Ceperley},\ and\ \citenamefont {Senatore}}]{Mo92}%
  \BibitemOpen
  \bibfield  {author} {\bibinfo {author} {\bibfnamefont {S.}~\bibnamefont {Moroni}}, \bibinfo {author} {\bibfnamefont {D.~M.}\ \bibnamefont {Ceperley}},\ and\ \bibinfo {author} {\bibfnamefont {G.}~\bibnamefont {Senatore}},\ }\bibfield  {title} {\bibinfo {title} {Static response from quantum {M}onte {C}arlo calculations},\ }\href {https://doi.org/10.1103/PhysRevLett.69.1837} {\bibfield  {journal} {\bibinfo  {journal} {Phys. Rev. Lett.}\ }\textbf {\bibinfo {volume} {69}},\ \bibinfo {pages} {1837} (\bibinfo {year} {1992})}\BibitemShut {NoStop}%
\bibitem [{\citenamefont {Gold}(1997)}]{Go97}%
  \BibitemOpen
  \bibfield  {author} {\bibinfo {author} {\bibfnamefont {A.}~\bibnamefont {Gold}},\ }\bibfield  {title} {\bibinfo {title} {The local-field correction for the interacting electron gas: many-body effects for unpolarized and polarized electrons},\ }\href {https://doi.org/10.1007/s002570050404} {\bibfield  {journal} {\bibinfo  {journal} {Z. Phys. B}\ }\textbf {\bibinfo {volume} {103}},\ \bibinfo {pages} {491} (\bibinfo {year} {1997})}\BibitemShut {NoStop}%
\bibitem [{\citenamefont {Lomba}\ \emph {et~al.}(2020)\citenamefont {Lomba}, \citenamefont {Weis}, \citenamefont {Guis\'andez},\ and\ \citenamefont {Torquato}}]{Lo20}%
  \BibitemOpen
  \bibfield  {author} {\bibinfo {author} {\bibfnamefont {E.}~\bibnamefont {Lomba}}, \bibinfo {author} {\bibfnamefont {J.-J.}\ \bibnamefont {Weis}}, \bibinfo {author} {\bibfnamefont {L.}~\bibnamefont {Guis\'andez}},\ and\ \bibinfo {author} {\bibfnamefont {S.}~\bibnamefont {Torquato}},\ }\bibfield  {title} {\bibinfo {title} {Minimal statistical-mechanical model for multihyperuniform patterns in avian retina},\ }\href {https://doi.org/10.1103/PhysRevE.102.012134} {\bibfield  {journal} {\bibinfo  {journal} {Phys. Rev. E}\ }\textbf {\bibinfo {volume} {102}},\ \bibinfo {pages} {012134} (\bibinfo {year} {2020})}\BibitemShut {NoStop}%
\bibitem [{\citenamefont {Zhao}\ \emph {et~al.}(2021)\citenamefont {Zhao}, \citenamefont {Li}, \citenamefont {Cao},\ and\ \citenamefont {Sun}}]{Zh21b}%
  \BibitemOpen
  \bibfield  {author} {\bibinfo {author} {\bibfnamefont {X.-Y.}\ \bibnamefont {Zhao}}, \bibinfo {author} {\bibfnamefont {L.-J.}\ \bibnamefont {Li}}, \bibinfo {author} {\bibfnamefont {L.}~\bibnamefont {Cao}},\ and\ \bibinfo {author} {\bibfnamefont {M.-J.}\ \bibnamefont {Sun}},\ }\bibfield  {title} {\bibinfo {title} {Bionic birdlike imaging using a multi-hyperuniform led array},\ }\href {https://doi.org/10.3390/s21124084} {\bibfield  {journal} {\bibinfo  {journal} {Sensors}\ }\textbf {\bibinfo {volume} {21}},\ \bibinfo {pages} {4084} (\bibinfo {year} {2021})}\BibitemShut {NoStop}%
\bibitem [{\citenamefont {Sakai}\ \emph {et~al.}(2022)\citenamefont {Sakai}, \citenamefont {Arita},\ and\ \citenamefont {Ohtsuki}}]{Sa22}%
  \BibitemOpen
  \bibfield  {author} {\bibinfo {author} {\bibfnamefont {S.}~\bibnamefont {Sakai}}, \bibinfo {author} {\bibfnamefont {R.}~\bibnamefont {Arita}},\ and\ \bibinfo {author} {\bibfnamefont {T.}~\bibnamefont {Ohtsuki}},\ }\bibfield  {title} {\bibinfo {title} {Quantum phase transition between hyperuniform density distributions},\ }\href {https://doi.org/10.1103/PhysRevResearch.4.033241} {\bibfield  {journal} {\bibinfo  {journal} {Phys. Rev. Res.}\ }\textbf {\bibinfo {volume} {4}},\ \bibinfo {pages} {033241} (\bibinfo {year} {2022})}\BibitemShut {NoStop}%
\bibitem [{\citenamefont {Cao}\ and\ \citenamefont {Voth}(1994)}]{Ca94}%
  \BibitemOpen
  \bibfield  {author} {\bibinfo {author} {\bibfnamefont {J.}~\bibnamefont {Cao}}\ and\ \bibinfo {author} {\bibfnamefont {G.~A.}\ \bibnamefont {Voth}},\ }\bibfield  {title} {\bibinfo {title} {{The formulation of quantum statistical mechanics based on the Feynman path centroid density. I. Equilibrium properties}},\ }\href {https://doi.org/10.1063/1.467175} {\bibfield  {journal} {\bibinfo  {journal} {J. Chem. Phys.}\ }\textbf {\bibinfo {volume} {100}},\ \bibinfo {pages} {5093} (\bibinfo {year} {1994})}\BibitemShut {NoStop}%
\bibitem [{\citenamefont {Bulutay}\ and\ \citenamefont {Tanatar}(2002)}]{Bu02}%
  \BibitemOpen
  \bibfield  {author} {\bibinfo {author} {\bibfnamefont {C.}~\bibnamefont {Bulutay}}\ and\ \bibinfo {author} {\bibfnamefont {B.}~\bibnamefont {Tanatar}},\ }\bibfield  {title} {\bibinfo {title} {Spin-dependent analysis of two-dimensional electron liquids},\ }\href {https://doi.org/10.1103/PhysRevB.65.195116} {\bibfield  {journal} {\bibinfo  {journal} {Phys. Rev. B}\ }\textbf {\bibinfo {volume} {65}},\ \bibinfo {pages} {195116} (\bibinfo {year} {2002})}\BibitemShut {NoStop}%
\bibitem [{\citenamefont {Craig}\ and\ \citenamefont {Manolopoulos}(2004)}]{Cr04}%
  \BibitemOpen
  \bibfield  {author} {\bibinfo {author} {\bibfnamefont {I.~R.}\ \bibnamefont {Craig}}\ and\ \bibinfo {author} {\bibfnamefont {D.~E.}\ \bibnamefont {Manolopoulos}},\ }\bibfield  {title} {\bibinfo {title} {{Quantum statistics and classical mechanics: Real time correlation functions from ring polymer molecular dynamics}},\ }\href {https://doi.org/10.1063/1.1777575} {\bibfield  {journal} {\bibinfo  {journal} {J. Chem. Phys.}\ }\textbf {\bibinfo {volume} {121}},\ \bibinfo {pages} {3368} (\bibinfo {year} {2004})}\BibitemShut {NoStop}%
\bibitem [{\citenamefont {Smith}\ \emph {et~al.}(2015)\citenamefont {Smith}, \citenamefont {Poulsen}, \citenamefont {Nyman},\ and\ \citenamefont {Rossky}}]{Sm15}%
  \BibitemOpen
  \bibfield  {author} {\bibinfo {author} {\bibfnamefont {K.~K.~G.}\ \bibnamefont {Smith}}, \bibinfo {author} {\bibfnamefont {J.~A.}\ \bibnamefont {Poulsen}}, \bibinfo {author} {\bibfnamefont {G.}~\bibnamefont {Nyman}},\ and\ \bibinfo {author} {\bibfnamefont {P.~J.}\ \bibnamefont {Rossky}},\ }\bibfield  {title} {\bibinfo {title} {{A new class of ensemble conserving algorithms for approximate quantum dynamics: Theoretical formulation and model problems}},\ }\href {https://doi.org/10.1063/1.4922887} {\bibfield  {journal} {\bibinfo  {journal} {J. Chem. Phys.}\ }\textbf {\bibinfo {volume} {142}},\ \bibinfo {pages} {244112} (\bibinfo {year} {2015})}\BibitemShut {NoStop}%
\bibitem [{\citenamefont {Zhang}\ and\ \citenamefont {Torquato}(2020)}]{Zh20}%
  \BibitemOpen
  \bibfield  {author} {\bibinfo {author} {\bibfnamefont {G.}~\bibnamefont {Zhang}}\ and\ \bibinfo {author} {\bibfnamefont {S.}~\bibnamefont {Torquato}},\ }\bibfield  {title} {\bibinfo {title} {Realizable hyperuniform and nonhyperuniform particle configurations with targeted spectral functions via effective pair interactions},\ }\href {https://doi.org/10.1103/PhysRevE.101.032124} {\bibfield  {journal} {\bibinfo  {journal} {Phys. Rev. E}\ }\textbf {\bibinfo {volume} {101}},\ \bibinfo {pages} {032124} (\bibinfo {year} {2020})}\BibitemShut {NoStop}%
\bibitem [{\citenamefont {Liu}\ and\ \citenamefont {Wu}(2014)}]{Li14b}%
  \BibitemOpen
  \bibfield  {author} {\bibinfo {author} {\bibfnamefont {Y.}~\bibnamefont {Liu}}\ and\ \bibinfo {author} {\bibfnamefont {J.}~\bibnamefont {Wu}},\ }\bibfield  {title} {\bibinfo {title} {{An improved classical mapping method for homogeneous electron gases at finite temperature}},\ }\href {https://doi.org/10.1063/1.4892587} {\bibfield  {journal} {\bibinfo  {journal} {J. Chem. Phys.}\ }\textbf {\bibinfo {volume} {141}},\ \bibinfo {pages} {064115} (\bibinfo {year} {2014})}\BibitemShut {NoStop}%
\bibitem [{\citenamefont {Gunnarsson}\ \emph {et~al.}(1979)\citenamefont {Gunnarsson}, \citenamefont {Jonson},\ and\ \citenamefont {Lundqvist}}]{Gu79}%
  \BibitemOpen
  \bibfield  {author} {\bibinfo {author} {\bibfnamefont {O.}~\bibnamefont {Gunnarsson}}, \bibinfo {author} {\bibfnamefont {M.}~\bibnamefont {Jonson}},\ and\ \bibinfo {author} {\bibfnamefont {B.~I.}\ \bibnamefont {Lundqvist}},\ }\bibfield  {title} {\bibinfo {title} {Descriptions of exchange and correlation effects in inhomogeneous electron systems},\ }\href {https://doi.org/10.1103/PhysRevB.20.3136} {\bibfield  {journal} {\bibinfo  {journal} {Phys. Rev. B}\ }\textbf {\bibinfo {volume} {20}},\ \bibinfo {pages} {3136} (\bibinfo {year} {1979})}\BibitemShut {NoStop}%
\bibitem [{\citenamefont {Perdew}\ and\ \citenamefont {Zunger}(1981)}]{Pe81}%
  \BibitemOpen
  \bibfield  {author} {\bibinfo {author} {\bibfnamefont {J.~P.}\ \bibnamefont {Perdew}}\ and\ \bibinfo {author} {\bibfnamefont {A.}~\bibnamefont {Zunger}},\ }\bibfield  {title} {\bibinfo {title} {Self-interaction correction to density-functional approximations for many-electron systems},\ }\href {https://doi.org/10.1103/PhysRevB.23.5048} {\bibfield  {journal} {\bibinfo  {journal} {Phys. Rev. B}\ }\textbf {\bibinfo {volume} {23}},\ \bibinfo {pages} {5048} (\bibinfo {year} {1981})}\BibitemShut {NoStop}%
\bibitem [{\citenamefont {Wang}\ \emph {et~al.}(2020)\citenamefont {Wang}, \citenamefont {Stillinger},\ and\ \citenamefont {Torquato}}]{Wa20}%
  \BibitemOpen
  \bibfield  {author} {\bibinfo {author} {\bibfnamefont {H.}~\bibnamefont {Wang}}, \bibinfo {author} {\bibfnamefont {F.~H.}\ \bibnamefont {Stillinger}},\ and\ \bibinfo {author} {\bibfnamefont {S.}~\bibnamefont {Torquato}},\ }\bibfield  {title} {\bibinfo {title} {Sensitivity of pair statistics on pair potentials in many-body systems},\ }\href {https://doi.org/10.1063/5.0021475} {\bibfield  {journal} {\bibinfo  {journal} {J. Chem. Phys.}\ }\textbf {\bibinfo {volume} {153}},\ \bibinfo {pages} {124106} (\bibinfo {year} {2020})}\BibitemShut {NoStop}%
\bibitem [{\citenamefont {Filippov}(2023)}]{Fi23}%
  \BibitemOpen
  \bibfield  {author} {\bibinfo {author} {\bibfnamefont {A.}~\bibnamefont {Filippov}},\ }\bibfield  {title} {\bibinfo {title} {Thermodynamic stability of a multicomponent non-ideal plasma},\ }\href {https://doi.org/10.1134/S1063780X22600967} {\bibfield  {journal} {\bibinfo  {journal} {Plasma Phys. Rep.}\ }\textbf {\bibinfo {volume} {49}},\ \bibinfo {pages} {49} (\bibinfo {year} {2023})}\BibitemShut {NoStop}%
\bibitem [{\citenamefont {Gann}\ \emph {et~al.}(1979)\citenamefont {Gann}, \citenamefont {Chakravarty},\ and\ \citenamefont {Chester}}]{Ga79}%
  \BibitemOpen
  \bibfield  {author} {\bibinfo {author} {\bibfnamefont {R.~C.}\ \bibnamefont {Gann}}, \bibinfo {author} {\bibfnamefont {S.}~\bibnamefont {Chakravarty}},\ and\ \bibinfo {author} {\bibfnamefont {G.~V.}\ \bibnamefont {Chester}},\ }\bibfield  {title} {\bibinfo {title} {Monte {C}arlo simulation of the classical two-dimensional one-component plasma},\ }\href {https://doi.org/10.1103/PhysRevB.20.326} {\bibfield  {journal} {\bibinfo  {journal} {Phys. Rev. B}\ }\textbf {\bibinfo {volume} {20}},\ \bibinfo {pages} {326} (\bibinfo {year} {1979})}\BibitemShut {NoStop}%
\bibitem [{\citenamefont {Ewald}(1921)}]{Ew21}%
  \BibitemOpen
  \bibfield  {author} {\bibinfo {author} {\bibfnamefont {P.~P.}\ \bibnamefont {Ewald}},\ }\bibfield  {title} {\bibinfo {title} {Die berechnung optischer und elektrostatischer gitterpotentiale},\ }\href {https://doi.org/https://doi.org/10.1002/andp.19213690304} {\bibfield  {journal} {\bibinfo  {journal} {Ann. Phys.}\ }\textbf {\bibinfo {volume} {369}},\ \bibinfo {pages} {253} (\bibinfo {year} {1921})}\BibitemShut {NoStop}%
\bibitem [{\citenamefont {Liu}\ and\ \citenamefont {Nocedal}(1989)}]{Liu89}%
  \BibitemOpen
  \bibfield  {author} {\bibinfo {author} {\bibfnamefont {D.~C.}\ \bibnamefont {Liu}}\ and\ \bibinfo {author} {\bibfnamefont {J.}~\bibnamefont {Nocedal}},\ }\bibfield  {title} {\bibinfo {title} {On the limited memory {BFGS} method for large scale optimization},\ }\href {https://doi.org/https://doi.org/10.1007/BF01589116} {\bibfield  {journal} {\bibinfo  {journal} {Math. Program.}\ }\textbf {\bibinfo {volume} {45}},\ \bibinfo {pages} {503} (\bibinfo {year} {1989})}\BibitemShut {NoStop}%
\bibitem [{\citenamefont {Wang}\ \emph {et~al.}(2022)\citenamefont {Wang}, \citenamefont {Stillinger},\ and\ \citenamefont {Torquato}}]{Wa22b}%
  \BibitemOpen
  \bibfield  {author} {\bibinfo {author} {\bibfnamefont {H.}~\bibnamefont {Wang}}, \bibinfo {author} {\bibfnamefont {F.~H.}\ \bibnamefont {Stillinger}},\ and\ \bibinfo {author} {\bibfnamefont {S.}~\bibnamefont {Torquato}},\ }\bibfield  {title} {\bibinfo {title} {{Realizability of iso-g$_2$ processes via effective pair interactions}},\ }\href {https://doi.org/10.1063/5.0130679} {\bibfield  {journal} {\bibinfo  {journal} {J. Chem. Phys.}\ }\textbf {\bibinfo {volume} {157}},\ \bibinfo {pages} {224106} (\bibinfo {year} {2022})}\BibitemShut {NoStop}%
\bibitem [{\citenamefont {Wang}\ and\ \citenamefont {Torquato}(2023)}]{Wa23}%
  \BibitemOpen
  \bibfield  {author} {\bibinfo {author} {\bibfnamefont {H.}~\bibnamefont {Wang}}\ and\ \bibinfo {author} {\bibfnamefont {S.}~\bibnamefont {Torquato}},\ }\bibfield  {title} {\bibinfo {title} {Equilibrium states corresponding to targeted hyperuniform nonequilibrium pair statistics},\ }\href {https://doi.org/10.1039/D2SM01294D} {\bibfield  {journal} {\bibinfo  {journal} {Soft Matter}\ }\textbf {\bibinfo {volume} {19}},\ \bibinfo {pages} {550} (\bibinfo {year} {2023})}\BibitemShut {NoStop}%
\bibitem [{\citenamefont {Norgaard}\ \emph {et~al.}(2008)\citenamefont {Norgaard}, \citenamefont {Ferkinghoff-Borg},\ and\ \citenamefont {Lindorff-Larsen}}]{No08b}%
  \BibitemOpen
  \bibfield  {author} {\bibinfo {author} {\bibfnamefont {A.~B.}\ \bibnamefont {Norgaard}}, \bibinfo {author} {\bibfnamefont {J.}~\bibnamefont {Ferkinghoff-Borg}},\ and\ \bibinfo {author} {\bibfnamefont {K.}~\bibnamefont {Lindorff-Larsen}},\ }\bibfield  {title} {\bibinfo {title} {Experimental parameterization of an energy function for the simulation of unfolded proteins},\ }\href {https://doi.org/10.1529/biophysj.107.108241} {\bibfield  {journal} {\bibinfo  {journal} {Biophys. J.}\ }\textbf {\bibinfo {volume} {94}},\ \bibinfo {pages} {182} (\bibinfo {year} {2008})}\BibitemShut {NoStop}%
\bibitem [{\citenamefont {Lenard}(1975)}]{Le75a}%
  \BibitemOpen
  \bibfield  {author} {\bibinfo {author} {\bibfnamefont {A.}~\bibnamefont {Lenard}},\ }\bibfield  {title} {\bibinfo {title} {States of classical statistical mechanical systems of infinitely many particles {I.}},\ }\href {https://doi.org/10.1007/BF00251601} {\bibfield  {journal} {\bibinfo  {journal} {Arch. Rational Mech. Anal.}\ }\textbf {\bibinfo {volume} {59}},\ \bibinfo {pages} {219} (\bibinfo {year} {1975})}\BibitemShut {NoStop}%
\bibitem [{\citenamefont {Crawford}\ \emph {et~al.}(2003)\citenamefont {Crawford}, \citenamefont {Torquato},\ and\ \citenamefont {Stillinger}}]{Cr03}%
  \BibitemOpen
  \bibfield  {author} {\bibinfo {author} {\bibfnamefont {J.~R.}\ \bibnamefont {Crawford}}, \bibinfo {author} {\bibfnamefont {S.}~\bibnamefont {Torquato}},\ and\ \bibinfo {author} {\bibfnamefont {F.~H.}\ \bibnamefont {Stillinger}},\ }\bibfield  {title} {\bibinfo {title} {Aspects of correlation function realizability},\ }\href {https://doi.org/10.1063/1.1606678} {\bibfield  {journal} {\bibinfo  {journal} {J. Chem. Phys.}\ }\textbf {\bibinfo {volume} {119}},\ \bibinfo {pages} {7065} (\bibinfo {year} {2003})}\BibitemShut {NoStop}%
\bibitem [{\citenamefont {Costin}\ and\ \citenamefont {Lebowitz}(2004)}]{Cos04}%
  \BibitemOpen
  \bibfield  {author} {\bibinfo {author} {\bibfnamefont {O.}~\bibnamefont {Costin}}\ and\ \bibinfo {author} {\bibfnamefont {J.}~\bibnamefont {Lebowitz}},\ }\bibfield  {title} {\bibinfo {title} {On the construction of particle distributions with specified single and pair densities},\ }\href {https://doi.org/10.1021/jp047793m} {\bibfield  {journal} {\bibinfo  {journal} {J. Phys. Chem. B.}\ }\textbf {\bibinfo {volume} {108}},\ \bibinfo {pages} {19614} (\bibinfo {year} {2004})}\BibitemShut {NoStop}%
\bibitem [{\citenamefont {Torquato}\ and\ \citenamefont {Stillinger}(2006)}]{To06b}%
  \BibitemOpen
  \bibfield  {author} {\bibinfo {author} {\bibfnamefont {S.}~\bibnamefont {Torquato}}\ and\ \bibinfo {author} {\bibfnamefont {F.~H.}\ \bibnamefont {Stillinger}},\ }\bibfield  {title} {\bibinfo {title} {New conjectural lower bounds on the optimal density of sphere packings},\ }\href {https://doi.org/10.1080/10586458.2006.10128964} {\bibfield  {journal} {\bibinfo  {journal} {Experimental Math.}\ }\textbf {\bibinfo {volume} {15}},\ \bibinfo {pages} {307} (\bibinfo {year} {2006})}\BibitemShut {NoStop}%
\bibitem [{\citenamefont {Henderson}(1974)}]{He74}%
  \BibitemOpen
  \bibfield  {author} {\bibinfo {author} {\bibfnamefont {R.~L.}\ \bibnamefont {Henderson}},\ }\bibfield  {title} {\bibinfo {title} {{A uniqueness theorem for fluid pair correlation functions}},\ }\href {https://doi.org/10.1016/0375-9601(74)90847-0} {\bibfield  {journal} {\bibinfo  {journal} {Phys. Lett. A}\ }\textbf {\bibinfo {volume} {49}},\ \bibinfo {pages} {197} (\bibinfo {year} {1974})}\BibitemShut {NoStop}%
\bibitem [{\citenamefont {Batten}\ \emph {et~al.}(2008)\citenamefont {Batten}, \citenamefont {Stillinger},\ and\ \citenamefont {Torquato}}]{Ba08}%
  \BibitemOpen
  \bibfield  {author} {\bibinfo {author} {\bibfnamefont {R.~D.}\ \bibnamefont {Batten}}, \bibinfo {author} {\bibfnamefont {F.~H.}\ \bibnamefont {Stillinger}},\ and\ \bibinfo {author} {\bibfnamefont {S.}~\bibnamefont {Torquato}},\ }\bibfield  {title} {\bibinfo {title} {Classical disordered ground states: {S}uper-ideal gases, and stealth and equi-luminous materials},\ }\href {https://doi.org/10.1063/1.2961314} {\bibfield  {journal} {\bibinfo  {journal} {J. Appl. Phys.}\ }\textbf {\bibinfo {volume} {104}},\ \bibinfo {pages} {033504} (\bibinfo {year} {2008})}\BibitemShut {NoStop}%
\bibitem [{\citenamefont {{Zhang}}\ \emph {et~al.}(2017)\citenamefont {{Zhang}}, \citenamefont {{Stillinger}},\ and\ \citenamefont {{Torquato}}}]{Zh17b}%
  \BibitemOpen
  \bibfield  {author} {\bibinfo {author} {\bibfnamefont {G.}~\bibnamefont {{Zhang}}}, \bibinfo {author} {\bibfnamefont {F.~H.}\ \bibnamefont {{Stillinger}}},\ and\ \bibinfo {author} {\bibfnamefont {S.}~\bibnamefont {{Torquato}}},\ }\bibfield  {title} {\bibinfo {title} {Classical many-particle systems with unique disordered ground states},\ }\href {https://doi.org/10.1103/PhysRevE.96.042146} {\bibfield  {journal} {\bibinfo  {journal} {Phys. Rev. E}\ }\textbf {\bibinfo {volume} {96}},\ \bibinfo {pages} {042146} (\bibinfo {year} {2017})}\BibitemShut {NoStop}%
\bibitem [{\citenamefont {Song}\ \emph {et~al.}(2012)\citenamefont {Song}, \citenamefont {Rachel}, \citenamefont {Flindt}, \citenamefont {Klich}, \citenamefont {Laflorencie},\ and\ \citenamefont {Le~Hur}}]{PhysRevB.85.035409}%
  \BibitemOpen
  \bibfield  {author} {\bibinfo {author} {\bibfnamefont {H.~F.}\ \bibnamefont {Song}}, \bibinfo {author} {\bibfnamefont {S.}~\bibnamefont {Rachel}}, \bibinfo {author} {\bibfnamefont {C.}~\bibnamefont {Flindt}}, \bibinfo {author} {\bibfnamefont {I.}~\bibnamefont {Klich}}, \bibinfo {author} {\bibfnamefont {N.}~\bibnamefont {Laflorencie}},\ and\ \bibinfo {author} {\bibfnamefont {K.}~\bibnamefont {Le~Hur}},\ }\bibfield  {title} {\bibinfo {title} {Bipartite fluctuations as a probe of many-body entanglement},\ }\href {https://doi.org/10.1103/PhysRevB.85.035409} {\bibfield  {journal} {\bibinfo  {journal} {Phys. Rev. B}\ }\textbf {\bibinfo {volume} {85}},\ \bibinfo {pages} {035409} (\bibinfo {year} {2012})}\BibitemShut {NoStop}%
\bibitem [{\citenamefont {Yatsenko}\ \emph {et~al.}(2004)\citenamefont {Yatsenko}, \citenamefont {Sambriski}, \citenamefont {Nemirovskaya},\ and\ \citenamefont {Guenza}}]{Ya04}%
  \BibitemOpen
  \bibfield  {author} {\bibinfo {author} {\bibfnamefont {G.}~\bibnamefont {Yatsenko}}, \bibinfo {author} {\bibfnamefont {E.~J.}\ \bibnamefont {Sambriski}}, \bibinfo {author} {\bibfnamefont {M.~A.}\ \bibnamefont {Nemirovskaya}},\ and\ \bibinfo {author} {\bibfnamefont {M.}~\bibnamefont {Guenza}},\ }\bibfield  {title} {\bibinfo {title} {{Analytical Soft-Core Potentials for Macromolecular Fluids and Mixtures}},\ }\href {https://doi.org/10.1103/PhysRevLett.93.257803} {\bibfield  {journal} {\bibinfo  {journal} {Phys. Rev. Lett.}\ }\textbf {\bibinfo {volume} {93}},\ \bibinfo {pages} {257803} (\bibinfo {year} {2004})}\BibitemShut {NoStop}%
\bibitem [{\citenamefont {Capone}\ \emph {et~al.}(2012)\citenamefont {Capone}, \citenamefont {Coluzza}, \citenamefont {LoVerso}, \citenamefont {Likos},\ and\ \citenamefont {Blaak}}]{Ca12}%
  \BibitemOpen
  \bibfield  {author} {\bibinfo {author} {\bibfnamefont {B.}~\bibnamefont {Capone}}, \bibinfo {author} {\bibfnamefont {I.}~\bibnamefont {Coluzza}}, \bibinfo {author} {\bibfnamefont {F.}~\bibnamefont {LoVerso}}, \bibinfo {author} {\bibfnamefont {C.~N.}\ \bibnamefont {Likos}},\ and\ \bibinfo {author} {\bibfnamefont {R.}~\bibnamefont {Blaak}},\ }\bibfield  {title} {\bibinfo {title} {Telechelic star polymers as self-assembling units from the molecular to the macroscopic scale},\ }\href {https://doi.org/10.1103/PhysRevLett.109.238301} {\bibfield  {journal} {\bibinfo  {journal} {Phys. Rev. Lett.}\ }\textbf {\bibinfo {volume} {109}},\ \bibinfo {pages} {238301} (\bibinfo {year} {2012})}\BibitemShut {NoStop}%
\bibitem [{\citenamefont {Chou}\ and\ \citenamefont {Das~Sarma}(2020)}]{chou2020nonmonotonic}%
  \BibitemOpen
  \bibfield  {author} {\bibinfo {author} {\bibfnamefont {Y.-Z.}\ \bibnamefont {Chou}}\ and\ \bibinfo {author} {\bibfnamefont {S.}~\bibnamefont {Das~Sarma}},\ }\bibfield  {title} {\bibinfo {title} {{Nonmonotonic plasmon dispersion in strongly interacting Coulomb Luttinger liquids}},\ }\href {https://doi.org/10.1103/PhysRevB.101.075430} {\bibfield  {journal} {\bibinfo  {journal} {Phys. Rev. B}\ }\textbf {\bibinfo {volume} {101}},\ \bibinfo {pages} {075430} (\bibinfo {year} {2020})}\BibitemShut {NoStop}%
\bibitem [{\citenamefont {Moudgil}\ \emph {et~al.}(1995)\citenamefont {Moudgil}, \citenamefont {Ahluwalia},\ and\ \citenamefont {Pathak}}]{Mo95}%
  \BibitemOpen
  \bibfield  {author} {\bibinfo {author} {\bibfnamefont {R.~K.}\ \bibnamefont {Moudgil}}, \bibinfo {author} {\bibfnamefont {P.~K.}\ \bibnamefont {Ahluwalia}},\ and\ \bibinfo {author} {\bibfnamefont {K.~N.}\ \bibnamefont {Pathak}},\ }\bibfield  {title} {\bibinfo {title} {Spin correlations in a two-dimensional electron gas},\ }\href {https://doi.org/10.1103/PhysRevB.51.1575} {\bibfield  {journal} {\bibinfo  {journal} {Phys. Rev. B}\ }\textbf {\bibinfo {volume} {51}},\ \bibinfo {pages} {1575} (\bibinfo {year} {1995})}\BibitemShut {NoStop}%
\end{thebibliography}
%

\end{document}